\documentclass[11pt]{article}
\usepackage[cp850]{inputenc}
\usepackage{colortbl}
\usepackage{epsf,psfig}
\usepackage{amsfonts,amssymb,latexsym}
\newcounter{fig}

\newcounter{gif}

\newenvironment{belaufz}{\begin{list}{}
{\usecounter{fig} \setlength{\leftmargin}{0.5cm} \setlength{\labelwidth}{-0.2cm} \setlength{\labelsep}{0.2cm} \setlength{\itemsep}{0ex plus0.2ex}
}}{\end{list}}
\newenvironment{belaufz2}{\begin{list}{}
{\usecounter{fig} \setlength{\leftmargin}{1.0cm} \setlength{\labelwidth}{0.3cm} \setlength{\labelsep}{0.3cm} \setlength{\itemsep}{0ex plus0.0ex}
}}{\end{list}}

\newcommand{\beq}{\begin{eqnarray*}}
\newcommand{\eeq}{\end{eqnarray*}}
\newcommand{\beqn}{\begin{eqnarray}}
\newcommand{\eeqn}{\end{eqnarray}}
\newcommand{\mref}[1]{$(\ref{#1})$}  % reference for mathematical formul\ae
\renewcommand{\cellcolor}[3]{\multicolumn{1}{>{\columncolor[#1]{#2}}c}{#3}}
\newcommand{\Z}{\mathbb{Z}}

\newcommand{\id}{1\!\!1}

\def\sqr#1#2{{\vcenter{\vbox{\hrule height.#2pt          % square
            \hbox{\vrule width.#2pt height#1pt \kern#1pt
                  \vrule width.#2pt}\hrule height.#2pt}}}}

\newcommand{\otimesf}{\otimes_{\mbox{\footnotesize f}}}
\newcommand{\hl}[1]{\textcolor{blue}{#1 ?} \,}
\newcommand{\hls}[1]{\textcolor{blue}{#1}}
\begin{document}

\title{Virasoro representations and fusion for general augmented minimal models}
\author{Holger Eberle\footnote{email: {\tt eberle@th.physik.uni-bonn.de}} \hspace{0.1cm}
and \hspace{-0.1cm} Michael Flohr\footnote{email: {\tt flohr@th.physik.uni-bonn.de}}\vspace{0.3cm} \\
Physikalisches Institut der Universit\"at Bonn\\ 
Nu\ss allee 12, 53115 Bonn, Germany}

\date{}

\maketitle

\vspace{-8cm}
\begin{flushright}
BONN-TH-2006-03\\
hep-th/0604097
\end{flushright}
\vspace{+7cm}

%\addcontentsline{toc}{section}{\protect\numberline{4}{...}           % Hinzuf\"ugen ins Inhaltsverzeichnis

%%%%%%%%%%%%%%%%%%%%%%%%%%%%%%%%%%%%%%%%%%%%%

\begin{abstract}
In this paper we present explicit results for the fusion
of irreducible and higher rank representations in two
logarithmically conformal models, the augmented $c_{2,3} = 0$
model as well as the augmented Yang--Lee model at $c_{2,5} = -22/5$.
We analyse their spectrum of representations
which is consistent with the symmetry and associati\-vity
of the fusion algebra. We also describe the first few higher rank
representations in detail.
In particular, we present the first examples of consistent
rank $3$ indecomposable representations and describe their
embedding structure.

Knowing these two generic models we also conjecture
the general representation content and fusion rules
for general augmented $c_{p,q}$ models.
\end{abstract}

%%%%%%%%%%%%%%%%%%%%%%%%%%%%%%%%%%%%%%%%%%%%%

\newpage
\tableofcontents

%%%%%%%%%%%%%%%%%%%%%%%%%%%%%%%%%%%%%%%%%%%%%
%%%%%%%%%%%%%%%%%%%%%%%%%%%%%%%%%%%%%%%%%%%%%
%%%%%%%%%%%%%%%%%%%%%%%%%%%%%%%%%%%%%%%%%%%%%

\section{Introduction}

Logarithmic conformal field theories (logarithmic CFTs)
have attracted quite a lot of attention in recent years.
These theories are a generalisation of standard CFT
which also allows for indecomposable action of the
Virasoro modes. There are already quite a number of
applications in such different fields as
statistical physics (e.g.\ \cite{Sal92,GFN97,Rue02,PRG01}),
string theory (e.g.\ \cite{GM01,Bakas02,LLM03}) and
Seiberg Witten theory (e.g.\ \cite{Flohr04}) which 
necessarily incorporate this generalisation of CFT.
See \cite{Flohr01,Gab01,Kawai02,Nich02,FG05} for an introduction to
the field and a more complete list of applications.
Nevertheless, studying logarithmic CFTs has only just
begun because we still know only few
logarithmic models explicitly and the efforts to disentangle
the general structure prove to be much
more complicated and tedious as in ordinary CFT (see e.g.\ 
\cite{Flohr97,Flohr00,Flohr01b,FK05,Mog00,Mog02}).
This paper hopes to give a considerable contribution to
this ongoing effort as it presents and explicitly discusses
the first examples of models with a much more complicated
indecomposable structure up to rank $3$.

The up to now most prominent examples of logarithmic CFTs
have emerged from studying a specific series of the
so-called minimal models in CFT, the $c_{p,1}$ models
\cite{GK96,GK96b,Roh96,GK98,Flohr95,Flohr96,FG05}.
As standard minimal models these actually emerge to be
trivial as they provide zero representation content.
On the other hand, if one takes into account representations
corresponding to an enlarged Kac table one encounters
non-trivial models which include representations with
indecomposable structure. This is the reason why we like
to call these ``augmented $c_{p,1}$ models''.
In the point of view of the Virasoro representation
theory this augmentation actually comprises representations
with weights in the whole infinite Kac table. On the other
hand it has been found that these theories exhibit
an enlarged triplet ${\cal W}$-algebra \cite{Kau90}.
Wrt to this enlarged symmetry
algebra the representation content of this theory
can be composed into a finite set of representations
associated to a finite standard cell of the Kac table.
In comparison to the original minimal model
of central charge $c_{p,1}$ this standard cell
is enlarged by a factor of $3$ on either side
as if it described $c_{3p,3}$.

But there is no reason not to consider a similar augmentation 
in the Kac table of the general $c_{p,q}$ minimal models.
This paper is devoted to the study of these general augmented
$c_{p,q}$ models. We will concentrate on the 
Virasoro representation theory here and, thus, regard
an augmentation to the whole infinite Kac table.
The question whether these models also exhibit
an enlarged ${\cal W}$-algebra has not been answered
yet, although we hope to use the findings of this paper
to settle this question soon \cite{EFN} 
(see also \cite{CF05,FGST05a,FGST05}).
The particular example of the augmented $c_{2,3}=0$
model has already been studied in \cite{FM05,EF05},
but up to this paper the precise structure
of the Virasoro representations has been out
of reach.

There have also been studies of Jordan cells with rank
higher than $2$ on the level of correlation functions and
Ward identities in the CFT literature 
\cite{CKT95,MS96,GK97,Flohr97,Flohr01b,FK05,Ras04a,Ras04b}. 
However, the models and their representations discussed in this paper
are the first ones where we can see explicitly
how a higher rank structure appears while generating
a representation as a Virasoro module from some groundstates. 

We want to explore the full spectrum of 
these augmented $c_{p,q}$ models
by successively fusing representations which we already
know we have to take into the spectrum. 
The concept of the fusion product lies at the heart
of conformal representation theory and has been
subject to many thorough mathematical 
studies (see e.g.\ \cite{FFK89,FFK90,Gab93,Lep05,FRS06}). It governs
the representation theoretic aspects of the operator
product expansion and, hence, puts severe constraints
on all $n$-point-functions in CFT. It follows
that the fusion product actually dictates 
which set of representations of the  Virasoro
algebra at a certain conformal weight
can be combined into a consistent CFT model.
Hence, the successive application of the fusion product
will actually lead us to the whole consistent representation
space of the augmented $c_{p,q}$ models.

The algorithm which we use to compute these fusion products
relies on the work of \cite{Nahm94,GK96}.
In \cite{Nahm94} W.\ Nahm showed that the main
information characterising a representation
can be found in a small quotient space of
this representation, called the special subspace,
which is finite for the large class of so-called 
quasirational CFTs. He actually proved
that the fusion of quasirational representations
leads again to a finite number of quasirational
representations. In \cite{GK96} M.\ Gaberdiel and H.\ Kausch
used the Nahm algorithm of the proof in
\cite{Nahm94} to propose a procedure how to efficiently
calculate a fusion pro\-duct of two quasirational representations.
This procedure was successfully applied to the
augmented $c_{p,1}$ models in \cite{GK96}.
In this paper we present results of fusion
products which have been calculated based on 
the implementation of the Nahm algorithm as well
as this procedure for the general $c_{p,q}$ models.

In section \ref{nahm} we give a short review of this
Nahm algorithm and the resulting procedure
how to calculate a fusion product. We also
comment on our specific computer implementation.
In section \ref{rep_intro} we shortly introduce
minimal models as well as their augmented generalisation.
In section \ref{discussion_c0} we then discuss the
easiest general augmented $c_{p,q}$ model which is not
contained in the $c_{p,1}$ model series, the
$c_{2,3} = 0$ model. This model exhibits a much 
more complicated indecomposable structure up to rank $3$
in comparison to the $c_{p,1}$ models. We describe
a number of rank $2$ and rank $3$ representations
explicitly and also discuss the representation
content which is consistent with the fusion product.
In section \ref{discussion_YL} we discuss
a second example of general augmented $c_{p,q}$ models,
the augmented Yang-Lee model at $c_{2,5} = -22/5$.
We actually rediscover all crucial features
observed in the $c_{2,3} = 0$ model. These two examples
substantiate our conjecture of the general
representation content and fusion rules for
general augmented $c_{p,q}$ models given in
section \ref{general}. In section \ref{conclusion}
we conclude and also give a short outlook on the
expected implications of these results. 
In appendix \ref{explicit_L0} we give the basis of
states which brings $L_0$ into Jordan diagonal
form for the rank $3$ representation ${\cal R}^{(3)} (0,1,2,5)$.
Finally, the 
appendices \ref{fusion_c0} and \ref{fusion_cminus22over5}
contain the explicit results for the
fusion product calculation for both examples
$c_{2,3} = 0$ and $c_{2,5} = -22/5$.

%%%%%%%%%%%%%%%%%%%%%%%%%%%%%%%%%%%%%%%%%%%%%
%%%%%%%%%%%%%%%%%%%%%%%%%%%%%%%%%%%%%%%%%%%%%
%%%%%%%%%%%%%%%%%%%%%%%%%%%%%%%%%%%%%%%%%%%%%

\section{How to calculate fusion products\label{nahm}}

Our calculations of fusion products are based on an algorithm first developed by
W.\ Nahm \cite{Nahm94} to prove that the fusion of quasirational representations
contains only finitely many quasirational subrepresentations and,
hence, that the category of quasirational representations is stable.
In \cite{GK96} this procedure was generalised
and it was shown how one can use this algorithm to get
(computationally usually sufficient) constraints
to fix the field content of the fusion product at a given level.
Although the presentation of the algorithm is already very precise and thorough
in \cite{GK96} we nevertheless want to give a short summary here
in order to make this paper self-contained. We will then present the 
specific properties of our particular implementation.

%%%%%%%%%%%%

\subsection{The Nahm algorithm}

The nice and short presentation of the Nahm algorithm in \cite{GK96}
relies on the coproduct formula. For a holomorphic field of conformal
weight $h$ and mode expansion
\beq
S(w) &=& \sum_{l\in \Z +h} w^{l-h} S_{-l}
\eeq
it is given by \cite{Gab93,Gab94}
\beq
\Delta_{z,\zeta} (S_n) &=& \sum_{m=1-h}^n \left( n+h-1 \atop m+h-1 \right) \; \zeta^{n-m} \; (S_m \otimes \id)  \nonumber \\
&& \qquad \qquad + \; \epsilon \sum_{l=1-h}^n \left( n+h-1 \atop l+h-1 \right) \; z^{n-l} \; (\id \otimes S_l) \nonumber \\
&=& \tilde{\Delta}_{z,\zeta} (S_n)  \quad \forall \, n \le 1-h \nonumber \\
\Delta_{z,\zeta} (S_{-n}) &=& \sum_{m=1-h}^{\infty} \left( n+m-1 \atop m+h-1 \right) \; (-1)^{m+h-1} \, \zeta^{-(n+m)} \; (S_m \otimes \id)  \nonumber \\
&& \qquad \qquad + \; \epsilon \sum_{l=n}^{\infty} \left( l-h \atop n-h \right) \; (-z)^{l-n} \; (\id \otimes S_{-l})  \quad \forall \, n \le h \nonumber \\
\tilde{\Delta}_{z,\zeta} (S_{-n}) &=& \sum_{m=n}^{\infty} \left( m-h \atop n-h \right) \; (-\zeta)^{m-n} \; (S_{-m} \otimes \id)  \nonumber \\
&& \qquad \qquad + \; \epsilon \sum_{l=1-h}^{\infty} \left( n+l-1 \atop l+h-1 \right) \; (-1)^{l+h-1} \; z^{-(n+l)} \; (\id \otimes S_l) \quad \forall \, n \le h \; ,
\eeq
where $\epsilon = -1$ if both $S_m$ and the first field in the tensor product 
it is applied to are fermionic and $\epsilon = +1$ otherwise.
Furthermore, $z$ and $\zeta$ are the positions of the two fields of the
tensor product which this fused operator is applied to.
Due to the symmetry of the fusion product there are two
alternative ways of writing the comultiplication, denoted $\Delta_{z,\zeta}$ and $\tilde{\Delta}_{z,\zeta}$.
Demanding that both $\Delta_{z,\zeta}$ and $\tilde{\Delta}_{z,\zeta}$ actually yield the same result
we get the fusion space of two representations ${\cal H}_i$ at positions $z_i$, $i=1,2$,
\beq
{\cal H}_1 \otimesf {\cal H}_2 := ({\cal H}_1 \otimes {\cal H}_2) / (\Delta_{z_1,z_2} - \tilde{\Delta}_{z_1,z_2}) \; .
\eeq
Throughout the paper we will use $\otimesf$ to denote the fusion product of two representations.

In this paper we will only look at representations wrt 
the Virasoro algebra ${\cal A} (L)$ which is generated
by the modes $L_n$ of the $h(L)=2$ Virasoro field $L$, 
the holomorphic energy momentum tensor of conformal field theory.
We need the following subalgebras
\beq
{\cal A}_-^0 (L) &:=& \langle L_{-n} | \, 0 < n < h(L) \rangle \nonumber \\
{\cal A}_{--} (L) &:=& \langle L_{-n} | \, n \ge h(L) \rangle \nonumber \\
{\cal A}_{\pm} (L) &:=& \langle L_{n} | \, \pm n > 0 \rangle
\eeq
as well as the subalgebra of words with length of at least $n$
\beq
{\cal A}_n (L) &=& \Big\langle \prod_{j=1}^m L_{-l_j}^{k_j} \Big| \, \sum_{j=1}^m \, l_j \ge n \Big\rangle \; .
\eeq
The essential information about a representation ${\cal H}$ is already encoded
in its ``special subspace'', the quotient space
\beq
{\cal H}^s &:=& {\cal H}\, /\, ({\cal A}_{--} (L) \: {\cal H}) \; .
\eeq
We also need the family of filtrations of $\cal H$ given as quotient spaces
\beq
{\cal H}^n &:=& {\cal H}\, /\, ({\cal A}_{n+1} (L) \: {\cal H}) \; .
\eeq
Especially for irreducible $\cal H$ this space is equal to the
set of descendants up to level $n$.

We want to restrict to a certain type of representations, the
``quasirational representations''. We use the definition of a quasirational
representation that it is a representation with finite special
subspace. For quasirational representations of the Virasoro algebra
it has been shown that \cite{Nahm94,GK96}
\beqn \label{inclusion}
\left( {\cal H}_1 \otimesf {\cal H}_2 \right)^n \subset {\cal H}_1^s \otimes {\cal H}_2^n \quad \wedge \quad
\left( {\cal H}_1 \otimesf {\cal H}_2 \right)^n \subset {\cal H}_1^n \otimes {\cal H}_2^s \; .
\eeqn
The proof uses the following Nahm algorithm which can be shown to map every state
of the tensor product ${\cal H}^{\cdot \cdot}_1 \otimes {\cal H}^{\cdot \cdot}_2$ to a state 
in the respective right hand side in \mref{inclusion}
in a finite number of steps.

We present the algorithm only for the first equation in \mref{inclusion}
as the other version works the same way by symmetry of the fusion product.
In the following, we regard the states of the tensor product to be
at positions $(z_1,z_2)=(1,0)$.
The two steps of the Nahm algorithm are then given by:
\begin{belaufz}
  \item[(A1)] A vector $\psi_1 \otimes \psi_2 \in {\cal H}_1 \otimes {\cal H}_2$ 
is rewritten in the form
\beq
\psi_1 \otimes \psi_2 &=& \sum_i \, \psi_1^i \otimes \psi_2^i + \Delta_{1,0}({\cal A}_{n+1} (L)) \, ({\cal H}_1 \otimes {\cal H}_2) 
\eeq
with $\psi_1^i \in {\cal H}_1^s$. This can be achieved by the following recursive 
procedure.

The crucial step is to use the nullvector
conditions on $\psi_1$ to re-express it in the form
\beq
\psi_1 &=& \sum_j \, \psi_j^s + \sum_k \, {\cal A}_{--} \, \chi_k^s \; ,
\eeq
with $\psi_j^s,\chi_k^s \in {\cal H}_1^s$.
We still need to get rid of the ${\cal A}_{--}$ action on the $\chi_k^s$.
We use the following formula derived from the comultiplication formula and its translation
properties for $m \le n$ \cite{GK96}
\beq
(L_{-m} \otimes \id) &=& \sum_{l=m}^n \left( l-h \atop m-h \right) \Delta_{1,0} (L_{-l}) \nonumber \\
&& \quad - \; \epsilon \; \sum_{l=1-h}^{\infty} \left( m+l-1 \atop m-h \right) \, (-1)^{h-m-1} \, (\id \otimes L_l)
+  \Delta_{1,0}({\cal A}_{n+1} (L) ) \; .
\eeq
This formula actually enables us to replace the
$(L_{-m} \otimes \id)$ action in ${\cal A}_{--} \, \chi_k^s \otimes \psi_2$
by terms where ${\cal A}_-^0$ or even the identity acts on the left hand
vector of the tensor product. This is true as in the range $m \le l \le n$ 
the comultiplication $\Delta_{1,0} (L_{-l})$
is actually of the simple form ${\cal A}_-^0 \otimes \id + \id \otimes {\cal A}_{--}$.
Now we have to take the result and repeat this procedure starting again
with the re-expression of the first fields $\psi_1$ in the tensor product.
A simple count of the strictly decreasing level of modes during the iteration
shows that this algorithm has to terminate \cite{Nahm94}.

  \item[(A2)] This step has to be applied to each term of the
resulting sum from step (A1) separately.
The input, a resulting tensor product from (A1) 
$\psi_1 \otimes \psi_2 \in {\cal H}_1^s \otimes {\cal H}_2$,
is rewritten as
\beq
\psi_1 \otimes \psi_2 &=& \sum_t \, \psi_1^t \otimes \psi_2^t + \Delta_{1,0}({\cal A}_{n+1} (L)) \, ({\cal H}_1 \otimes {\cal H}_2) \; ,
\eeq
where now $\psi_1^t \in {\cal A}^0_{-} \, {\cal H}_1^s$ and $\psi_2^t \in {\cal H}_2^n$. 

This is achieved by repeatedly using
\beq
\Delta_{1,0} (L_{-I}) &=& (\id \otimes L_{-l} ) + \sum_k \, c_k \, ({\cal A}_-^0 \otimes L_{-I_k})
\eeq
for a word $L_{-I} = L_{-i_1}\, L_{-i_2} \dots$ of negative Virasoro modes with level $|I|$ and constant $c_k$. 
This recursion 
has to finish as the Virasoro monomials $L_{-I_k}$ are of strictly lower
level $|I_k| < |I|$. This formula is just the result of repeated use
of the comultiplication formula for a monomial of modes of the same field and
for the special coordinates $(z,\zeta) = (1,0)$.
\end{belaufz}

As we want to have the states in the fusion product which are projected
to the subspace $({\cal H}_1 \otimesf {\cal H}_2)^n$
we do not have to care about contributions $\Delta_{1,0}({\cal A}_{n+1} (L)) \, ({\cal H}_1 \otimes {\cal H}_2)$.
It is then easy to see that iterated application of steps (A1) and (A2)
will finally yield the required result \mref{inclusion}.
This algorithm actually terminates in a finite number of steps as
the number of modes on both fields strictly decreases
when re-expressing $\psi_1$ in step (A1) using its nullvector
condition and does not increase in step (A2).

%%%%%%%%%%%%

\subsection{Constraints for the fusion algebra}

By \mref{inclusion} we know that $\left( {\cal H}_1 \otimesf {\cal H}_2 \right)^n$
is actually embedded in the easily constructed space
${\cal F} := {\cal H}_1^s \otimes {\cal H}_2^n$.
Hence, we want to find the full set of constraints which
describes $\left( {\cal H}_1 \otimesf {\cal H}_2 \right)^n$
in $\cal F$.

The important idea of \cite{GK96} was that one can find
nontrivial constraints by applying ${\cal A}_{n+1}$
to states in $\cal F$. We then have to use the Nahm algorithm
in order to map the resulting descendant states into our ``standard'' space $\cal F$.
By definition these descendant states are divided out
of  $\left( {\cal H}_1 \otimesf {\cal H}_2 \right)^n$ and,
hence, are supposed to vanish. Thus, their mapping to $\cal F$
should evaluate to zero---if we acquire non-trivial expressions
this simply yields the desired constraints by imposing their vanishing.

This procedure is even improved if we use the nontrivial
nullvector conditions on the second field to replace
the action of certain Virasoro monomials before performing
the Nahm algorithm. This introduces the information about
the nullvector structure on the second representation
of the tensor product into the game; the information about
the nullvector structure on the first representation of
the tensor product has already been used in the Nahm
algorithm itself.

As we noticed during our calculation it even improves the
situation to include the nullvector conditions on the tensor
factors in the space $\cal F$ itself.

In the following
we will denote the level $n$ at which we perform
the computation with $L$. Certainly one cannot perform 
this calculation for all of ${\cal A}_{L+1}$. 
We hence restricted our computation to the application
of Virasoro monomials
\beq
\Big\langle \prod_{j=1}^m L_{-l_j}^{k_j} | \, \sum_{j=1}^m \, l_j = \tilde{L} \Big\rangle \; .
\eeq
of equal level $\tilde{L}$. Usually we performed the
calculation from $\tilde{L} = L+1$ up to a 
maximal $\tilde{L}_{\mbox{\footnotesize max}}$.
Both $L$ and $\tilde{L}_{\mbox{\footnotesize max}}$ are given
for the respective calculations in the appendix.

As we are limited to the calculation of a finite
number of constraints this procedure is only able 
to give a lower bound on the number of constraints
and, hence, an upper bound on the number of states
in the fusion product at that level.
On the other hand, these constraints seem to be highly
non-trivial such that already a very low  
$\tilde{L}_{\mbox{\footnotesize max}} > L$, often
even $\tilde{L}_{\mbox{\footnotesize max}} = L+1$,
is sufficient to gain all constraints which
yield representations in a consistent fusion algebra.
This already worked very well in \cite{GK96} for
the $c_{p,1}$ model case and as we will see
it also works very well in the general augmented
$c_{p,q}$ model case.

Now, it is especially interesting to observe the action
of positive Virasoro modes on $\left( {\cal H}_1 \otimesf {\cal H}_2 \right)^L$.
The positive Virasoro modes, however,
induce an action
\beq
L_m : \left( {\cal H}_1 \otimesf {\cal H}_2 \right)^L \rightarrow 
\left( {\cal H}_1 \otimesf {\cal H}_2 \right)^{L-m} \qquad \forall \, m \le L \; .
\eeq
It is important to note that the $L_m$ map to a space
of respective lower maximal level. Hence, we need to
construct all spaces of lower maximal level
$0 \le n < L$. To achieve this we start with $\left( {\cal H}_1 \otimesf {\cal H}_2 \right)^L$
and successively impose the constraints which
arise in the above described way from the vanishing
of the action of Virasoro monomials 
of level $m$ with $n < m \le L$ on $\cal F$.

%%%%%%%%%%%%

\subsection[Implementation for the $c_{p,q}$ models]{Implementation for the $\mathbf{c_{p,q}}$ models}

We have implemented the main calculational tasks for this paper,
especially the Nahm algorithm and the calculation of the constraints,
in C++ using the computer algebra package GiNaC \cite{Ginac}.
We constructed new classes for the algebraic objects fields,
fieldmodes, products of fieldmodes, descendant fields as well as
tensor products of fields which are the basic ingredients in this
algorithm. (Some of the classes have already been used in \cite{EF05}.)

As GiNaC does not support factorisation we used the JordanForm
package of the computer algebra system Maple in order
to get the Jordan diagonal form of the $L_0$ matrix on the resulting space
as well as the matrices of base change.
This Maple calculation is performed via command-line during the
run of the C++ programme.

As we will see in section \ref{rep_intro} some irreducible
representations
in the general augmented $c_{p,q}$ models have more than one 
nullvector (two, to be precise) which are completely independent, i.e.\ such that
none of these nullvectors can be written as a descendant of the 
others. Hence, it is important to include both independent nullvectors
into the nullvector lists which are used for replacements in
the calculation as explained above. This is needed to provide the
full information about the nullvector structure of the original
representations which are to be fused. Sometimes one even needs
to choose an $\tilde{L}_{\mbox{\footnotesize max}}$ large enough
such that the second nullvector can also become effective.
The special subspace, however, is nevertheless determined by the
level $l$ of the lowest nullvector
\beq
\langle \psi, L_{-1} \psi,  L_{-1}^2 \psi, \dots , L_{-1}^{l-1} \psi \rangle \; .
\eeq

Besides the fusion of two irreducible representations
we also implemented the possibility of fusing an
irreducible representation with a rank $2$ representation.
Actually this generalisation is quite straight forward.
Instead of one state which generates an irreducible
representation we now need two generating states. However,
we have to be careful because the second generating
state, the logarithmic partner, is not primary. Hence,
we implemented the indecomposable action on this second
generating field as additional conditions proprietary 
to that field (as already
done in \cite{EF05}). We also have to be careful to calculate
the correct nullvector structure which includes besides
an ordinary nullvector on the primary field the first
logarithmic nullvector of the whole indecomposable representation.
We have calculated these logarithmic nullvectors
using the algorithm described in \cite{EF05}.

In order to speed up the algorithm we widely used hashing tables.
This measure actually resulted in a quite equal use of
computing time and memory (on a standard PC with up to 4GB memory);
calculations which are on the edge of using up the memory
have run times between half a day and a few days.

The performance of the implemented algorithm is, however, 
quite hard to benchmark
as it varies very much with different input and output.
Concerning the input the computing time rises with the 
level of the nullvectors---especially 
the nullvector level of the first tensor factor is crucial. 
We also need much more time to compute fusion products with
representation on weights that are strictly rational than
corresponding ones with integer weights.
And then the performance of course depends exponentially on $L$ as well
as $\tilde{L}$, although the dependence on $L$ is much stronger.
Concerning the output the computational power of Maple
is frequently the limiting factor if we have a large resulting $L_0$ matrix.

We have also checked the correct implementation of the algorithm
by reproducing quite some fusion rules of the $c_{p,1}$
models given in \cite{GK96}. In particular we reproduced
the Virasoro matrices for the example given in the appendix
of \cite{GK96}. We also noted that the algorithm is very
sensitive and fragile such that an only small change
in the parameters or the programme yields completely unreasonable results.

In contrast to the lowest $c_{p,1}$ models we have to cover
a much larger parameter space with states of higher nullvectors
already for the easiest general augmented $c_{p,q}$ model,
the augmented $c_{2,3}=0$ model. We hence decided to calculate 
the fusion of the lowest representations at $L=6$
in order to be able to get results at the same $L$ for
a large parameter space. For the higher fusion as well as the fusion
with rank $2$ representations we had to reduce $L$.
Details as well as the results are given in the appendices 
\ref{fusion_c0} and \ref{fusion_cminus22over5}.

%%%%%%%%%%%%%%%%%%%%%%%%%%%%%%%%%%%%%%%%%%%%%
%%%%%%%%%%%%%%%%%%%%%%%%%%%%%%%%%%%%%%%%%%%%%
%%%%%%%%%%%%%%%%%%%%%%%%%%%%%%%%%%%%%%%%%%%%%

\section[Virasoro representation theory for (augmented) minimal models]{Virasoro representation theory for minimal models and their extensions\label{rep_intro}}

In this section we want to give a short overview about the representation theory
of minimal models \cite{BPZ84,DMS97} as well as the augmented $c_{p,1}$ models \cite{Flohr95,GK96}.

In the following we will exclusively regard representations of the Virasoro algebra
which is spanned by the modes of the (holomorphic) $h=2$ stress energy tensor
$L_n$ obeying
\beq
[L_m,L_n] &=& (m-n) \, L_{m+n} + \frac{c}{12} \, (m-1) \, m \, (m+1) \, \delta_{m+n,0} \qquad m,n \in \Z \; .
\eeq
To any highest weight $h$ the application of the negative Virasoro modes $L_n, \; n<0,$
freely generates the Verma module 
\beq
{\cal M} (h) &=& \{ L_{-n_1} \dots L_{-n_k} v_h | n_1 \ge \dots \ge n_k > 0, \; k \in \Z^+ \}
\eeq
where $v_h$ is the highest weight state to $h$. In order to find the irreducible
or at least indecomposable representations we need to identify the largest
true subrepresentations of ${\cal M} (h)$ which decouple from the rest of the
representation and need to construct the respective factor module. 

A subrepresentation can be generated from any singular vector $v$ 
in ${\cal M} (h)$, i.e.\ a vector
which obeys $L_p v = 0 \; \forall p >0$ and which is hence a highest weight state
of its own; and certainly we can also have unions of such representations.
On the other hand a subrepresentation only decouples from the rest of the
representation and can hence be factored out if it consists of nullvectors, i.e.\
vectors which are orthogonal to all other vectors in the Fock space
of states wrt the natural sesquilinear Shapovalov form on this space. 
As long as we do not encounter any indecomposable structure in our representation
singular vectors are at the same time nullvectors and generate
subrepresentations which are null in Fock space.
However, the interrelation between singular vectors and nullvectors becomes
much more intricate as soon as we deal with indecomposable 
representations.

The Kac determinant actually parametrises
the relation between conformal charges $c$ and spectra
of conformal weights $h_{r,s}$ at which we encounter nullvectors.
It is hence an ingenious tool to explore interesting conformal field
theories with relatively few and small representations.
We are especially interested in this series of conformal field theories
which emerge from the study of the Kac determinant and which, hence,
exhibit a rich nullvector structure to be described below. These theories
are parametrised by the conformal charges (see e.g.\ \cite{DMS97})
\beq
c = c_{p,q} = 1-6 \, \frac{p-q}{pq} \qquad 1 \le p,q \in \Z \; ,
\eeq
where $p$ and $q$ do not have a common divisor;
their highest weight spectrum is given by the weights in the
Kac table
\beq
h_{r,s} &=& \frac{(pr-qs)^2 - (p-q)^2}{4pq} \quad 1 \le r \in \Z, \;  1 \le s \in \Z \; .
\eeq
An extract of the (infinite) Kac table for $c_{2,3} = 0$
is given in table \ref{Kac0}.

\renewcommand{\arraystretch}{0.03}
\begin{table} \caption{Kac table for $c_{2,3}=0$}
\begin{center}
\begin{tabular}{cc | ccccc}
\multicolumn{7}{c}{\rule[-.6em]{0cm}{1.6em} $\qquad\quad  s$} \\
&\rule[-.6em]{0cm}{1.6em} & $1$ & $2$ & $3$ & $4$ & $5$ \\ \cline{2-7}
& & \multicolumn{5}{c}{} \\
&\rule[-.6em]{0cm}{1.6em} $1$ & $0$ & \cellcolor{gray}{.9}{$\frac{5}{8}$} & $2$ & \cellcolor{gray}{.9}{$\frac{33}{8}$} & $7$ \\
&\rule[-.6em]{0cm}{1.6em} $2$ & $0$ & \cellcolor{gray}{.9}{$\frac{1}{8}$} & $1$ & \cellcolor{gray}{.9}{$\frac {21}{8}$} & $5$ \\
&\rule[-.6em]{0cm}{1.6em} $3$ & \multicolumn{1}{>{\columncolor[gray]{.9}[0.205cm][0.22cm]}c}{$\frac{1}{3}$} 
& \cellcolor{gray}{.7}{$-\frac{1}{24}$} 
& \cellcolor{gray}{.9}{$\frac{1}{3}$} & \cellcolor{gray}{.7}{$\frac {35}{24}$} & \cellcolor{gray}{.9}{$\frac{10}{3}$} \\
&\rule[-.6em]{0cm}{1.6em} $4$ & $1$ & \cellcolor{gray}{.9}{$\frac{1}{8}$} & $0$ & \cellcolor{gray}{.9}{$\frac{5}{8}$} & $2$ \\
&\rule[-.6em]{0cm}{1.6em} $5$ & $2$ & \cellcolor{gray}{.9}{$\frac{5}{8}$} & $0$ & \cellcolor{gray}{.9}{$\frac{1}{8}$} & $1$ \\
$r$ &\rule[-.6em]{0cm}{1.6em} $6$ & \multicolumn{1}{>{\columncolor[gray]{.9}[0.205cm][0.22cm]}c}{$\frac{10}{3}$}
& \cellcolor{gray}{.7}{$\frac{35}{24}$} 
\rule[-.6em]{0cm}{1.6em}& \cellcolor{gray}{.9}{$\frac{1}{3}$} & \cellcolor{gray}{.7}{$-\frac{1}{24}$} & \cellcolor{gray}{.9}{$\frac{1}{3}$} \\
&\rule[-.6em]{0cm}{1.6em} $7$ & $5$ & \cellcolor{gray}{.9}{$\frac {21}{8}$} & $1$ & \cellcolor{gray}{.9}{$\frac{1}{8}$} & $0$ \\
&\rule[-.6em]{0cm}{1.6em} $8$ & $7$ &\cellcolor{gray}{.9}{ $\frac {33}{8}$} & $2$ & \cellcolor{gray}{.9}{$\frac{5}{8}$} & $0$ \\
&\rule[-.6em]{0cm}{1.6em} $9$ & \multicolumn{1}{>{\columncolor[gray]{.9}[0.205cm][0.22cm]}c}{$\frac{28}{3}$}
& \cellcolor{gray}{.7}{$\frac{143}{24}$} 
\rule[-.6em]{0cm}{1.6em}& \cellcolor{gray}{.9}{$\frac{10}{3}$} & \cellcolor{gray}{.7}{$\frac{35}{24}$} & \cellcolor{gray}{.9}{$\frac{1}{3}$} \\
&\rule[-.6em]{0cm}{1.6em} $10$ & $12$ & \cellcolor{gray}{.9}{$\frac {65}{8}$} & $5$ & \cellcolor{gray}{.9}{$\frac{21}{8}$} & $1$ \\
&\rule[-.6em]{0cm}{1.6em} $11$ & $15$ & \cellcolor{gray}{.9}{$\frac {85}{8}$} & $7$ & \cellcolor{gray}{.9}{$\frac{33}{8}$} & $2$ \\
\end{tabular}
\end{center} \label{Kac0}
\end{table}
\renewcommand{\arraystretch}{1.3}

The so-called minimal models are a series of such conformal field theories which
manage to extract the smallest possible representation
theory from the Kac table of some central charge $c_{p,q}$ by relating all weights
to some standard cell $\{(r,s) | 1 \le r < q, \;  1 \le s < p\}$ subject to the relation \cite{BPZ84,DMS97}
\beqn \label{minmod_relations}
h_{r,s} &=& h_{q-r,p-s} \; .
\eeqn
All larger higher weights are related to this standard cell by the addition
of integers according to the relations \cite{BPZ84,DMS97}
\beqn \label{minmod_relations2}
h_{r,s} &=& h_{r+q,s+p}  \nonumber \\
h_{r,s} + rs &=& h_{q+r,p-s} = h_{q-r,p+s} \nonumber \\
h_{r,s} + (q-r)(p-s) &=& h_{r,2p-s} = h_{2q-r,s}
\eeqn
as long as they are in the bulk and not on the border or corners of this standard cell Kac table, 
i.e.\ as long as their indices do not obey $r=i\, q$ or $s=j\, p$
for some $i,j \in \Z$.

\begin{figure}
\centering \leavevmode
\psfig{file=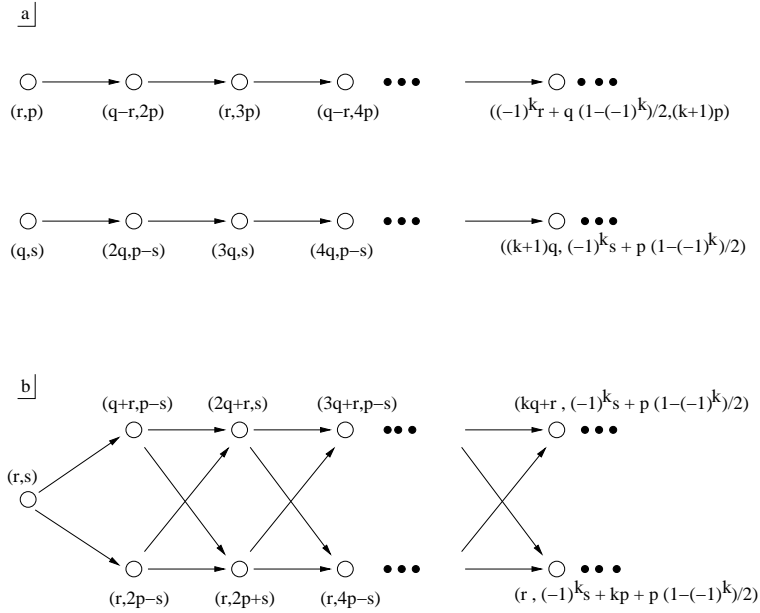,width=10cm}
\caption{Nullvector embedding structure \cite{DMS97}}
\label{embedding}
\end{figure}

These larger weights in the Kac table bulk are exactly the weights of the nullvector
descendants of the highest weights in the above standard cell.
To be precise we actually find that the maximal subrepresentation
of ${\cal M} (h)$ for $h$ in the bulk of the Kac table is generated by two singular
vectors $v_1$, $v_2$. The highest weight representations generated
on $v_1$ and $v_2$, however, each contain two subrepresentations
which are again both generated from two singular vectors; but actually both
subrepresentations of ${\cal M} (v_1)$ and ${\cal M} (v_2)$
coincide. We therefore arrive at an embedding structure or ``embedding cascade''
of nullvectors as depicted in figure \ref{embedding}b \cite{FF82,DMS97}
whose weights are exactly the integer shifted weights appearing in the
Kac table. The corresponding characters have been calculated in \cite{Roch85}.

The irreducible representations ${\cal V}_{(r,s)}$ with weights $h_{r,s}$ in
this standard cell and the described nullvector embedding structure
have been shown to close under the following so-called BPZ fusion rules \cite{BPZ84}
\beq
{\cal V}_{(r_1,s_1)} \otimesf {\cal V}_{(r_2,s_2)}
&=& \sum_{r_3 = |r_1-r_2|+1, \; \mbox{\footnotesize step}\: 2}^{\min (r_1+r_2-1,2q-r_1-r_2-1)} \; \;  
\sum_{s_3 = |s_1-s_2|+1, \; \mbox{\footnotesize step}\: 2}^{\min (s_1+s_2-1,2p-s_1-s_2-1)} {\cal V}_{(r_3,s_3)} \; ,
\eeq
where $\otimesf$ denotes the fusion product.
We notice that the above excluded weights for $r=i\, q$ or $s=j\, p$ with $i,j \in \Z$,
do not pop up in these fusion rules; they are hence simply ignored in
these minimal models.

On the other hand, augmenting the theory with representations beyond this standard cell,
especially with irreducible representations of the above excluded weights,
has also led to the construction of consistent CFTs. These contain
representations with non-trivial Jordan blocks and are thus examples of logarithmic CFTs.
For the Virasoro representation theory of these models
one actually needs the full Kac table to describe its different
representations, subject only to the relation \mref{minmod_relations}.

To make the terminology more precise we will call, following \cite{EF05},
weights whose indices obey $r=i\, q$ and $s=j\, p$ ($i,j \in \Z$)
``on the corners of the Kac table'' and weights whose indices obey (exclusively)
either $r=i\, q$ or $s=j\, p$ ($i,j \in \Z$) ``on the borders of
the Kac table''. All other weights which already appear in the minimal models
we call ``in the bulk of the Kac table''.
In table \ref{Kac0}, the Kac table for $c_{2,3} = 0$, we have indicated the borders 
as areas with lighter shade and the corners
as areas with darker shade; the bulk consists of the unshaded areas.
Actually it is just this inclusion of irreducible representations with
weights on the corners and the border which forces us
to include states of weights of the whole Kac table into the theory.

The only well studied models up to now are contained in the
series $c_{p,1}$, $p=2,3,\dots$ (see e.g. \cite{Flohr01,GK96,GK96b,Roh96,GK98,Flohr95,Flohr96}). 
But as these models do not contain any bulk in their
Kac table, we do not expect them to be generic; indeed
as we will see in this paper the existence of representations
with weights in the Kac table bulk actually induces an even richer
structure with indecomposable representations up to rank $3$.

The nullvector embedding structure stays of course
the same for representations corresponding to weights in the bulk.
However, as already explained in \cite{GK96} the 
nullvector embedding structure actually collapses to
a string for representations with weights on the
corners or on the border---as depicted in figure \ref{embedding}a.
It is very important to keep in mind that the
nullvectors corresponding to these higher weights 
in figure \ref{embedding} are only true
nullvectors within representations that are generated as
a Virasoro module from one (!) singular vector,
i.e.\ irreducible representations.
This picture changes as soon as there appears 
indecomposable structure within the representation \cite{GK96,EF05}.
Nevertheless these vectors keep their prominent
role even within higher rank representations.

For later use we need to define the notion of a ``weight chain'' 
for conformal weights on the border or in the bulk.
These weight chains are supposed to be
a handy storage of information about the
weights on successive embedding levels
in the above discussed embedding structures.
A weight chain for weights on the border is a list of all weights 
which differ just by integers, ordered by size without multiplicity
(see figure \ref{embedding}a):
\beq
W_{(r,p)}^{\mbox{\scriptsize border}} &:=& \{ h_{r,p},h_{q-r,2p},h_{r,3p}, ... \} \qquad \forall r<q \; , \nonumber \\
W_{(q,s)}^{\mbox{\scriptsize border}} &:=& \{ h_{q,s},h_{2q,p-s},h_{3q,s}, ... \} \qquad \forall s<p  \; .
\eeq
To form a weight chain for weights in the bulk
we take a likewise list of weights differing just by integers, ordered by size without multiplicity.
Then we map this list into a list of sets, 
the first set just consisting of the lowest weight, then
every next set consisting of the next two weights.
Regarding figure \ref{embedding}b we get
\beq
W_{(r,s)}^{\mbox{\scriptsize bulk}} := \{ h_{r,s},\{h_{r,2p-s},h_{q+r,p-s}\},\{h_{r,2p+s},h_{2q+r,s}\}, ... \}  \qquad \forall r<q, s<p\; .
\eeq

In order to understand the generic features of augmented
$c_{p,q}$ models we will study the two easiest candidates
for augmented models with non-empty bulk of the Kac table
in this paper,
the augmented $c_{2,3}=0$ model with Kac table given in table \ref{Kac0} 
as well as the augmented Yang-Lee model with $c_{2,5}=-22/5$.
Possible candidates for higher rank representations in the 
augmented $c_{2,3}=0$ model have already been explored in \cite{EF05}
by calculation 
of logarithmic nullvectors. We will show that these calculations
are consistent with the findings in this article and will
furthermore use these calculational tools for the explicit
computation of fusion with rank $2$ representations.

For the $c_{p,1}$ models it was shown that they actually
have a larger ${\cal W}$ algebra as symmetry algebra,
the triplet algebra ${\cal W} (2,2p-1,2p-1,2p-1)$.
We have strong hints that such a larger ${\cal W}$ algebra
is also the underlying symmetry algebra of the generic augmented
$c_{p,q}$ models \cite{EFN}.
The effective Kac table of these theories can then again be
reduced to a standard cell, which is though larger
as their minimal model counterpart.
The standard cell is then given by
$\{(r,s) | 1 \le r < n\,q , \;  1 \le s < n\,p \}$
with $n$ usually an odd integer larger than $1$, e.g.\
$3$ for the above mentioned triplet algebras of the 
$c_{p,1}$ models.
In this article, however, we want to concentrate on the pure
Virasoro representation theory and, hence, do not
restrict our Kac table in any way.

%%%%%%%%%%%%%%%%%%%%%%%%%%%%%%%%%%%%%%%%%%%%%
%%%%%%%%%%%%%%%%%%%%%%%%%%%%%%%%%%%%%%%%%%%%%
%%%%%%%%%%%%%%%%%%%%%%%%%%%%%%%%%%%%%%%%%%%%%

\section{Explicit discussion of the augmented $\mathbf{c_{2,3}=0}$ model\label{discussion_c0}}

In this section we explicitly discuss our calculations
of the fusion product of representations in the $c_{2,3}=0$ augmented model
which lead us to the conjectured general fusion rules of section \ref{general}.
We present examples for the newly appearing higher rank
representations and also elaborate the consistency conditions
for the fusion product in this case.

%%%

\renewcommand{\arraystretch}{1.2}
\begin{table} \caption{Specific properties of rank $2$ representations in $c_{2,3}=0$}
\begin{center}
\footnotesize
\begin{tabular}{l || c | c | c | c | c | c}
& $\beta_1$ & $\beta_2$ & level of & level of first & level of first & type\\
&&& log.\ partner & nullvector & log.\ nullvector & \\ \hline
${\cal R}^{(2)} (5/8,5/8)$ & -- & -- & $0$ & $2$ & $10$ & A \\
${\cal R}^{(2)} (1/3,1/3)$ & -- & -- & $0$ & $3$ & $9$ & A \\
${\cal R}^{(2)} (1/8,1/8)$ & -- & -- & $0$ & $4$ & $8$ & A \\
${\cal R}^{(2)} (5/8,21/8)$ & $-5$ & -- & $2$ & $10$ & $16$ & B \\
${\cal R}^{(2)} (1/3,10/3)$ & $140/27$ & -- & $3$ & $9$ & $18$ & B \\
${\cal R}^{(2)} (1/8,33/8)$ & $-700/81$ & -- & $4$ & $8$ & $20$ & B \\ \hline
${\cal R}^{(2)} (0,1)_5$ & $1/3$ & -- & $1$ & $2$ & $7$ & C \\
${\cal R}^{(2)} (0,1)_7$ & $-1/2$ & -- & $1$ & $2$ & $5$ & D \\
${\cal R}^{(2)} (0,2)_5$ & -- & $-5/8$ & $2$ & $1$ & $7$ & E \\
${\cal R}^{(2)} (0,2)_7$ & -- & $5/6$ & $2$ & $1$ & $5$ & F \\
${\cal R}^{(2)} (1,5)$ & $2800/9$ & -- & $4$ & $6$ & $14$ & C\\
${\cal R}^{(2)} (1,7)$ & $30800/27$ & $1100/9$ & $6$ & $4$ & $14$ & E \\
${\cal R}^{(2)} (2,5)$ & $-420$ & -- & $3$ & $5$ & $10$ & D \\
${\cal R}^{(2)} (2,7)$ & $-880$ & $-440/3$ & $5$ & $3$ & $10$ & F
\end{tabular}
\end{center} \label{rank2}
\end{table}
\normalsize

\subsection{Higher rank representations}

\subsubsection[Representations of rank $2$]{Representations of rank $\mathbf{2}$\label{discussion_c0_rank2}}

Table \ref{rank2} gives an overview over the specific properties
of all rank $2$ representations we have calculated for this model.
The two parameters of the rank $2$ representation ${\cal R}^{(2)}$
give the lowest weight and the weight of the logarithmic partner
in this representation, i.e.\ the weights of the
two states which generate this representation. The additional index
will be explained when we discuss these particular representations.

\begin{figure}
\centering \leavevmode
\psfig{file=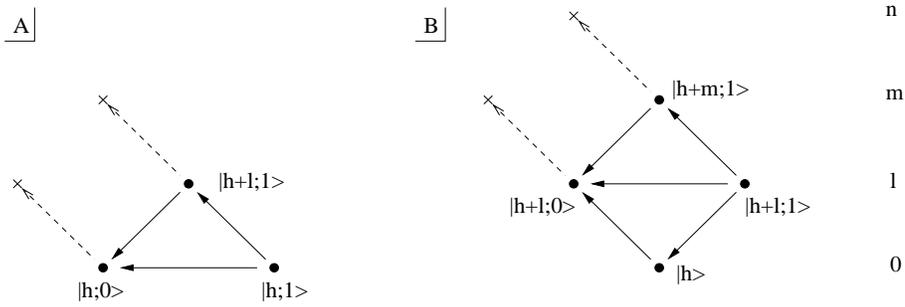,width=12cm}
\caption{Rank $2$ representations for weights on the border}
\label{rep_on_border}
\end{figure}

The first block contains the rank $2$ representations to the three different 
weight chains lying on the border of the Kac table, i.e.\ 
$W_{(1,2)}^{\mbox{\scriptsize border}} := \{5/8, 21/8, 85/8, \dots \}$,
$W_{(3,1)}^{\mbox{\scriptsize border}} := \{1/3, 10/3, 28/3, \dots \}$ and
$W_{(2,2)}^{\mbox{\scriptsize border}} := \{1/8, 33/8, 65/8, \dots \}$.
Both different types of rank $2$ representations are depicted
in figure \ref{rep_on_border}.
They actually exhibit precisely the same structure
as the rank $2$ representations of the augmented $c_{p,1}$ models
described in \cite{GK96}. This has already been conjectured
in \cite{EF05} by calculation of their first logarithmic
nullvectors.
Throughout this paper we stick to the graphical conventions of previous
publications, see e.g.\ \cite{GK96,EF05}. As e.g.\ in figure \ref{rep_on_border}
we take black dots to denote (sub-)singular vectors which are not null,
crosses to denote nullvectors, horizontal arrows to denote indecomposable
action (of $L_0$), arrows pointing upwards to denote a descendant
relation and arrows pointing downwards to denote a non-trivial action of positive
Virasoro modes. Furthermore, we indicate the levels on the
right hand side of each picture.

The first three representations in table \ref{rank2} are the groundstate rank $2$
re\-presentation visualised in figure \ref{rep_on_border} A.
They exhibit an indecomposable Jordan cell
already on the zeroth level. In this case the logarithmic
partner state $| h; 1 \rangle$ of the irreducible ground state 
$| h; 0 \rangle$ is logarithmic primary, i.e.\
it is primary with the exception of an indecomposable action of $L_0$
\beq
L_0 | h; 1 \rangle &=& h \; | h; 1 \rangle + | h; 0 \rangle \; .
\eeq
$| h; 0 \rangle$ actually spans an irreducible subrepresentation;
its first nullvector (depicted as a cross in the figure)
is hence found on the level of the next weight $h+l$ in the
corresponding weight chain. The descendants of $| h; 1 \rangle$,
however, do not form to give a nullvector at $h+l$. In order
to find the first nullvector involving descendants of $| h; 1 \rangle$
we have to go one weight further in the weight chain.
These representations are uniquely generated by the two groundstates
$| h; i \rangle, \; i = 0,1$; there is no need of an additional
parameter to describe them.

The next three representations in table \ref{rank2} 
are the first excited representations of
these three weight chains depicted in figure \ref{rep_on_border} B. 
The logarithmic partner state  $| h+l; 1 \rangle$ lies on that
level $l$ at which we would expect the first nullvector of the groundstate
$|h \rangle$ if the groundstate were to span an irreducible representation.
Hence, the subre\-presentation generated by $|h \rangle$ is not
irreducible, but only indecomposable. Actually the indecomposable action
of $L_0$ just maps $| h+l; 1 \rangle$ to the singular level $l$ descendant
of $|h \rangle$ which we call $| h+l; 0 \rangle$ and
which normally would be the first nullvector of an irreducible $h$
representation
\beq
L_0 | h+l; 1 \rangle &=& (h+l) \; | h+l; 1 \rangle + | h+l; 0 \rangle \; .
\eeq
$| h+l; 0 \rangle$ on the other hand spans an irreducible subrepresentation
and yields its first nullvector on the level of the second weight
after $h$ in the weight chain, called $h+m$.
In order to find the first logarithmic nullvector we have to go
even one weight further in the weight chain to $h+n$.
Now, the logarithmic partner field can certainly be not logarithmic
primary; but as discussed in \cite{GK96} it can still be made to 
at least vanish under all $L_p, \; p>2$. 
This induces one characteristic parameter $\beta = \beta_1$ in this representation due to
the action of $L_1$; we take $\beta$ to parametrise the following equation
\beq
(L_1)^l \; | h+l; 1 \rangle &=& \beta \; | h \rangle \; .
\eeq

The structure visualised in figure \ref{rep_on_border} B is actually
thought to be the generic type of rank $2$ representation for weights on the border.
It should be found for every two adjacent weights in a border weight
chain. As in the $c_{p,1}$ model case the representations of the type
depicted in figure \ref{rep_on_border} A can only be found for the
first and hence lowest weight of a weight chain. 
Furthermore, we want to stress that we can think of these
rank $2$ representations as being constructed by indecomposably
connecting several irreducible representations. This is already
suggested by figures \ref{rep_on_border} A and \ref{rep_on_border} B where
the black dots represent these ``towers of states'' which
have been irreducible representations before their indecomposable
connection to a rank $2$ representation. Generically,
these towers of states are
not irreducible subrepresentations any more, but they resemble
their irreducible counterparts in terms of number of states
and singular vectors. This point of view has also been worked
out in \cite{GK96} for rank $2$ representations in the augmented
$c_{p,1}$ models.

We have also successfully checked for the existence of the first
logarithmic nullvectors at the levels given in table \ref{rank2}
using the algorithm of \cite{EF05}. Even the level $20$
logarithmic nullvector was now accessible to our computational power
due to our explicit knowledge of $\beta$.

\begin{figure}
\centering \leavevmode
\psfig{file=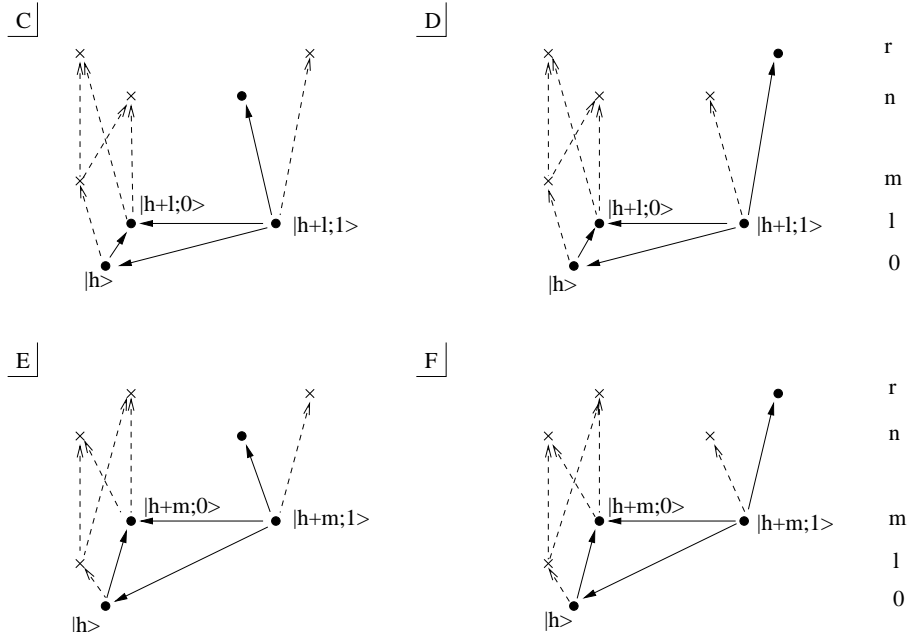,width=12cm}
\caption{Rank $2$ representations for weights in the bulk}
\label{rep_on_bulk2}
\end{figure}

The second block of table \ref{rank2} contains the specific
properties of the lowest rank $2$ representations
which we found for weights in the weight chain of
the bulk of the Kac table, i.e.\
$W_{(1,1)}^{\mbox{\scriptsize bulk}} := \{\{0\}, \{1,2\}, \{5,7\}, \dots \}$.
There are actually four different types of rank $2$
representations depicted in figure \ref{rep_on_bulk2}
which appear to be a generalisation of the situation on 
the border for the case of figure \ref{rep_on_border} B.
This generalisation has to take into account the more
complicated embedding structure of representations
with weights in the bulk---the linear picture of 
figure \ref{embedding}a has to be replaced by the 
two string twisted picture of figure \ref{embedding}b.
As we now have two possible nullvectors on every step
of the weight chain, of which only one is rendered non-null
by the existence of a logarithmic partner state of the
same level, we actually encounter cases E and F where
there is a true nullvector on a level lower than the
logarithmic partner state.
This is new and makes the description of these
particular cases more complicated.

Let us first describe the cases C and D. Starting with the
lowest weight state $| h \rangle$ the first possible
nullvectors are given by the next set in the weight chain
at levels $l$ and $m$. In the present cases the corresponding
singular descendant $| h+l;0 \rangle$ on $| h \rangle$ at the lower 
of these two levels $l$
is rendered to be non-null by the existence
of a logarithmic partner state $| h+l;1 \rangle$. The
indecomposable action of $L_0$ on $| h+l;1 \rangle$
is again given by
\beq
L_0 | h+l; 1 \rangle &=& (h+l) \; | h+l; 1 \rangle + | h+l; 0 \rangle \; .
\eeq
Furthermore, the argument of \cite{GK96} still applies
that due to the absence of a nullvector on $| h \rangle$
on a level lower than the Jordan cell we can transform
$| h+l;1 \rangle$ by the addition of level $l$
descendants of $| h \rangle$ such that it is annihilated
by $L_p \; \forall p \ge 2$. Then $L_1$ maps
$| h+l;1 \rangle$ to the unique level $l-1$ descendant
of $| h \rangle$ which is annihilated
by $L_p \; \forall p \ge 2$. As for representations with
weights on the border we take the resulting one
parameter $\beta = \beta_1$ to parametrise the equation
\beq
(L_1)^l \; | h+l; 1 \rangle &=& \beta \; | h \rangle \; .
\eeq
The second state corresponding to this set in the weight chain,
the one at level $m$, actually stays null in the rank $2$
representation. This fixes the nullvector structure on $| h \rangle$
as the embedding structure of figure \ref{embedding} b
tells us that the nullvectors of the next set in the
weight chain, at levels $n$ and $r$, are joint descendants
of the states corresponding to the previous set of weights.
But as the singular state at level $m$ is already a nullvector the singular
states at level $n$ and $r$ have to be null as well.
The situation is somewhat different for the descendants
of the logarithmic partner as $| h+l;1 \rangle$ is the starting point
of its embedding structure. The cases C and D correspond
to the two possibilities of having the first logarithmic nullvector
on level $r$ respectively $n$. There is, however, no additional
nullvector at the respective other weight.
Examples for the case C are ${\cal R}^{(2)} (0,1)_5$ and
${\cal R}^{(2)} (1,5)$, for the case D ${\cal R}^{(2)} (0,1)_7$ and
${\cal R}^{(2)} (2,5)$. Again the lowest representations
play a special role as both cases are realised for
lowest weight $0$. Hence, we indicate the level of the logarithmic
descendant which is promoted to be non-null as an index.
It is important to note, however, that both cases are already
distinguished by their different $\beta$ values.

Turning to the cases E and F they exhibit very much
the same structure as the cases C and D. The crucial
difference is the existence of a nullvector already on a level
lower than the level of the logarithmic partner.
This fact prevents us from applying the above argument
how to describe the representation by only one parameter.
The special cases of ${\cal R}^{(2)} (0,2)_5$ and 
${\cal R}^{(2)} (0,2)_7$ can nevertheless be reduced to
one parameter quite easily. As there is no non-null
descendant of the lowest weight state $| 0 \rangle$
at level $1$ the only positive Virasoro mode
which can map the logarithmic partner $| 2;1 \rangle$
to a non-zero state is $L_2$. Hence, we
take the one parameter $\beta = \beta_2$ to parametrise the equation
\beq
L_2 \; | 2; 1 \rangle &=& \beta \; | 0 \rangle \; .
\eeq
This behaviour is, however, not generic. We find that we
need at least two parameters to describe these two kinds
of rank $2$ representations in general. To see this let us regard all
normal ordered monomials in Virasoro modes of length $m$. We are
able to transform $| h+m;1 \rangle$ by addition of
level $m$ descendants of $| h \rangle$ in such a way
that only the application of two such Virasoro monomials
does not annihilate $| h+m;1 \rangle$. In particular we have:
\begin{itemize}
  \item For ${\cal R}^{(2)} (1,7)$ we find the two parameters
$\beta_1 = 30800/27$ and $\beta_2 = 1100/9$ parametrising
\beq
(L_1)^6 \; |7;1 \rangle &=& \beta_1 \,  |1 \rangle \nonumber \\
(L_1)^3\, L_3 \; |7;1 \rangle &=& \beta_2 \,  |1 \rangle \; .
\eeq
${\cal R}^{(2)} (1,7)$ has been parametrised in such a way that the
monomials $(L_1)^6$ and $(L_1)^3\, L_3$ are the only ones 
with a non-trivial action on $|7;1 \rangle$.
This yields the following mappings of single Virasoro modes
\beq
L_1 \;  |7;1 \rangle &=& \frac{11}{729} \, \left( \frac{857}{2} L_{-5}
- 473\, L_{-4}\, L_{-1}- 721\, L_{-3}\, L_{-2} +\frac{4279}{12} \, L_{-3}\, L_{-1}^2 \right. \nonumber \\
&& \qquad \left. +\frac{809}{2} \, L_{-2}^2 \, L_{-1} - \frac{1295}{12} \, L_{-2}\, L_{-1}^3 \right) |1 \rangle \nonumber \\
L_2 \;  |7;1 \rangle &=& 0  \nonumber \\
L_3 \;  |7;1 \rangle &=& \frac{275}{324} \, \left( L_{-1}^3 
+ 12\, L_{-2}\, L_{-1} - 24\, L_{-3} \right)  \, |1 \rangle \nonumber \\
L_p \;  |7;1 \rangle &=& 0  \qquad  \forall p \ge 4 \; .
\eeq
\item  For ${\cal R}^{(2)} (2,7)$ the two parameters are
$\beta_1 = -880$ and $\beta_2 = -440/3$ parametrising
\beq
(L_1)^5 \; |7;1 \rangle &=& \beta_1 \,  |2 \rangle \nonumber \\
(L_1)^2\, L_3 \; |7;1 \rangle &=& \beta_2 \,  |2 \rangle \; .
\eeq
This yields the following mappings of single Virasoro modes
\beq
L_1 \;  |7;1 \rangle &=& \frac{5}{17} \, \left( - \frac{143}{3} L_{-4}
+ \frac{44}{3}\, L_{-3}\, L_{-1} +\frac{49}{3} \, L_{-2}^2 - 6 \, L_{-2}\, L_{-1}^2 \right) |2 \rangle \nonumber \\
L_2 \;  |7;1 \rangle &=& 0  \nonumber \\
L_3 \;  |7;1 \rangle &=&  -\frac{20}{3}\left( L_{-1}^2 - \frac{3}{2} \, L_{-2} \right) \, |2 \rangle \nonumber \\
L_p \;  |7;1 \rangle &=& 0  \qquad   \forall p \ge 4 \; .
\eeq
\end{itemize}
In both cases $L_3$ actually maps $| h+m;1 \rangle$
to a multiple of the unique descendant of $| h \rangle$
on level $l-1$ which is annihilated by $L_p \; \forall p \ge 2$.

We conjecture that it actually suffices to have two parameters
in order to characterise rank $2$ representations of type
E and F.
As we have a nullvector already on the lower level $l$
this unique state on level $m-1$ which is annihilated by 
$L_p \; \forall p \ge 2$ is a descendant of this nullvector.
Hence, we want to lift the restrictions by incorporating
one further non-zero mapping, the mapping by $L_{m-(l-1)}$
to the unique state on level $l-1$ which is annihilated 
by $L_p \; \forall p \ge 2$, in order to ensure that we do not
map into pure descendants of the lower nullvector.
This is indeed equivalent to demanding that the application
of all normal ordered positive Virasoro monomials of length $m$ annihilates
$| h+m;1 \rangle$ except for $L_1^m$ and $L_1^{l-1} \, L_{m-(l-1)}$.

Again, we want to stress that also these rank $2$ representations
can be thought of to be constructed by an indecomposable
connection of several (to be precise four) irreducible
representations. As for the rank $2$ representations on the 
border the former irreducible representations are
signified by the black dots in the corresponding figures
\ref{rep_on_bulk2} C-F.

For all bulk rank $2$ representations listed in table \ref{rank2}
we were actually able to see the level of the first logarithmic
nullvector already in the calculated fusion spectrum.
We also confirmed this lowest
logarithmic nullvector using the algorithm of \cite{EF05}.
It is remarkable to see that for the
bulk rank $2$ representations ${\cal R}^{(2)} (0,1)_i$
and ${\cal R}^{(2)} (0,2)_i$, $i=5,7$, we encounter the 
first nullvectors already on lower levels
than the ones proposed in \cite{EF05}. However,
the solutions in \cite{EF05} are given for general
$\beta$; it is only for the special $\beta$s in 
table \ref{rank2} that we encounter solutions even on lower
levels.

\subsubsection[Representations of rank $3$]{Representations of rank $\mathbf{3}$\label{discussion_c0_rank3}}

Representations of rank $3$ only appear for weights
in the bulk. In the following we will discuss the
three lowest examples explicitly.

\begin{table} \caption{Number of states for ${\cal R}^{(3)} (0,0,1,1)$}
\begin{center}
\renewcommand{\arraystretch}{1.2}
\begin{tabular}{c || c | c | c || c | c}
level &\multicolumn{3}{c||}{number of states} & total number & new null \\
& Jordan & Jordan & Jordan & of state & subrepresentations \\
& level $0$ & level $1$ & level $2$ & & \\ \hline
$0$ & $1$ & $1$ & -- & $2$ & -- \\
$1$ & $2$ & $1$ & $1$ & $4$ & -- \\
$2$ & $3$ & $1$ & $1$ & $5$ & $1$ \\
$3$ & $5$ & $2$ & $2$ & $9$ & -- \\
$4$ & $8$ & $3$ & $3$ & $14$ & -- \\
$5$ & $11$ & $5$ & $4$ & $20$ & $2$ \\
$6$ & $17$ & $7$ & $6$ & $30$ & -- \\
$7$ & $23?$ & $11?$ & $8?$ & $42?$ & $2$
\end{tabular}
\end{center} \label{rank3_states}
\end{table}

\vspace{0.2cm}
\noindent
{$\bf {\cal R}^{(3)} (0,0,1,1)$:} Although we only
encounter a rank $3$ indecomposable structure in this
representation, we nevertheless need four states to
generate it. We will see that this is necessary and
natural by two different ways of visualising
the nullvector structure of ${\cal R}^{(3)} (0,0,1,1)$.

\begin{figure}
\centering \leavevmode
\psfig{file=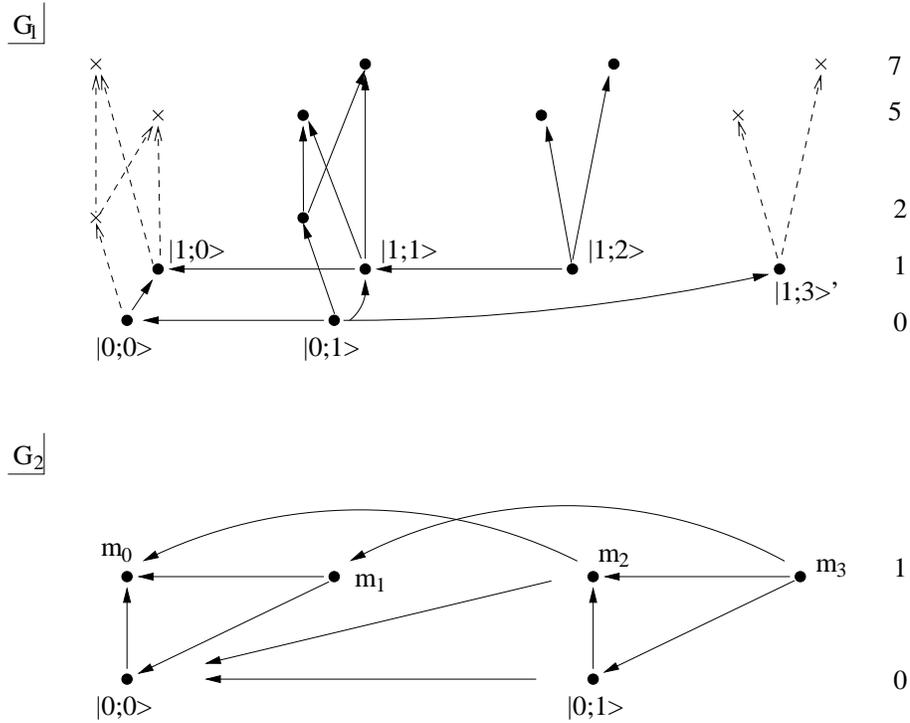,width=12cm}
\caption{Two ways to visualise ${\cal R}^{(3)} (0,0,1,1)$}
\label{rep_rank3a}
\end{figure}

The first way starts out with the Jordan diagonalisation
of the representation and is shown in figure \ref{rep_rank3a} $\mbox{G}_1$.
We have two groundstates $|0;i \rangle$, $i=0,1$, 
at level $0$ which are
interrelated by the rank $2$ indecomposable action of $L_0$
\beq
L_0 \, |0;1 \rangle = |0;0 \rangle &\qquad& L_0 \, |0;0 \rangle = 0 \; .
\eeq
On level $1$, the level of the first possible nullvector
on $|0;i \rangle$, the Jordan cell is enhanced to rank $3$
\beq
L_0 \, |1;2 \rangle &=& |1;2 \rangle + |1;1 \rangle \nonumber \\
L_0 \, |1;1 \rangle &=& |1;1 \rangle + |1;0 \rangle \nonumber \\
L_0 \, |1;0 \rangle &=& |1;0 \rangle \; .
\eeq
A further fourth state of weight $1$ decouples in the $L_0$ action
\beq
L_0 \, |1;3 \rangle^{\prime} &=& |1;3 \rangle^{\prime} \; ;
\eeq
but this seeming decoupling is deceiving as the singular
descendant of $|0;1 \rangle$ is actually composed of the
sum of the Jordan cell state $|1;1 \rangle$ and the ``decoupling'' state
$|1;3 \rangle^{\prime}$; indeed the action of $L_{-1}$ on $|0;i \rangle$, $i=0,1$, 
is given by
\beq
L_{-1} \, |0;0 \rangle &=& |1;0 \rangle \nonumber \\
L_{-1} \, |0;1 \rangle &=& |1;1 \rangle + |1;3 \rangle^{\prime} \; .
\eeq
The further nullvector structure is also depicted in figure \ref{rep_rank3a} $\mbox{G}_1$. 
We were able to calculate all states up to level $6$ explicitly.
The total number of states is also attainable for level $7$, but in
an indirect way via the second view on this representation to be
discussed below. We give a list of the total number of states in
table \ref{rank3_states}; we split the number according to the
position of the state in the Jordan cell, which we also call the
``Jordan level'' of that state in the cell
(following \cite{Flohr01} the enumeration starts with $0$). E.g.\
for level $5$ there are four rank $3$ Jordan cells,
one additional rank $2$ Jordan cell plus the four
which are subcells of a rank $3$ cell as well as
six additional single eigenvalues.

Now we find one nullvector on level $2$, the singular descendant $L_{-2} \,  |0;0 \rangle$.
This nullvector has two singular descendants on level $5$ and $7$
which are also nullvectors of $|1;0 \rangle = L_{-1} \, |0;0 \rangle$ 
due to the bulk embedding structure. The ``decoupling'' state $|1;3 \rangle^{\prime}$
generates an irreducible representation with null singular vectors
on level $5$ and $7$. These vectors as well as their descendants
are the only nullvectors up to level $7$. We do not yet
encounter a logarithmic nullvector up to this level.

In a second way of visualising this representation we can actually
view it as an indecomposable combination of rank $2$ representations
whose structure is given in much the same way as in figure \ref{rep_on_border} A;
we only have to replace the two lowest black dots by the two rank $2$
representations ${\cal R}^{(2)} (0,1)_5$ and the higher by 
${\cal R}^{(2)} (2,7)$. Surprisingly, we can also put ${\cal R} (0,0,1,1)$
in a likewise form with the lower two black dots replaced by
${\cal R}^{(2)} (0,1)_7$ and the higher with ${\cal R}^{(2)} (2,5)$.
Let us see how this comes about. 

We choose the setup for the lowest levels
as depicted in figure \ref{rep_rank3a} $\mbox{G}_2$: we have two se\-parate rank $2$
representations with lowest states $|0;0 \rangle$ respectively
$|0;1 \rangle$. $|0;0 \rangle$ has a logarithmic partner at level $1$, called
$m_1$, to its singular descendant $m_0 := L_{-1}\, |0;0 \rangle = |1;0 \rangle$.
Likewise, $|0;1 \rangle$ has a logarithmic partner at level $1$, called
$m_3$, to its singular descendant 
$m_2 := L_{-1}\, |0;1 \rangle = |1;1 \rangle + |1;3 \rangle^{\prime}$.
Furthermore, both representations are connected by the indecomposable $L_0$ action
on $|0;1 \rangle$. This directly promotes to the indecomposable action
of $L_0$ on their $L_{-1}$ descendants
\beq
L_0\, m_0 = |1;0 \rangle &\qquad& L_0\, m_2 = |1;2 \rangle +|1;0 \rangle
\eeq
as well as the consistent $L_1$ action
\beq
L_1\, m_0 = 0   &\qquad&   L_1\, m_2  =  2\, |0;0 \rangle  \; .
\eeq
Now we still have to check whether we can find $m_1$ and $m_3$ which fit this setup.
We express $m_1$ and $m_3$ as linear combinations of the level $1$ states $|1;j \rangle$, $j=1,2,3$
and impose the following conditions and parameters:
\begin{itemize}
  \item $L_0\, m_1 = m_1 +m_0$,
  \item $L_1\, m_1 = \xi_1 |0;0 \rangle$,
  \item $L_0\, m_3 = m_3 +m_2+r\, m_1+s\, m_0$,
  \item $L_1\, m_3 = \xi_2 |0;1 \rangle + \xi_3 |0;0 \rangle$.
\end{itemize}
We find that $s$ and $\xi_3$ are actually irrelevant parameters which
can be set to $0$ using the residual freedom. $r$ and $\xi_2$ are functions of
$\xi_1$. Hence there is one free parameter in this representation.
It is most useful to just express $r$ in terms of $\xi_1$ and then
study the relation between $\xi_1$ and $\xi_2$ as these parameters
are the defining parameters of the rank $2$ representations
we started with. The relation is given by
\beq
(12\, \xi_1 +1) \, (12\, \xi_2 +1) &=& 25 \; .
\eeq 
There are only two solutions which fit the bulk rank $2$ spectrum
discussed above; they also seem to be the most natural ones
\beq
\xi_1 = \xi_2 &=& -\,\frac{1}{2} \nonumber \\
\xi_1 = \xi_2 &=& \frac{1}{3} \; .
\eeq
These two solutions give exactly the lower level rank $2$ representations
which we proposed above to be the two lower dots in figure \ref{rep_on_border} A.
In order to get the respective representation corresponding to the higher dot
of figure \ref{rep_on_border} A
we just have to count the number of states and compare these.
As this higher rank $2$ representation only ``fills up'' states
which would be null in a pure rank $2$ setting, but are not null in
this rank $3$ setting (e.g.\ $L_{-2} |0;1 \rangle$) we do not need any 
further parameters to describe this representation.

Therefore there is only the one additional parameter $r$ besides the parameters
of the ingredient rank $2$ representations which we need
to describe ${\cal R} (0,0,1,1)$ which is given in terms of $\xi_1$
\beq
r &=& \frac{-25}{12\, \xi_1 +1} \; .
\eeq

Finally, we still need to explain how to determine the number of states
for level $7$ (as in table \ref{rank3_states}). If we want to use the total
number of states to determine this higher black dot in the above setting
we can decide this question knowing the total number of states of ${\cal R}^{(3)} (0,0,1,1)$
up to level $5$ (${\cal R}^{(2)} (2,5)$ and ${\cal R}^{(2)} (2,5)$ already differ at their
third level). But then we can in turn easily use this setting in order
to determine the number of states for any higher level.

\vspace{0.2cm}
\noindent
{$\bf {\cal R}^{(3)} (0,0,2,2)$:} The rank $3$ representation
${\cal R}^{(3)} (0,0,2,2)$ looks much the same as the previous one.
Due to the nullvector at level $1$ which appears at a lower
level than the final rank $3$ structure we have to take
some more care in fixing the freedom.

There are again two ways to decompose this rank $3$ representation
into rank $2$ representations. Either we find twice ${\cal R}^{(2)} (0,2)_5$ 
and ${\cal R}^{(2)} (1,7)$ or twice
${\cal R}^{(2)} (0,2)_7$ and ${\cal R}^{(2)} (1,5)$.

\begin{table} \caption{Number of states for ${\cal R}^{(3)} (0,1,2,5)$}
\begin{center}
\renewcommand{\arraystretch}{1.2}
\begin{tabular}{c || c | c | c || c | c}
level &\multicolumn{3}{c||}{number of states} & total number & new null \\
& Jordan & Jordan & Jordan & of state & subrepresentations \\
& level $0$ & level $1$ & level $2$ & & \\ \hline
$0$ & $1$ & -- & -- & $1$ & -- \\
$1$ & $1$ & $1$ & -- & $2$ & -- \\
$2$ & $2$ & $2$ & -- & $4$ & -- \\
$3$ & $3$ & $3$ & -- & $6$ & -- \\
$4$ & $5$ & $5$ & -- & $10$ & -- \\
$5$ & $8$ & $7$ & $1$ & $16$ & --
\end{tabular}
\end{center} \label{rank3_statesb}
\end{table}

\vspace{0.2cm}
\noindent
{$\bf {\cal R}^{(3)} (0,1,2,5)$:} The rank $3$ representation
${\cal R}^{(3)} (0,1,2,5)$ is the only higher rank $3$ representation
which was accessible to our calculations up to that level
at which the rank $3$ structure appears. From our knowledge
of the other towers of representations it should nevertheless be fair
to conjecture that most of the generic features of rank $3$
representations in these $c_{p,q}$ models are already
visible in this example.

\begin{figure}
\centering \leavevmode
\psfig{file=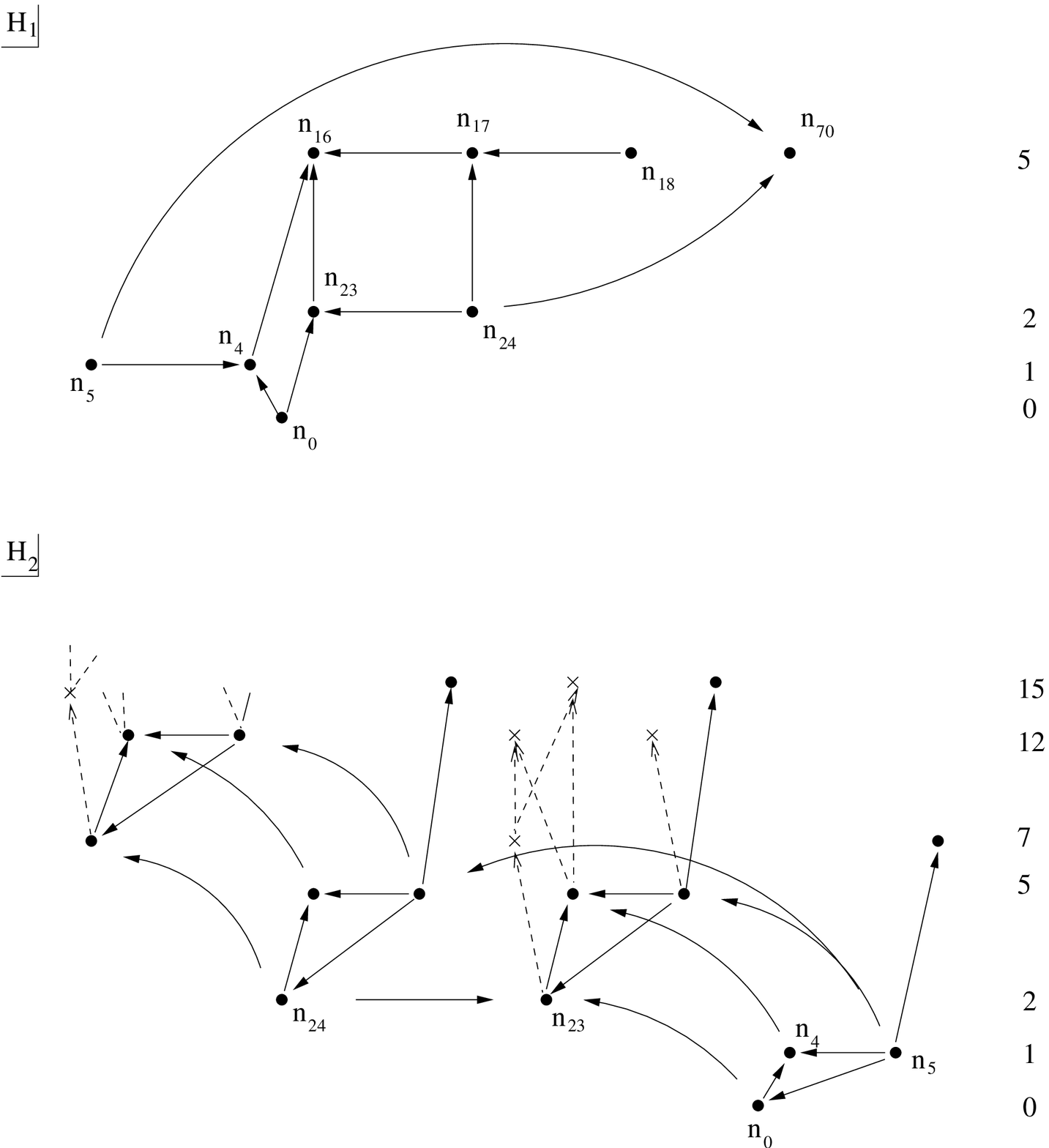,width=12cm}
\caption{Two ways to visualise ${\cal R}^{(3)} (0,1,2,5)$}
\label{rep_rank3b}
\end{figure}

In table \ref{rank3_statesb} we list the number of states
as calculated. We have also included
the basis of states which brings $L_0$ into Jordan diagonal
form in appendix \ref{explicit_L0}.

Again we find two ways to visualise the embedding structure.
The Jordan diagonalisation of $L_0$ gives an embedding
structure of the form depicted in figure \ref{rep_rank3b} $\mbox{H}_1$.
As the situation is even more complicated as for ${\cal R}^{(3)} (0,0,1,1)$
we have labeled the states according to the indexed basis
which is chosen by the computer and listed in appendix \ref{explicit_L0}.
We can see that both singular vectors on $n_0$, i.e.\ $n_4$ at level
$1$ and $n_{23}$ at level $2$, are incorporated into rank $2$
Jordan cells. But nevertheless we do not encounter a first rank $3$
cell until level $5$. The first vector of this rank $3$ Jordan cell, $n_{16}$,
which is a true eigenvector is given by the joint singular vector
on $n_4$ and $n_{23}$. As in the ${\cal R}^{(3)} (0,0,1,1)$ case there is
a vector at level $5$ which seems to decouple from the representation.
But again this decoupling in terms of the $L_0$ action is deceiving
as this vector is a sum of descendants of $n_5$ and $n_{24}$ (see
appendix  \ref{explicit_L0} for the explicit expressions).
Hence, as for ${\cal R}^{(3)} (0,0,1,1)$ this rank $3$ representation
needs four generating states.

As a second way to visualise this representation we conjecture
that we can construct this representation by an indecomposable
combination of the four rank $2$ representations
${\cal R}^{(2)} (0,1)_7$, twice ${\cal R}^{(2)} (2,5)$ and 
${\cal R}^{(2)} (7,12)$ assembled as in figure \ref{rep_on_border} B;
as before we replace the black dots in figure \ref{rep_on_border} B
by the respective
rank $2$ representations. The picture that emerges is depicted
in figure \ref{rep_rank3b} $\mbox{H}_2$. What are the hints at such
a deconstruction of ${\cal R}^{(3)} (0,1,2,5)$? First of all
we find that
\beq
L_1 \, n_5 &=& -\frac{1}{2} n_0 \; ;
\eeq
this equation reproduces the defining $\beta$ parameter
of ${\cal R}^{(2)} (0,1)_7$. Hence, the two lowest generating
states of ${\cal R}^{(3)} (0,1,2,5)$, $n_0$ and $n_5$, generate
a tower of states which resembles the rank $2$ representation
${\cal R}^{(2)} (0,1)_7$. The only difference to a true subrepresentation
${\cal R}^{(2)} (0,1)_7$ is that the usual first nullvector on $n_0$, 
$n_{23} = L_{-2}\, n_0$, is rendered non-null by its inclusion into
an indecomposable rank $2$ cell with
\beqn \label{n23_n24}
L_0\, n_{23} &=& 2\, n_{23} \nonumber \\
L_0\, n_{24} &=& 2\, n_{24} + n_{23} \; .
\eeqn
In the  ${\cal R}^{(3)} (0,0,1,1)$ case we have seen an example
that the indecomposable connection of two rank $2$ cells of the
same type of rank $2$ representation produces a rank $3$ cell
and a seemingly decoupling further state. But this is exactly
the structure we discover in this case at level $5$---a rank $3$
cell
\beq
L_0\, n_{16} &=& 2\, n_{16} \nonumber \\
L_0\, n_{17} &=& 2\, n_{17} + n_{16} \nonumber \\
L_0\, n_{18} &=& 2\, n_{18} + n_{17}
\eeq
as well as a seemingly decoupling state $n_{70}$. Hence, we
conjecture that $n_{23}$ and $n_{24}$ are actually both the lower generators
of towers of states which both resemble ${\cal R}^{(2)} (2,5)$,
but which are indecomposably connected according to
equation \mref{n23_n24}. This structure up to level $5$
is in perfect agreement with the total count of states given
in table \ref{rank3_statesb}.
Unfortunately, we cannot say anything about the embedding
structure or the count of states for higher levels and, thus,
the inclusion of the fourth rank $2$
representation ${\cal R}^{(2)} (7,12)$ is highly conjectural.
It is only lead by the intuition that in any rank $2$ structure
the first nullvectors on the true eigenstate have to have non-null
corresponding states on the side of the logarithmic partner
due to the non-degeneracy of the Shapovalov form.

There is, however, one further way of looking at the situation.
Let us regard the fusion equation
\beq
{\cal V} (1/3) \otimesf {\cal R}^{(2)} (1,5) &=& {\cal R}^{(3)} (0,1,2,5) \; .
\eeq
But as described before
we can also view ${\cal R}^{(2)} (1,5)$ as a combination of
four towers of states which resemble their irreducible counterparts
in terms of numbers of states and singular vectors but which
are indecomposably connected to form ${\cal R}^{(2)} (1,5)$. 
In this case, we can think of ${\cal R}^{(2)} (1,5)$ as being
constructed by indecomposably connecting ${\cal V} (1)$, twice
${\cal V} (5)$ as well as  ${\cal V} (12)$. But the fusion
of its single constituents should be consistent with
the fusion of ${\cal R}^{(2)} (1,5)$ itself. Therefore,
inspecting the fusion rules (the first two calculated, see
appendix \ref{fusion_c0}, the third inferred from the fusion rules of
section \ref{general})
\beq
{\cal V} (1/3) \otimesf {\cal V} (1) &=& {\cal R}^{(2)} (0,1)_7 \nonumber \\
{\cal V} (1/3) \otimesf {\cal V} (5) &=& {\cal R}^{(2)} (2,5) \nonumber \\
{\cal V} (1/3) \otimesf {\cal V} (12) &=& {\cal R}^{(2)} (7,12)
\eeq
we are again lead to the conjecture that we can build 
${\cal R}^{(3)} (0,1,2,5)$ by indecomposably connecting
${\cal R}^{(2)} (0,1)_7$, twice ${\cal R}^{(2)} (2,5)$ and 
${\cal R}^{(2)} (7,12)$.

A similar inspection of the fusion equation
\beq
{\cal V} (5/8) \otimesf {\cal R}^{(2)} (5/8,21/8) &=& {\cal R}^{(3)} (0,1,2,5)
\eeq
leads to the conjecture that we can also construct
${\cal R}^{(3)} (0,1,2,5)$ by indecomposably connecting
${\cal R}^{(2)} (0,2)_7$, twice ${\cal R}^{(2)} (1,5)$ and 
${\cal R}^{(2)} (7,15)$. This is also in perfect agreement
with the total count of states up to the accessible level $5$
and, furthermore, a nice generalisation of the ${\cal R}^{(3)} (0,0,1,1)$
case, which can also be constructed out of rank $2$ representations
in two ways. To see the embedding of the lowest rank $2$
representation ${\cal R}^{(2)} (0,2)_7$ is a bit more tricky
this time. For the non-trivial action of the positive
Virasoro modes on the new state $n_{24}$ at level $2$ we find
(after removal of some residual freedom)
\beqn \label{n24}
L_1\, n_{24} &=& 3\, n_5 \nonumber \\
L_2\, n_{24} &=& -\frac{17}{12}\, n_0 \; . 
\eeqn
This state $n_{24}$ is the logarithmic partner of the $n_0$
descendant $n_{23}=L_{-2}\, n_0$. From the description of
${\cal R}^{(2)} (0,2)_7$ we are used to these two states
spanning the lowest Jordan cell. But the non-trivial mapping
of $L_1$ in equation \mref{n24} is in clear contradiction to
${\cal R}^{(2)} (0,2)_7$ having a first nullvector and, hence, no
state at level $1$. Furthermore, we do not recover the correct
$\beta$ value in equation \mref{n24}.
But we have forgotten to take into account that due to the 
absence of this nullvector on level $1$ the level $2$ singular
vector has shifted to
\beq
n_{23}^{\prime} &=& n_{23} - \frac{3}{2}\, n_4 \; .
\eeq
The correct logarithmic partner to $n_{23}^{\prime}$ is given by
\beq
n_{24}^{\prime} &=& n_{24} + \frac{9}{8}\, n_4  - \frac{3}{2}\, n_5 \; .
\eeq
Indeed, if we calculate the action of the positive Virasoro modes
for $n_{24}^{\prime}$ we find the desired properties
\beq
L_1\, n_{24}^{\prime} &=& 0 \nonumber \\
L_2\, n_{24}^{\prime} &=& \frac{5}{6}\, n_0 \; . 
\eeq

\vspace{0.2cm}
\noindent
{$\bf {\cal R}^{(3)} (0,1,2,7)$:} Although the rank $3$
representation ${\cal R}^{(3)} (0,1,2,7)$ is not accessible
to our computational power we can nevertheless tackle
its decomposition in the same way as the last method in the
preceding case. For this argument we take the
appearances of ${\cal R}^{(3)} (0,1,2,7)$ for granted as
conjectured in appendix \ref{fusion_c0}. Then looking
at the fusion equation
\beq
{\cal V} (5/8) \otimesf {\cal R}^{(2)} (1/8,33/8) &=& {\cal R}^{(3)} (0,1,2,7)
\eeq
we conjecture that  ${\cal R}^{(3)} (0,1,2,7)$ can be constructed
by indecomposably connecting the four rank $2$ representations
${\cal R}^{(2)} (0,1)_5$, twice ${\cal R}^{(2)} (2,7)$ and 
${\cal R}^{(2)} (5,12)$. Similarly, looking at
\beq
{\cal V} (1/3) \otimesf {\cal R}^{(2)} (1/3,10/3) &=& {\cal R}^{(3)} (0,1,2,7) \oplus {\cal R}^{(2)} (1/3,10/3)
\eeq
we can think of ${\cal R}^{(3)} (0,1,2,7)$ to be composed of
${\cal R}^{(2)} (0,2)_5$, twice ${\cal R}^{(2)} (1,7)$ and 
${\cal R}^{(2)} (5,15)$.

%%%

\subsection{Explicit calculation of the fusion products}

We have calculated the fusion products of a large variety
of representations in the augmented $c_{2,3} = 0$ model.
To do this we have used the Nahm algorithm described in section \ref{nahm}
to determine the fusion product of irreducible
and rank $1$ representations with themselves as well as with the
lowest lying and first excited rank $2$ representations.
In order to show that the fusion algebra indeed closes
we have used the symmetry and associativity of the
fusion product in order to calculate the fusion of higher
rank representations. We have also used these conditions
in order to perform consistency checks as described in the
next subsection.
The results itself are listed in appendix \ref{fusion_c0}.

In section \ref{general} we want to propose a generalisation
of the BPZ and $c_{p,1}$ fusion rules which is applicable
to all augmented $c_{p,q}$ models and, hence, also describes
in a unifying way the fusion of this model. But as these general
rules look quite complex it is also possible to find 
simplified versions for the augmented $c_{2,3} = 0$ model,
e.g.\
\beq
{\cal V} (5/8) \otimesf {\cal W} (1/3|i) &=& {\cal W} (-1/24|i) \; ,
\eeq
where ${\cal W} (h|i)$ signifies the $i$th element in the weight chain
starting with $h$.

%%%

\subsection{Consistency of fusion products\label{discussion_c0_consistency}}

A consistent fusion product has to
obey two main properties, symmetry and associati\-vi\-ty. We have used both
these properties for consistency checks of the chosen spectrum and the
performed calculation as well as for the determination of the fusion
product of higher rank representations.

%Several examples of such consistency checks are given in the lists in appendix \ref{fusion_c0}. 
The main implication of this consistency, however,
is the absence of an irreducible representation of weight $h=0$ in the
spectrum, call it ${\cal V} (0)$. The representation ${\cal V} (0)$ would
be endowed with nullvectors on level $1$ and $2$ and would, hence, only consist
of the one groundstate. Performing the Nahm algorithm of section \ref{nahm}
we get the following fusion products
\beqn
{\cal V} (0) \otimesf {\cal V} (0) &=& {\cal V} (0) \label{first_line} \\
{\cal V} (0) \otimesf {\cal V} (h) &=& 0 \quad \forall \, h \in \left\{\frac{5}{8},\frac{1}{3},\frac{1}{8},\frac{-1}{24},
\frac{33}{8},\frac{10}{3},\frac{21}{8},2,1,7,5\right\} \; . \label{second_line}
\eeqn
On the other hand using just the equations \mref{second_line} and the fusion rules
of appendix \ref{fusion_c0} we arrive at
\beq
\Big( {\cal V} (0) \otimesf \Big({\cal V} (2)  \otimesf {\cal V} (2)\Big) \Big) &=&
\Big({\cal V} (0)  \otimesf {\cal V} (0)\Big) \oplus \Big({\cal V} (0)  \otimesf {\cal V} (2)\Big) \nonumber \\
&=& {\cal V} (0)  \otimesf {\cal V} (0)
\eeq
as well as by associativity at
\beq
\Big( \Big({\cal V} (0) \otimesf {\cal V} (2)\Big)  \otimesf {\cal V} (2) \Big) &=&
0  \otimesf {\cal V} (2) \nonumber \\
&=& 0 \; .
\eeq
(Similar equations can be obtained involving the other ${\cal V} (h)$ with $h$ from \mref{second_line}.)
This argument thus implies
\beq
{\cal V} (0)  \otimesf {\cal V} (0) &=& 0 \; .
\eeq
But this is in clear contradiction to \mref{first_line}. Fortunately, however,
${\cal V} (0)$ completely decouples from the rest of the fusion (as one can see in appendix
\ref{fusion_c0}). Hence, the contradiction is easily solved by simply excluding
${\cal V} (0)$ from the spectrum. 

On the other hand the representations ${\cal R}^{(2)} (0,1)_5$ and  ${\cal R}^{(2)} (0,1)_7$
contain a state with weight $0$ which generates a subrepresentation ${\cal R}^{(1)} (0)_1$.
This subrepresentation is indecomposable but neither is it irreducible nor
does it exhibit any higher rank behaviour. It only exists as a subrepresentation
as it needs the embedding into the rank $2$ representation in order not to
have nullvectors at both levels $1$ and $2$.
But, nevertheless, being a subrepresentation of a representation in the spectrum
it has to be included into the spectrum, too. Similarly, the representations
${\cal R}^{(2)} (0,2)_5$ and  ${\cal R}^{(2)} (0,2)_7$ contain a rank $1$ subrepresentation
${\cal R}^{(1)} (0)_2$. 
However, looking at the fusion rules which we calculate for these two rank $1$ representations
(see appendix \ref{fusion_c0}), especially
\beq
{\cal R}^{(1)} (0)_2 \otimesf {\cal V} (h) &=& {\cal V} (h)\quad \forall \, h \in \left\{\frac{5}{8},\frac{1}{3},\frac{1}{8},\frac{-1}{24},
\frac{33}{8},\frac{10}{3},\frac{21}{8},2,1,7,5\right\} \; ,
\eeq
we see that the situation is after all not really too bad: 
${\cal R}^{(1)} (0)_2$ behaves much more like the true vacuum representation as the
expected ${\cal V} (0)$ (only regard the behaviour in \mref{second_line}).

We now want to present one nice example how to use the symmetry and associativity
of the fusion product in order to determine the higher rank fusion.
Let us assume that we already know the fusion of irreducible representations
with themselves as well as with rank $2$ representations, i.e.\ the results
we actually calculated using the Nahm algorithm (see appendix \ref{fusion_c0}).
Then we start off with
\beq
\left( \Big({\cal V} (1/8) \otimesf {\cal V} (1/3) \Big) \otimesf {\cal R}^{(2)} (1/3,1/3) \right) &=& 
{\cal R}^{(2)} (1/8,1/8) \otimesf {\cal R}^{(2)} (1/3,1/3) \; .
\eeq
Using associativity we can also calculate
\beq
\lefteqn{\!\!\!\!\!\!\!\!\!\!\!\!\!\!\!\! \left( {\cal V} (1/8) \otimesf \left(({\cal V} (1/3) \otimesf {\cal R}^{(2)} (1/3,1/3)\right) \right) } \nonumber \\ 
&=& \left( {\cal V} (1/8) \otimesf \left({\cal R}^{(3)} (0,0,2,2) \oplus {\cal R}^{(2)} (1/3,1/3)\right) \right) \nonumber \\
&=& 2\, {\cal R}^{(2)} (1/8,1/8) \oplus {\cal R}^{(2)} (5/8,21/8) \oplus  \left( {\cal V} (1/8) \otimesf {\cal R}^{(3)} (0,0,2,2) \right) \; .
\eeq
On the other hand we can use the symmetry as well as the associativity once more
to get
\beq
\lefteqn{\left( {\cal V} (1/3) \otimesf \left({\cal V} (1/8) \otimesf {\cal R}^{(2)} (1/3,1/3)\right) \right) } \nonumber \\
&=& \left( {\cal V} (1/3) \otimesf \left(2\, {\cal R}^{(2)} (1/8,1/8) \oplus {\cal R}^{(2)} (5/8,21/8) \right) \right) \nonumber \\
&=& 4\, {\cal R}^{(2)} (1/8,1/8) \oplus 2\, {\cal R}^{(2)} (5/8,21/8) \oplus {\cal V} (-1/24) \oplus 2\, {\cal V} (35/24) \oplus {\cal V} (143/24) \; .
\eeq
By comparison we thus arrive at already two new higher rank fusion products
\beq
\lefteqn{{\cal R}^{(2)} (1/8,1/8) \otimesf {\cal R}^{(2)} (1/3,1/3)} \nonumber \\
&=& 4\, {\cal R}^{(2)} (1/8,1/8) \oplus 2\, {\cal R}^{(2)} (5/8,21/8) \oplus {\cal V} (-1/24) \oplus 2\, {\cal V} (35/24) \oplus {\cal V} (143/24) \nonumber \\
\lefteqn{{\cal V} (1/8) \otimesf {\cal R}^{(3)} (0,0,2,2)} \nonumber \\
&=& 2\, {\cal R}^{(2)} (1/8,1/8) \oplus {\cal R}^{(2)} (5/8,21/8) \oplus {\cal V} (-1/24) \oplus 2\, {\cal V} (35/24) \oplus {\cal V} (143/24) \; .
\eeq
The complete list of the higher rank fusion products which we calculated is given
in appendix \ref{fusion_c0}.

%%%%%%%%%%%%%%%%%%%%%%%%%%%%%%%%%%%%%%%%%%%%%
%%%%%%%%%%%%%%%%%%%%%%%%%%%%%%%%%%%%%%%%%%%%%
%%%%%%%%%%%%%%%%%%%%%%%%%%%%%%%%%%%%%%%%%%%%%

\section{Explicit discussion of the augmented Yang--Lee model\label{discussion_YL}}

\renewcommand{\arraystretch}{0.03}
\begin{table} \caption{Kac table for $c_{2,5}=-22/5$}
\begin{center}
\begin{tabular}{cc | ccccc}
\multicolumn{7}{c}{\rule[-.6em]{0cm}{1.6em} $\qquad\quad  s$} \\
&\rule[-.6em]{0cm}{1.6em} & $1$ & $2$ & $3$ & $4$ & $5$ \\ \cline{2-7}
& & \multicolumn{5}{c}{} \\
&\rule[-.6em]{0cm}{1.6em} $1$ & $0$ & \cellcolor{gray}{.9}{$\frac{11}{8}$} & $4$ & \cellcolor{gray}{.9}{$\frac{63}{8}$} & $13$ \\
&\rule[-.6em]{0cm}{1.6em} $2$ & $-\frac{1}{5}$ & \cellcolor{gray}{.9}{$\frac{27}{40}$} & $\frac{14}{5}$ & \cellcolor{gray}{.9}{$\frac{247}{40}$} & $\frac{54}{5}$ \\
&\rule[-.6em]{0cm}{1.6em} $3$ & $-\frac{1}{5}$ & \cellcolor{gray}{.9}{$\frac{7}{40}$} & $\frac{9}{5}$ & \cellcolor{gray}{.9}{$\frac{187}{40}$} & $\frac{44}{5}$ \\
&\rule[-.6em]{0cm}{1.6em} $4$ & $0$ & \cellcolor{gray}{.9}{$-\frac{1}{8}$} & $1$ & \cellcolor{gray}{.9}{$\frac {27}{8}$} & $7$ \\
&\rule[-.6em]{0cm}{1.6em} $5$ & \multicolumn{1}{>{\columncolor[gray]{.9}[0.205cm][0.22cm]}c}{$\frac{2}{5}$} 
& \cellcolor{gray}{.7}{$-\frac{9}{40}$} 
& \cellcolor{gray}{.9}{$\frac{2}{5}$} & \cellcolor{gray}{.7}{$\frac {91}{40}$} & \cellcolor{gray}{.9}{$\frac{27}{5}$} \\
&\rule[-.6em]{0cm}{1.6em} $6$ & $1$ & \cellcolor{gray}{.9}{$-\frac{1}{8}$} & $0$ & \cellcolor{gray}{.9}{$\frac{11}{8}$} & $4$ \\
$r$ &\rule[-.6em]{0cm}{1.6em} $7$ & $\frac{9}{5}$ & \cellcolor{gray}{.9}{$\frac{7}{40}$} & $-\frac{1}{5}$ & \cellcolor{gray}{.9}{$\frac{27}{40}$} & $\frac{14}{5}$ \\
&\rule[-.6em]{0cm}{1.6em} $8$ & $\frac{14}{5}$ & \cellcolor{gray}{.9}{$\frac{27}{40}$} & $-\frac{1}{5}$ & \cellcolor{gray}{.9}{$\frac{7}{40}$} & $\frac{9}{5}$ \\
&\rule[-.6em]{0cm}{1.6em} $9$ & $4$ & \cellcolor{gray}{.9}{$\frac{11}{8}$} & $0$ & \cellcolor{gray}{.9}{$-\frac{1}{8}$} & $1$ \\
&\rule[-.6em]{0cm}{1.6em} $10$ & \multicolumn{1}{>{\columncolor[gray]{.9}[0.205cm][0.22cm]}c}{$\frac{27}{5}$}
& \cellcolor{gray}{.7}{$\frac{91}{40}$} 
\rule[-.6em]{0cm}{1.6em}& \cellcolor{gray}{.9}{$\frac{2}{5}$} & \cellcolor{gray}{.7}{$-\frac{9}{40}$} & \cellcolor{gray}{.9}{$\frac{2}{5}$} \\
&\rule[-.6em]{0cm}{1.6em} $11$ & $7$ & \cellcolor{gray}{.9}{$\frac {27}{8}$} & $1$ & \cellcolor{gray}{.9}{-$\frac{1}{8}$} & $0$ \\
&\rule[-.6em]{0cm}{1.6em} $12$ & $\frac{44}{5}$ & \cellcolor{gray}{.9}{$\frac{187}{40}$} & $\frac{9}{5}$ & \cellcolor{gray}{.9}{$\frac{7}{40}$} & $-\frac{1}{5}$ \\
&\rule[-.6em]{0cm}{1.6em} $13$ & $\frac{54}{5}$ & \cellcolor{gray}{.9}{$\frac{247}{40}$} & $\frac{14}{5}$ & \cellcolor{gray}{.9}{$\frac{27}{40}$} & $-\frac{1}{5}$ \\
&\rule[-.6em]{0cm}{1.6em} $14$ & $13$ &\cellcolor{gray}{.9}{ $\frac {63}{8}$} & $4$ & \cellcolor{gray}{.9}{$\frac{11}{8}$} & $0$
\end{tabular}
\end{center} \label{Kacminus22over5}
\end{table}
\renewcommand{\arraystretch}{1.3}

Unfortunately a complete exploration of the low lying spectrum
of the next easiest general augmented model, the augmented Yang--Lee model
at $c_{2,5} = -22/5$, is not yet possible due to limitations
on the computational power. Nevertheless, we were able to compute
most of the crucial features which we observed in the fusion of the augmented
$c_{2,3} = 0$ model, including the lowest rank $2$ and rank $3$ representations
as well as the absence of irreducible representations corresponding
to the original minimal model.
We also give quite some examples of fusion products which confirm
the general fusion rules conjectured in section \ref{general}.
The explicit results are listed in appendix \ref{fusion_cminus22over5}.

The Kac table of $c_{2,5}=-22/5$ is depicted in table \ref{Kacminus22over5}. 
We encounter two bulk weight chains
\beq
W_{(1,1)}^{\mbox{\scriptsize bulk, YL}} &:=& \{\{0\}, \{1,4\}, \{7,13\}, \dots \} \nonumber \\
W_{(2,1)}^{\mbox{\scriptsize bulk, YL}} &:=& \left\{\left\{-\frac{1}{5} \right\},\left \{\frac{9}{5},\frac{14}{5}\right\}, 
\left\{\frac{44}{5},\frac{54}{5}\right\}, \dots \right\} 
\eeq
as well as five border weight chains
\beq
W_{(1,2)}^{\mbox{\scriptsize border, YL}} &:=& \left\{\frac{11}{8},\frac{27}{8}, \frac{155}{8},  \dots \right\} \nonumber \\
W_{(2,2)}^{\mbox{\scriptsize border, YL}} &:=& \left\{\frac{27}{40}, \frac{187}{40}, \frac{667}{40}, \dots \right\} \nonumber \\
W_{(3,2)}^{\mbox{\scriptsize border, YL}} &:=& \left\{\frac{7}{40}, \frac{247}{40}, \frac{567}{40}, \dots \right\} \nonumber \\
W_{(4,2)}^{\mbox{\scriptsize border, YL}} &:=& \left\{-\frac{1}{8}, \frac{63}{8}, \frac{95}{8}, \dots \right\} \nonumber \\
W_{(5,1)}^{\mbox{\scriptsize border, YL}} &:=& \left\{\frac{2}{5}, \frac{27}{5}, \frac{77}{5}, \dots \right\}
\eeq
and a chain $\{-9/40,91/40,391/40,\dots\}$ of weights on the corners.
In table \ref{rank2_YL} we present all rank $2$ representations which we found in
our sample calculations (see appendix \ref{fusion_cminus22over5})
as well as their defining parameters. This is indeed the complete
spectrum of lowest rank $2$ representations to be expected according to
the general considerations of section \ref{general}.

\renewcommand{\arraystretch}{1.2}
\begin{table} \caption{Specific properties of rank $2$ representations in $c_{2,5}=-22/5$}
\begin{center}\footnotesize
\begin{tabular}{l || c | c | c | c | c | c}
& $\beta_1$ & $\beta_2$ & level of & level of first & level of first & type\\
&&& log.\ partner & nullvector & log.\ nullvector & \\ \hline
${\cal R}^{(2)} (11/8,11/8)$ & -- & -- & $0$ & $2$ & $18$ & A \\
${\cal R}^{(2)} (27/40,27/40)$ & -- & -- & $0$ & $4$ & $16$ & A \\
${\cal R}^{(2)} (2/5,2/5)$ & -- & -- & $0$ & $5$ & $15$ & A \\
${\cal R}^{(2)} (7/40,7/40)$ & -- & -- & $0$ & $6$ & $14$ & A \\
${\cal R}^{(2)} (-1/8,-1/8)$ & -- & -- & $0$ & $8$ & $12$ & A \\ \hline
${\cal R}^{(2)} (0,1)_7$ & $3/5$ & -- & $1$ & $4$ & $13$ & C \\
${\cal R}^{(2)} (0,1)_{13}$ & $-3/2$ & -- & $1$ & $4$ & $7$ & D \\
${\cal R}^{(2)} (0,4)_7$ & $0$ & $231/50$ & $4$ & $1$ & $13$ & E \\
${\cal R}^{(2)} (0,4)_{13}$ & $0$ & $231/25$ & $4$ & $1$ & $7$ & F \\
${\cal R}^{(2)} (-1/5,9/5)_9$ & $-42/125$ & -- & $2$ & $3$ & $11$ & C \\
${\cal R}^{(2)} (-1/5,9/5)_{11}$ & $21/50$ & -- & $2$ & $3$ & $9$ & D \\
${\cal R}^{(2)} (-1/5,14/5)_9$ & $21/125$ & $21/50$ & $3$ & $2$ & $11$ & E \\
${\cal R}^{(2)} (-1/5,14/5)_{11}$ & $-126/625$ & $-63/125$ & $3$ & $2$ & $9$ & F
\end{tabular}
\end{center} \label{rank2_YL}
\end{table}
\normalsize

As in the $c_{2,3} = 0$ model case the lowest border rank $2$ representations,
given in the first block of table \ref{rank2_YL},
do not need a further parameter for characterisation.
Their structure is again represented by figure \ref{rep_on_border} A.

The structure of the lowest bulk rank $2$ representations,
given in the second block of table \ref{rank2_YL},
is depicted in figure \ref{rep_on_bulk2}; their respective special type
is given in the last column.
The representations of type C and D exhibit their Jordan cell
on the level of the first possible nullvector of the groundstate
and can thus again be described by just one parameter $\beta = \beta_1$
\beq
(L_1)^l \; | h+l; 1 \rangle &=& \beta \; | h \rangle \; ;
\eeq
$l$ denotes the level of the Jordan cell. Besides this we have
$L_p \; | h+l; 1 \rangle = 0$ for all $p \ge 2$ and hence
all other Virasoro monomials
of length $l$ vanish applied to $| h+l; 1 \rangle$.

The representations of type E and F, however, have to accommodate
a first nullvector on the groundstate already below the level
of the Jordan cell. The same difficulties as in the $c_{2,3} = 0$ model
apply. We again need two parameters $\beta_1$ and $\beta_2$
parametrising
\beq
(L_1)^l \; | h+l; 1 \rangle &=& \beta_1 \; | h \rangle \nonumber \\
P(L) \; | h+l; 1 \rangle &=& \beta_2 \; | h \rangle \; ,
\eeq
where we have taken $P(L) = L_4$ for ${\cal R}^{(2)} (0,4)_7$ 
and ${\cal R}^{(2)} (0,4)_{13}$ 
as well as $P(L) = L_2\, L_1$ for ${\cal R}^{(2)} (-1/5,14/5)_9$ and
${\cal R}^{(2)} (-1/5,14/5)_{11}$. All other Virasoro monomials
of length $l$ vanish applied to $| h+l; 1 \rangle$.
This behaviour actually confirms our conjecture of section \ref{discussion_c0_rank2}
that we only need two parameters for this type of representation;
furthermore, the above presented parametrisation is
performed exactly in the proposed way.

Again we checked for the appearance of the lowest
logarithmic nullvector using the algorithm of \cite{EF05}.
This could be successfully done in all cases
listed in table \ref{rank2_YL}.
For this $c_{2,5}=-22/5$ model these nullvector calculations
were actually a nice and necessary check of our proposed
fusion rules as this information was usually not
directly accessible in the fusion spectrum due to
the computational limits on $L$.

As the last issue in the discussion of the augmented
Yang--Lee model we want to have a look at the irreducible
representations with weights in the Kac-table of the corresponding
non-augmented minimal model. There are two possible representations
of this kind in this model, ${\cal V} (h=0)$ with first nullvectors
on levels $1$ and $4$ as well as ${\cal V} (h=-1/5)$ with first
nullvectors on levels $2$ and $3$. Explicit calculations with
the Nahm algorithm lead to
\beqn
{\cal V} (-1/5) \otimesf {\cal V} (-1/5) &=& {\cal V} (0) \oplus {\cal V} (-1/5) \label{first_line_YL} \\
{\cal V} (-1/5) \otimesf {\cal V} (h) &=& 0 \quad \forall \, h \in \left\{\frac{9}{5},\frac{14}{5},4 \right\} \; . \label{second_line_YL}
\eeqn
But again using only the equations of \mref{second_line_YL} and the fusion rules
of appendix \ref{fusion_cminus22over5} we arrive at the contradicting results
\beq
\lefteqn{\Big( {\cal V} (-1/5) \otimesf \Big({\cal V} (14/5)  \otimesf {\cal V} (14/5)\Big) \Big) } \nonumber \\
&=& \Big({\cal V} (-1/5)  \otimesf {\cal V} (0)\Big) \oplus \Big({\cal V} (-1/5)  \otimesf {\cal V} (4)\Big)  
\oplus \Big({\cal V} (-1/5)  \otimesf {\cal V} (-1/5)\Big)  \nonumber \\
&& \qquad \oplus \Big({\cal V} (-1/5)  \otimesf {\cal V} (9/5)\Big) \nonumber \\
&=& \Big({\cal V} (-1/5)  \otimesf {\cal V} (0)\Big) \oplus \Big({\cal V} (-1/5)  \otimesf {\cal V} (-1/5)\Big) 
\eeq
as well as
\beq
\Big( \Big({\cal V} (-1/5) \otimesf {\cal V} (14/5)\Big)  \otimesf {\cal V} (14/5) \Big) 
&=& 0  \otimesf {\cal V} (14/5) \nonumber \\
&=& 0 \; 
\eeq
which lead to
\beq
{\cal V} (-1/5)  \otimesf {\cal V} (0) &=& 0  \nonumber \\
{\cal V} (-1/5)  \otimesf {\cal V} (-1/5) &=& 0 \; .
\eeq
A similar calculation applies again to ${\cal V} (0)$. Thus, we have to discard
${\cal V} (0)$ and ${\cal V} (-1/5)$ from the spectrum.

But with the same reasoning as in the augmented $c_{2,3} = 0$ model case
we encounter rank $1$ subrepresentations of rank $2$ representations
which are generated by states with weights $h=0$ and $h=-1/5$. We
therefore have to include the four rank $1$ indecomposable (but not
irreducible) representations ${\cal R}^{(1)} (0)_1$, ${\cal R}^{(1)} (0)_4$,
${\cal R}^{(1)} (-1/5)_2$ and ${\cal R}^{(1)} (-1/5)_3$ into the spectrum.
${\cal R}^{(1)} (0)_4$ actually acquires the role of the vacuum representation.

%%%%%%%%%%%%%%%%%%%%%%%%%%%%%%%%%%%%%%%%%%%%%
%%%%%%%%%%%%%%%%%%%%%%%%%%%%%%%%%%%%%%%%%%%%%
%%%%%%%%%%%%%%%%%%%%%%%%%%%%%%%%%%%%%%%%%%%%%

\section[Representations and fusion for general augmented $\mathbf{c_{p,q}}$ models]{Representations and fusion product for general augmented $\mathbf{c_{p,q}}$ models\label{general}}

In this section we conjecture fusion rules for general augmented $c_{p,q}$ models;
we furthermore discuss their spectrum of representations 
which is to be consistent with the symmetry and associativity of the fusion product.
This conjecture is mainly substantiated by the thorough exploration of the full fusion algebra 
for the lower lying representations of the augmented model $c_{2,3} = 0$
in section \ref{discussion_c0}. In addition, we have checked the proposed rules as well
as the existence of the lowest rank $2$ and rank $3$ indecomposable representations
for a considerable number of fusion products in the augmented Yang-Lee model
at $c_{2,5} = -22/5$ in section \ref{discussion_YL}. The
explicitly calculated fusion products are listed in the appendices 
\ref{fusion_c0} respectively \ref{fusion_cminus22over5}.

%%%

\subsection{The spectrum of representations}

For weights on the border and corners of the Kac table the situation looks
much the same as in the $c_{p,1}$ augmented model case \cite{GK96}.
States corresponding to weights on corners only generate irreducible
representations. States with weights on the Kac-table border, however, also form rank $2$
indecomposable representations in addition to irreducible ones.
These rank $2$ indecomposable representations are of the same form
as described in \cite{GK96,Roh96}. They are either generated by two
groundstates whose weights are given by the lowest weight of one of the weight chains
corresponding to the Kac table border, see figure \ref{rep_on_border} A,
or by two successive weights in the weight chain, as in figure \ref{rep_on_border} B.
For case A we actually do not need a further parameter to describe
the representation; in case B one further parameter $\beta$ is sufficient---it
is taken to parametrise the equation
\beqn \label{beta}
(L_1)^l \; | h+l; 1 \rangle &=& \beta \; | h \rangle \; .
\eeqn

As we can see in figure \ref{rep_on_border} these representations can actually 
be thought of to consist of several towers of states each generated
from a basic state by application of negative Virasoro modes; these towers of states are very close
to the corresponding irreducible representations of the same weight
as they exhibit the same number of nullvectors at the levels given by the Kac table
and, hence, the same number of states. They (usually) are, however, not irreducible; 
but we can think of them as former irreducible representations which have been 
indecomposably connected 
to form an indecomposable representation. This indecomposable connection
expresses itself by the indecomposable action of $L_0$ 
as well as the non-trivial action of
a few of the positive Virasoro modes. We need
three such towers to build the indecomposable representation
in case A, four in case B.

The weights in the bulk of the Kac table exhibit an even richer
structure as they also form rank $3$ representations of the
Virasoro algebra. Let us first describe the rank $2$ representations.
All possibilities of choosing one weight each from two adjacent sets in the weight chain
will deliver the weights of the two generating states of a bulk rank $2$
representation which is generically unique. 
Only in case that this set of two weights $h(r_1,s_1)$, $h(r_2,s_2)$ 
contains the lowest weight 
of this weight chain, let it be $h(r_1,s_1)$, there
are still two possible representations with this set of weights. The additional 
index is given by the level $w$ of the weight in the second set of the weight chain which is
not(!) the first possible nullvector on $h(r_2,s_2)$, i.e.\ which is not equal to
$h(r_2,s_2) + r_2\, s_2$. The nullvector structure of these bulk
rank $2$ representations is a generalisation of the one on the border.
There are four different types of nullvector structure, depicted
in figure \ref{rep_on_border}. The types C and D are very similar to the
case on the border and also need only one additional parameter $\beta$
taken to parametrise equation \mref{beta}.
The novel feature of types E and F is that they exhibit a first
nullvector already below the level of the first logarithmic partner.
This makes it necessary to at least present two non-zero parameters
to describe these representations. They have to be chosen as described
for the examples in section \ref{discussion_c0}. Conjecturally,
two is also a sufficient number of parameters.
Figure \ref{rep_on_border} also shows that all four possible types
of bulk rank $2$ representations can
be split up into four towers of states in the above described spirit.

But the really new feature of the possible bulk representations
are rank $3$ representations. These need four generating states 
and are found in two different types. The first of these two types
is generated by two states
of the lowest weight of this weight chain as well as two states
which are of the same weight in the second set of this weight chain.
This type is realised for the two lowest rank $3$ representations 
for a given bulk weight chain. The example ${\cal R}^{(3)} (0,0,1,1)$ 
is described in section \ref{discussion_c0_rank3} and depicted
in figure \ref{rep_rank3a}. As discussed there one can actually
build this representation by connecting three rank $2$ bulk
representations roughly according to figure \ref{rep_on_border} A.
Then there is only one additional parameter which is necessary to
describe the representation.

The second type, which is supposedly the generic one, is generated by states of weights
from three adjacent sets in a bulk weight chain---we have to take 
both weights from the middle set
as well as one from each of the other two. This renders all possible
rank $3$ representations. They are uniquely described by this set of
four generating weights. The example ${\cal R}^{(3)} (0,1,2,5)$,
its defining parameters and its composition of rank $2$ representations
have been discussed in section \ref{discussion_c0_rank3}.
This representation could be composed by indecomposably connecting
four bulk rank $2$ representations, roughly following figure
\ref{rep_on_border} B.
Unfortunately, this is the only example of this supposedly generic
type of rank $3$ representation which is at our grasp up to the
level where the rank $3$ behaviour appears for the first time.
Hence, these generalisations have to be taken with the necessary caution,
but nevertheless they are motivated by similar generalisations for
the rank $2$ representations. However, we expect the necessity to introduce
more defining parameters in cases where there appear more nullvectors
on levels lower than the first rank $3$ Jordan cell.

Knowing how to compose the rank $3$ representations of rank $2$ representations
we automatically get the split-up of the whole rank $3$ representation
into towers of states as discussed above for the rank $2$ case.

Last but not least we still have to describe the spectrum of irreducible
representations corresponding to weights in the Kac table bulk.
The calculations for  $c_{2,3} = 0$ and  $c_{2,5} = -22/5$ 
(see sections \ref{discussion_c0_consistency} and \ref{discussion_YL}) 
show that it is
not possible to include irreducible representations corresponding to
weights in the bulk which appear in the Kac table of the corresponding 
non-augmented minimal model $c_{p,q}$,
i.e.\ weights in the Kac table segment $1\le r < q$, $1\le s < p$.
These weights correspond exactly to the lowest weight of each bulk weight chain.
As shown in the examples the inclusion of these irreducible representations would simply 
violate the symmetry and associativity of the fusion product.
On the other hand the fusion of the border and corner irreducible representations
produces rank $2$ representations ${\cal R}^{(2)} (h(r_1,s_1),h(r_2,s_2))_w$
which contain the lowest weights of the corresponding bulk weight chain, let it be $h(r_1,s_1)$ in this case,
as a generating state. This lowest weight state even generates
an indecomposable subrepresentation of ${\cal R}^{(2)} (h(r_1,s_1),h(r_2,s_2))_w$
which does not exhibit higher rank behaviour. 
As they are subrepresentations of existing representations this kind of representations have to appear
in the spectrum as well and will be denoted by ${\cal R}^{(1)} (h(r_1,s_1))_{h(r_2,s_2)-h(r_1,s_1)}$.

In the fusion rules $(r_1,s_1)$ appears to be that entry in the minimal model Kac table 
which does not have its nullvector at $h(r_2,s_2)$. Hence, $(r_1,s_1)$ seems to be sufficient
to describe this rank $1$ representation and we can set 
\beq
{\cal R}^{(1)}_{(r_1,s_1)} (h(r_1,s_1)) &:=& {\cal R}^{(1)} (h(r_1,s_1))_{h(r_2,s_2)-h(r_1,s_1)} \; .
\eeq

This is the first example of a case where a rank $1$ indecomposable representation on
a weight $h$ appears in the spectrum although the corresponding irreducible
representation on $h$ cannot be included consistently.
The other weights in the bulk of the Kac table induce ordinary irreducible
representations which are consistent with the fusion algebra.

%%%

\subsection{The fusion of irreducible representations}

Consider first the fusion of two irreducible representations.
If we describe an irreducible representation ${\cal V}^{i}_{(r,s)} (h)$ corresponding to a conformal weight $h$
on a corner or a border in the Kac table, we choose the index $(r,s)$
with the smallest product $rs$. If there are two pairs with the same product, 
we choose the one with larger $r$ (although this last point is mere convention
and does not effect the result).
Then the fusion product for $i,j \in \{\mbox{corner},\,\mbox{border} \}$ simply amounts 
to the untruncated BPZ-rules
\beqn \label{bpz_irr}
\left({\cal V}^{i}_{(r_1,s_1)} (h_1) \otimes 
{\cal V}^{j}_{(r_2,s_2)} (h_2) \right)_{\mbox{\small f}}
&=& \sum_{r_3 = |r_1-r_2|+1, \; \mbox{\small step}\: 2}^{r_1+r_2-1} \; \;  
\sum_{s_3 = |s_1-s_2|+1, \; \mbox{\small step}\: 2}^{s_1+s_2-1} \tilde{{\cal V}}_{(r_3,s_3)} \Big|_{\mbox{\tiny rules}} \!\!\! . \phantom{mmm}
\eeqn
On the right hand side, however, we do not simply encounter a sum over irreducible representations
again. Some of the resulting $\tilde{{\cal V}}_{(r,s)}$ are automatically combined into
rank $2$ or rank $3$ representations. 
The corresponding {\bf rules} how to do this combination (indicated as constraints in the
equation) are given by:
\begin{enumerate}
  \item For $(r,s)$ on a corner $\tilde{{\cal V}}_{(r,s)}$ is simply replaced by 
the corresponding irreducible representation ${\cal V}^{\mbox{\tiny corner}}_{(r,s)} (h(r,s))$.
  \item Concerning the set of all $(r,s)$ on the right hand side of \mref{bpz_irr} 
which correspond to weights on the border
there are two possibilities how to encounter rank $2$ representations. If we find
twice the same weight  $h(r_1,s_1) = h(r_2,s_2)$ 
which is the lowest of a weight chain these two need to be
replaced by the rank $2$ representation ${\cal R}^{(2)} (h(r_1,s_1),h(r_2,s_2))$.
Then, if we find two weights $h(m_1,m_1)$, $h(m_2,m_2)$ 
of the same weight chain adjacent to each other in the chain, these
need to be replaced by ${\cal R}^{(2)} (h(m_1,n_1),h(m_2,n_2))$.
All other weights in this set simply form irreducible representations
and are replaced by ${\cal V}_{(r,s)}^{\mbox{\tiny border}} (h(r,s))$.
  \item For the set of all $(r,s)$ on the right hand side of \mref{bpz_irr} 
which correspond to weights in the bulk
we first need to identify the rank $3$ representations. Rank $3$
re\-presentations need four generating states of the two possible types
described in the preceding subsection: either the two lowest weights
and twice the same weight of the second set of a bulk weight chain;
or weights from three adjacent sets of a bulk weight chain---both weights of the middle set
and one from each of the other two. The last is the generic case.
If we encounter a set of four weights in this manner we need to replace them
by the rank $3$ representation  ${\cal R}^{(3)} (h(r_1,s_1),h(r_2,s_2),h(r_3,s_3),h(r_4,s_4))$.
  \item Rank $2$ representations need two generating states. For bulk representations
these consist of one weight each from two adjacent sets in the weight chain.
If this set of two weights $h(r_1,s_1)$, $h(r_2,s_2)$ contains the lowest weight 
of this weight chain, let it be $h(r_1,s_1)$, there
are still two possible representations with this set of weights. The additional 
index is given by the level $w$ of the weight in the second tuple of the weight chain which is
not (!) the first possible nullvector on $h(r_2,s_2)$, i.e.\ which is not equal to
$h(r_2,s_2) + r_2\, s_2$. Each two weights of this form have to be replaced by
the rank $2$ representation ${\cal R}^{(2)} (h(r_1,s_1),h(r_2,s_2))_w$.
  \item After extraction of all rank $3$ and rank $2$ representations from the set of
all $(r,s)$ within the bulk we find to each lowest weight $h(r_1,s_1)$ of a bulk weight chain
a corresponding weight $h(r_2,s_2)$ of the second tuple of this weight chain. These form
the rank $1$ indecomposable representation ${\cal R}^{(1)} (h(r_1,s_1))_{h(r_2,s_2)-h(r_1,s_1)}$.
  \item All $(r,s)$ corresponding to weights in the bulk which have not been used up
in the three preceding points have to be replaced by ${\cal V}_{(r,s)} (h(r,s))$.
\end{enumerate}

This set of rules requires the following remarks:
\begin{itemize}
  \item Unfortunately one cannot write down the fusion rules for general augmented theories in such
a nice form as for the augmented $c_{p,1}$ models \cite{GK96}. First of all, we do not have 
border representations just on one side such that a restriction to an infinite strip of the Kac-table
nicely promotes the first border conformal weights beyond that strip to corresponding rank $2$
representations. Nevertheless, this property that the second conformal weight to a border
conformal weight (ordered by the level of their first nullvectors, resp.\ products $rs$) somehow represents
the rank $2$ representation seems to persist. Secondly, for the bulk there are simply not enough 
entries in the Kac table in order to uniquely label all different irreducible, rank $2$
and rank $3$ representations.
  \item The fifth rule which describes the appearance of ${\cal R}^{(1)} (h(r_1,s_1))_{h(r_2,s_2)-h(r_1,s_1)}$ 
is not needed for the
fusion of border and corner representations with itself. Actually, the examples show that 
starting with border and corner representations the whole
fusion closes without the inclusion of any irreducible or rank $1$ representation
corresponding to weights in the bulk. It is only the appearance of these 
irreducible or rank $1$ representations as subrepresentations of rank $2$
and rank $3$ representations that makes us include these in the theory.
We advocate that it is only this setting they should be thought of to exist in.
This point of view is strongly stressed by the existence of the above
discussed rank $1$ representations on the lowest weights of the bulk weight chains.
Indeed, only their embedding in a rank $2$ representation makes the absence of one
of the first two nullvectors possible as it prevents this singular vector (which itself
spans an irreducible subrepresentation of the rank $1$ representation) to
be a nullvector in the whole space of states.
\end{itemize}

If we want to write down the fusion product with representations corresponding to conformal 
weights in the bulk we need more information than
the first nullvector. For these we choose as index the set of the two pairs $\{(r_1,s_1),(r_2,s_2)\}$
with the two lowest products $r_1\, s_1$, $r_2\, s_2$. 
As there are exactly as many entries in the Kac table for these bulk 
representations as there are nullvectors in the nullvector cascade and as all these nullvectors
in the cascade are mutually distinct, this choice is unique.
Using $A(r,m) := \{ |r-m|+1, |r-m|+3, \dots , r+m-1 \}$ we define the following two sets
\beq
\lefteqn{\!\!\!\!\! S_{\mbox{\tiny eb}} (r,s|m_1,n_1,m_2,n_2)} \nonumber \\ 
&:=& \{ (p,q) | p \in A(r,m_1), q \in A(s,n_1) \} 
\cap \{ (p,q) | p \in A(r,m_2), q \in A(s,n_2) \}
\eeq
as well as
\beq
\lefteqn{S_{\mbox{\tiny bb}} (r_1,s_1,r_2,s_2|m_1,n_1,m_2,n_2)} \nonumber \\ 
&:=& \Big( \{ (p,q) | p \in A(r_1,m_1), q \in A(s_1,n_1) \}  \cap \{ (p,q) | p \in A(r_2,m_2), q \in A(s_2,n_2) \} \Big) \nonumber \\
&\cup& \Big( \{ (p,q) | p \in A(r_2,m_1), q \in A(s_2,n_1) \}  \cap \{ (p,q) | p \in A(r_1,m_2), q \in A(s_1,n_2) \} \Big) \; .
\eeq
These define the parameter ranges in the following fusion rules with bulk representations, $i \in \{\mbox{corner},\,\mbox{border} \}$,
\beq
\left({\cal V}^{i}_{(r,s)} (h_1) \otimes 
{\cal V}^{\mbox{\tiny bulk}}_{\{(m_1,n_1),(m_2,n_2)\}} (h_2) \right)_{\mbox{\small f}}
&=& \sum_{(s,t) \in S_{\mbox{\tiny eb}} (r,s|m_1,n_1,m_2,n_2)} \tilde{{\cal V}}_{(s,t)} \Big|_{\mbox{\tiny rules}} \nonumber \\
\left({\cal V}^{\mbox{\tiny bulk}}_{\{(r_1,s_1),(r_2,s_2)\}} (h_1) \otimes 
{\cal V}^{\mbox{\tiny bulk}}_{\{(m_1,n_1),(m_2,n_2)\}} (h_2) \right)_{\mbox{\small f}}
&=& \sum_{(s,t) \in S_{\mbox{\tiny bb}} (r_1,s_1,r_2,s_2|m_1,n_1,m_2,n_2)} 
\!\!\!\!\!\!\!\! \tilde{{\cal V}}_{(s,t)} \Big|_{\mbox{\tiny rules}} \; .
\eeq

The fusion rules for the rank $1$ indecomposable representations actually
look much the same as for border and corner representations, $i \in \{\mbox{corner},\,\mbox{border} \}$,
\beq
\left({\cal R}^{(1)}_{(r_1,s_1)} (h_1) \otimes 
{\cal V}^{i}_{(r_2,s_2)} (h_2) \right)_{\mbox{\small f}}
&=& \sum_{r_3 = |r_1-r_2|+1, \; \mbox{\small step}\: 2}^{r_1+r_2-1} \; \;  
\sum_{s_3 = |s_1-s_2|+1, \; \mbox{\small step}\: 2}^{s_1+s_2-1} \tilde{{\cal V}}_{(r_3,s_3)} \Big|_{\mbox{\tiny rules'}} \nonumber \\
\left({\cal R}^{(1)}_{(r_1,s_1)} (h_1) \otimes 
{\cal R}^{(1)}_{(r_2,s_2)} (h_2) \right)_{\mbox{\small f}}
&=& \sum_{r_3 = |r_1-r_2|+1, \; \mbox{\small step}\: 2}^{r_1+r_2-1} \; \;  
\sum_{s_3 = |s_1-s_2|+1, \; \mbox{\small step}\: 2}^{s_1+s_2-1} \tilde{{\cal V}}_{(r_3,s_3)} \Big|_{\mbox{\tiny rules'}} \nonumber \\
\left({\cal R}^{(1)}_{(r,s)} (h_1) \otimes 
{\cal V}^{\mbox{\tiny bulk}}_{\{(m_1,n_1),(m_2,n_2)\}} (h_2) \right)_{\mbox{\small f}}
&=& \sum_{(s,t) \in S_{\mbox{\tiny eb}} (r,s|m_1,n_1,m_2,n_2)} \tilde{{\cal V}}_{(s,t)} \Big|_{\mbox{\tiny rules'}} \; .
\eeq
It is only the rule (5) we have to modify slightly in this case:
\begin{belaufz2}
  \item[5'.] After extraction of all rank $3$ and rank $2$ representations from the set of
all $(r,s)$ each lowest weight $h(r_1,s_1)$ of a bulk weight chain corresponds to
the rank $1$ indecomposable representation ${\cal R}^{(1)}_{(r_1,s_1)} (h(r_1,s_1))$.
\end{belaufz2}
The reason is that treating the rank $1$ indecomposable representations as generated
from one state we will also get only one generating state for each rank $1$
indecomposable representation in the fusion process.

%%%

\subsection{The fusion with higher rank representations}

Now, we can give the fusion rules with higher rank representations
based on the above calculated fusion rules.
These higher rank fusion rules make use of the above elaborated fact
that all higher rank representations can be constructed
by indecomposably connecting a number of representations of one rank lower.
Certainly, these constituents are usually no true subrepresentations.
But nevertheless they exhibit in sum the same number of states
as the composed higher representation and even their nullvector
structure survives in a certain way as ``special'' vectors of the
new larger representation.
E.g.\ generically one needs four ${\cal V}_{(r,s)}$ representations 
to compose a rank $2$ representation. 

The following general rules for fusion
with higher rank representations apply successively with rising rank:
\begin{enumerate}
  \item First split one higher rank representation into its constituents
of one rank lower.
  \item Calculate the fusion rules with these constituents as known
before (or iterate this process until you reach a level where the fusion 
rules are already known).
  \item If you now find the right constituents for a higher rank
representation within the result (e.g.\ all four $\tilde{\cal V}$
for a generic rank $2$ representation) you have to combine them
to this higher rank representation. All other parts of the total
result remain as they are. This reintroduces the indecomposable
structure which was broken up in step one where possible.
  \item The one exception to the last step applies to the indecomposable
rank $1$ representations ${\cal R}^{(1)} (h(r_1,s_1))_{h(r_2,s_2)-h(r_1,s_1)}$
on the lowest weight $h(r_1,s_1)$ of a weight chain. 
Whenever we find that instead of ${\cal R}^{(1)} (h(r_1,s_1))_{h(r_2,s_2)-h(r_1,s_1)}$
the corresponding subrepresentation ${\cal V} (h(r_2,s_2))$
can be used as a building block of a rank $2$ representation we
need to replace ${\cal R}^{(1)} (h(r_1,s_1))_{h(r_2,s_2)-h(r_1,s_1)}$
by  ${\cal V} (h(r_2,s_2))$. We then proceed as described in the preceding
step.
\end{enumerate}
This last case is only encountered if we fuse irreducible bulk representations
with higher rank bulk representations. Its artificiality stems from the
fact that the irreducible bulk representations only
exist as subrepresentations of the rank $2$ representations---it is
these rank $2$ representations we actually have to look
at when considering fusion rules.

Let us give two examples of the $c_{2,3}=0$ fusion to
show how these rules work
\beq
{\cal V} (5/8) \otimesf {\cal R}^{(2)} (5/8,5/8)  &\stackrel{\mbox{\tiny split-up}}{\longrightarrow}&
{\cal V} (5/8) \otimesf \Big( 2\, {\cal V} (5/8) \oplus {\cal V} (21/8) \Big) \nonumber \\
&=& 2\, {\cal R}^{(2)} (0,2)_7 \oplus {\cal R}^{(2)} (1,5) \nonumber \\
&\stackrel{\mbox{\tiny re-combination}}{\longrightarrow}& {\cal R}^{(3)} (0,0,2,2) \nonumber \\
{\cal V} (5/8) \otimesf {\cal R}^{(2)} (1/3,1/3)  &\stackrel{\mbox{\tiny split-up}}{\longrightarrow}&
{\cal V} (5/8) \otimesf \Big( 2\, {\cal V} (1/3) \oplus {\cal V} (10/3) \Big) \nonumber \\
&=& 2\, {\cal V} (-1/24) \oplus {\cal V} (35/24) \nonumber \\
&\stackrel{\mbox{\tiny re-combination}}{\longrightarrow}& 2\, {\cal V} (-1/24) \oplus {\cal V} (35/24) \; .
\eeq

%%%%%%%%%%%%%%%%%%%%%%%%%%%%%%%%%%%%%%%%%%%%%
%%%%%%%%%%%%%%%%%%%%%%%%%%%%%%%%%%%%%%%%%%%%%
%%%%%%%%%%%%%%%%%%%%%%%%%%%%%%%%%%%%%%%%%%%%%

\section{Conclusion and outlook\label{conclusion}}

In this paper we have given the explicit fusion rules
for two examples of general augmented minimal models,
the augmented $c_{2,3}=0$ and the augmented Yang-Lee
model. We have shown that these models exhibit an even richer
indecomposable structure in comparison to the $c_{p,1}$
models with representations up to rank $3$.
We have elaborated a number of these newly appearing
representations of rank $2$ and $3$ in detail. Especially,
we have shown for some of these examples that one can construct
rank $3$ representations by indecomposably connecting
rank $2$ representations. This is a very interesting 
generalisation of the rank $2$ representation case
which can be composed by indecomposably connecting
several irreducible representations. 

We have also shown
that the fusion rule consistency conditions of
symmetry and associativity actually restrict
the representation spectrum---irreducible
representations corresponding to weights
of the original non-augmented minimal model
cannot be included into the spectrum consistently.
We actually find that the vacuum representation is
given by an indecomposable but not irreducible
rank $1$ subrepresentation of a rank $2$
representation. Especially for the $c_{2,3} = 0$
theory this settles the long standing puzzle of
how to construct a consistent vacuum representation and
character, at least on the level of pure Virasoro
representations (see e.g.\ \cite{FM05}).

These two examples already reveal a concise general 
structure which we think exhibits almost all of the generic
structure of general augmented $c_{p,q}$ models. Accordingly,
we have extrapolated this structure to a 
conjecture of the general representation
content and fusion rules of general augmented
$c_{p,q}$ models.
These conjectured fusion rules still inherit much of the BPZ 
fusion structure. The main generalisation lies in the fact that
we do not interpret the BPZ fusion rules ``minimally'' for these
models. This ``minimal'' refers to the BPZ fusion rules
for the non-augmented minimal models where one takes 
a section over all ways of applying the
untruncated BPZ rules. The augmented fusion rules
allow for definitely more representation 
content.
This is, however, in perfect accordance with and directly
reduces to the already known rules for the special case
of augmented $c_{p,1}$ models \cite{GK96} although
the general rules do not look as nicely compact as
the special ones in \cite{GK96}.

Concerning our special example of $c_{2,3} = 0$ we actually
find that the numerical results of \cite{PRG01}, table 1,  
which are supposed to give the low level state content
of CFT representations at $c=0$, perfectly match our representations
${\cal R}^{(1)} (0)_2$, ${\cal R}^{(1)} (0)_1$
and ${\cal V} (1/3)$ respectively. Although the numerical results
give only relatively few levels we conjecture that they
are correctly described by the above mentioned $c_{2,3} = 0$
representations.

We have also shown that the findings in this paper
are in perfect agreement with the logarithmic nullvector calculations
performed in \cite{EF05}. Using the technique of the 
Nahm algorithm we have actually been able to pinpoint
the precise structure of higher rank Virasoro representations
in $c_{2,3} = 0$; various types of possible rank $2$ representations
had already been conjectured in \cite{EF05}.

On the other hand, the question of whether these models
exhibit a larger ${\cal W}$-algebra and how the Virasoro
representations combine to representations of this
larger symmetry algebra still remains open. There are,
however, strong hints at the existence of such an
enhanced symmetry algebra coming from inspections of 
the corresponding representations of the modular
group \cite{EFN}. And even the mere fact that the Virasoro
fusion rules so nicely generalise from the $c_{p,1}$
model case where we know a triplet ${\cal W}$-algebra
to exist is a further good hint. This may be exemplified
by the existence of a representation like ${\cal V} (7)$
in the $c_{2,3} = 0$ model whose fusion rules behave in just the same
way as the ones of the ``spin representation'' ${\cal V}_{1,p}$
in the $c_{p,1}$ models. Indeed, in the $c_{p,1}$ models 
this spin representation behaves
roughly like a square root of the ${\cal W}$-representation.

The representations of the
modular group are already very restrictive such that
for e.g.\ $c_{2,3} = 0$ it suffices to know the field
content up to level $7$ in order to precisely know
the full ${\cal W}$-representation.
We believe that this restrictiveness together with 
our new knowledge of the precise
structure of the Virasoro representations
should lead to a consistent set of modular functions
which represent the right ${\cal W}$-characters and
produce the right ${\cal W}$-fusion rules. This is 
subject to ongoing research \cite{EFN}.

%%%%%%%%%%%%%%%%%%%%%%%%%%%%%%%%%%%%%%%%%%%%%
%%%%%%%%%%%%%%%%%%%%%%%%%%%%%%%%%%%%%%%%%%%%%
%%%%%%%%%%%%%%%%%%%%%%%%%%%%%%%%%%%%%%%%%%%%%

\vspace{0.7cm}
\noindent
{\bf Acknowledgements.}
We are very grateful to Werner Nahm and Alexander Nichols
for useful discussions. Furthermore, we would like to thank
the Bonn Condensed Matter Theory Group for the kind
permission to use their computer cluster.
This work was supported in part by 
the DFG Schwerpunktprogramm no.\ 1096 ``Stringtheorie''.
HE acknowledges support by the Studienstiftung des Deutschen Volkes.
MF is also partially supported by  the European Union network
HPRN-CT-2002-00325 (EUCLID).

%%%%%%%%%%%%%%%%%%%%%%%%%%%%%%%%%%%%%%%%%%%%%
%%%%%%%%%%%%%%%%%%%%%%%%%%%%%%%%%%%%%%%%%%%%%
%%%%%%%%%%%%%%%%%%%%%%%%%%%%%%%%%%%%%%%%%%%%%

\begin{appendix}

\section{Explicit Jordan diagonalisation of $\mathbf{L_0}$ for $\mathbf{{\cal R}^{(3)} (0,1,2,5)}$\label{explicit_L0}}

In this appendix we want to present a basis
of states for the lowest five levels of ${\cal R}^{(3)} (0,1,2,5)$
which brings the $L_0$ matrix in a Jordan diagonal form. The basis is denoted
by $n_i$ with a running index as assigned by the computer programme. 
Vectors which are not shown to be equal to descendants of some
other vectors are understood to be generating states.
These are $n_0$, $n_5$, $n_{24}$ and $n_{18}$ at
levels $0$, $1$, $2$, $5$ respectively. On the right hand side we have
denoted only the Jordan block in $L_0$ for the respective states. Different
Jordan blocks are separated by lines. All other $L_0$ entries are zero.

\scriptsize
\vspace{0.8cm}
\renewcommand{\arraystretch}{1.2}
\begin{tabular}{l @{$\; = \; $} l | c c c}
\multicolumn{2}{l|}{states} & \multicolumn{3}{c}{$L_0$ matrix}\\ \hline \hline
\multicolumn{2}{l|}{$n_0$} & $0$ & & \\ \hline
$n_4$ & $L_{-1}\, n_0$ & $1$ & $1$ & \\ 
\multicolumn{2}{l|}{$n_5$} & $0$ & $1$ & \\ \hline
$n_6$ & $L_{-1}^2\, n_0 = L_{-1}\, n_4$ & $2$ & $1$ & \\ 
$n_7$ & $L_{-1}\, n_5$ & $0$ & $2$ & \\ \hline
$n_{23}$ & $L_{-2}\, n_0$ & $2$ & $1$ & \\ 
\multicolumn{2}{l|}{$n_{24}$} & $0$ & $2$ & \\ \hline
$n_8$ & $L_{-2}L_{-1}\, n_0 = L_{-2}\, n_4$ & $3$ & $1$ & \\ 
$n_9$ & $L_{-2}\, n_5$ & $0$ & $3$ & \\ \hline
$n_{25}$ & $L_{-1}^3\, n_0 = L_{-1}^2\, n_4$ & $3$ & $1$ & \\ 
$n_{26}$ & $L_{-1}^2\, n_5$ & $0$ & $3$ & \\ \hline
$n_{38}$ & $(L_{-3}+L_{-2}L_{-1}) n_0 = L_{-1}\, n_{23}$ & $3$ & $1$ & \\ 
$n_{39}$ & $L_{-1}\, n_{24}$ & $0$ & $3$ & \\ \hline
$n_{10}$ & $L_{-3}L_{-1}\, n_0 = L_{-3}\, n_4$ & $4$ & $1$ & \\ 
$n_{11}$ & $L_{-3}\, n_5$ & $0$ & $4$ & \\ \hline
$n_{27}$ & $L_{-2}L_{-1}^2\, n_0 = L_{-2}L_{-1}\, n_4$ & $4$ & $1$ & \\ 
$n_{28}$ & $L_{-2}L_{-1}\, n_5$ & $0$ & $4$ & \\ \hline
$n_{40}$ & $L_{-2}^2\, n_0 = L_{-2}\, n_{23}$ & $4$ & $1$ & \\ 
$n_{41}$ & $L_{-2}\, n_{24}$ & $0$ & $4$ & \\ \hline
$n_{48}$ & $L_{-1}^4\, n_0 = L_{-1}^3\, n_4$ & $4$ & $1$ & \\ 
$n_{49}$ & $L_{-1}^3\, n_5$ & $0$ & $4$ & \\ \hline
$n_{56}$ & $(2L_{-4}+2L_{-3}L_{-1}+L_{-2}L_{-1}^2) n_0 = L_{-1}^2\, n_{23}$ & $4$ & $1$ & \\ 
$n_{57}$ & $L_{-1}^2\, n_{24}$ & $0$ & $4$ & \\ \hline
$n_{16}$ & $\frac{4655}{31758} (L_{-4}L_{-1}-L_{-3}L_{-1}^2+\frac{5}{3}L_{-2}L_{-1}^3-L_{-2}^2L_{-1}-\frac{1}{4}L_{-1}^5) n_0$ & $5$ & $1$ & $0$ \\ 
$n_{17}$ & {\tiny $\frac{4655}{63516} \Big((L_{-4} - \frac{5}{2}L_{-3}L_{-1}-L_{-2}^2+\frac{19}{6}L_{-2}L_{-1}^2-\frac{1}{2}L_{-1}^4) n_5+(L_{-3}-L_{-2}L_{-1}+\frac{1}{6}L_{-1}^3)n_{24} \Big)$} & $0$ & $5$ & $1$ \\ 
\multicolumn{2}{l|}{$n_{18}$} & $0$ & $0$ & $5$\\ \hline
$n_{29}$ & $L_{-3}L_{-1}^2\, n_0 = L_{-3}L_{-1}\, n_4$ & $5$ & $1$ & \\ 
$n_{30}$ & $L_{-3}L_{-1}\, n_5$ & $0$ & $5$ & \\ \hline
$n_{42}$ & $L_{-3}L_{-2}\, n_0 = L_{-3}\, n_{23}$ & $5$ & $1$ & \\ 
$n_{43}$ & $L_{-3}\, n_{24}$ & $0$ & $5$ & \\ \hline
$n_{50}$ & $L_{-2}^2L_{-1}\, n_0 = L_{-2}^2\, n_4$ & $5$ & $1$ & \\ 
$n_{51}$ & $L_{-2}^2\, n_5$ & $0$ & $5$ & \\ \hline
$n_{58}$ & $L_{-2}L_{-1}^3\, n_0 = L_{-2}L_{-1}^2\, n_4$ & $5$ & $1$ & \\ 
$n_{59}$ & $L_{-2}L_{-1}^2\, n_5$ & $0$ & $5$ & \\ \hline
$n_{63}$ & $L_{-1}^5\, n_0 = L_{-1}^4\, n_4$ & $5$ & $1$ & \\ 
$n_{64}$ & $L_{-1}^4\, n_5$ & $0$ & $5$ & \\ \hline
$n_{68}$ & $(6L_{-5}+6L_{-4}L_{-1}+3L_{-3}L_{-1}^2+L_{-2}L_{-1}^3) n_0 = L_{-1}^3\, n_{23}$ & $5$ & $1$ & \\ 
$n_{69}$ & $L_{-1}^3\, n_{24}$ & $0$ & $5$ & \\ \hline
$n_{70}$ & $(L_{-4} + \frac{1}{2}L_{-3}L_{-1}-L_{-2}^2+\frac{1}{6}L_{-2}L_{-1}^2) n_5-(L_{-3}-L_{-2}L_{-1}+\frac{1}{6}L_{-1}^3)n_{24}$ & $5$ & & \\ \hline
\end{tabular}
\normalsize

%%%%%%%%%%%%%%%%%%%%%%%%%%%%%%%%%%%%%%%%%%%%%
%%%%%%%%%%%%%%%%%%%%%%%%%%%%%%%%%%%%%%%%%%%%%
%%%%%%%%%%%%%%%%%%%%%%%%%%%%%%%%%%%%%%%%%%%%%

\section{Explicit fusion rules for $\mathbf{c_{2,3}=0}$\label{fusion_c0}}

The following table contains the results of our explicit calculations of the
fusion product of irreducible and rank $1$ representations 
with each other in the 
augmented $c_{2,3}=0$ model. The first column gives the
representations to be fused. Then the level $L$ at which we
calculated the new representation as well as the maximal level
$\tilde{L}$ up to which we took the corresponding constraints into
account are given. The last column contains the result.

The computational power restricts the level $L$ at which
the fused representations are calculated, unfortunately 
quite severely for the higher representations. Hence, it
was sometimes not possible to reach a high enough $L$
to actually see the indecomposable structure of all
component representations of the result. We nevertheless
denoted the result as we would expect it according to the
proposed fusion rules---to discern this guessed higher
representations from the explicit results we indicated
them by blue colour and a question mark. But certainly
our results are always in agreement with these possible
higher rank representations up to the level $L$ given in the table.

\footnotesize
\vspace{0.8cm}
\renewcommand{\arraystretch}{1.2}
\begin{tabular}{c @{$\; \otimesf \; $} c | c | c || l}
\multicolumn{2}{c|}{} & $L$ & $\tilde{L}_{\mbox{\footnotesize max}}$ & Fusion product \\ \hline
${\cal V} (5/8)$ & ${\cal V} (5/8)$ & $6$ & $11$ & ${\cal R}^{(2)} (0,2)_7$ \\
& ${\cal V} (1/3)$ & $6$ & $11$ & ${\cal V} (-1/24)$ \\
& ${\cal V} (1/8)$ & $7$ & $8$ & ${\cal R}^{(2)} (0,1)_5$ \\
& ${\cal V} (-1/24)$ & $6$ & $9$ & ${\cal R}^{(2)} (1/3,1/3)$ \\
& ${\cal V} (33/8)$ & $6$ & $11$ & ${\cal R}^{(2)} (2,7)$ \\
& ${\cal V} (10/3)$ & $6$ & $7$ & ${\cal V} (35/24)$ \\
& ${\cal V} (21/8)$ & $6$ & $11$ & ${\cal R}^{(2)} (1,5)$ \\
& ${\cal V} (35/24)$ & $5$ & $7$ & ${\cal R}^{(2)} (1/3,10/3)$ \\
& ${\cal V} (2)$ & $6$ & $9$ & ${\cal V} (5/8)$ \\
& ${\cal V} (1)$ & $6$ & $9$ & ${\cal V} (1/8)$ \\
& ${\cal V} (7)$ & $6$ & $9$ & ${\cal V} (33/8)$ \\
& ${\cal V} (5)$ & $6$ & $9$ & ${\cal V} (21/8)$ \\ \hline
${\cal V} (1/3)$ & ${\cal V} (1/3)$ & $6$ & $11$ & ${\cal R}^{(2)} (0,2)_5 \oplus {\cal V} (1/3)$ \\
& ${\cal V} (1/8)$ & $6$ & $9$ & ${\cal R}^{(2)} (1/8,1/8)$ \\
& ${\cal V} (-1/24)$ & $6$ & $9$ & ${\cal R}^{(2)} (5/8,5/8) \oplus {\cal V} (-1/24)$ \\
& ${\cal V} (33/8)$ & $6$ & $9$ & ${\cal V} (35/24)$ \\
& ${\cal V} (10/3)$ & $6$ & $9$ & ${\cal R}^{(2)} (1,7) \oplus {\cal V} (10/3)$ \\
& ${\cal V} (21/8)$ & $6$ & $9$ & ${\cal R}^{(2)} (5/8,21/8)$ \\
& ${\cal V} (35/24)$ & $5$ & $7$ & ${\cal R}^{(2)} (1/8,33/8) \oplus {\cal V} (35/24)$ \\
& ${\cal V} (2)$ & $6$ & $9$ & ${\cal V} (1/3)$ \\
& ${\cal V} (1)$ & $6$ & $9$ & ${\cal R}^{(2)} (0,1)_7$ \\
& ${\cal V} (7)$ & $6$ & $9$ & ${\cal V} (10/3)$ \\
& ${\cal V} (5)$ & $6$ & $9$ & ${\cal R}^{(2)} (2,5)$ \\ \hline
\end{tabular}

\renewcommand{\arraystretch}{1.2}
\begin{tabular}{c @{$\; \otimesf \; $} c | c | c || l}
\multicolumn{2}{c|}{} & $L$ & $\tilde{L}_{\mbox{\footnotesize max}}$ & Fusion product \\ \hline
${\cal V} (1/8)$ & ${\cal V} (1/8)$ & $6$ & $8$ & ${\cal R}^{(2)} (1/3,1/3) \oplus {\cal R}^{(2)} (0,2)_7$ \\
& ${\cal V} (-1/24)$ & $6$ & $7$ & ${\cal R}^{(3)} (0,0,1,1)$ \\
& ${\cal V} (33/8)$ & $6$ & $9$ & ${\cal R}^{(2)} (1,5)$ \\
& ${\cal V} (10/3)$ & $6$ & $8$ & ${\cal R}^{(2)} (5/8,21/8)$ \\
& ${\cal V} (21/8)$ & $5$ & $7$ & ${\cal R}^{(2)} (1/3,10/3) \oplus {\cal R}^{(2)} (2,7)$ \\
& ${\cal V} (35/24)$ & $5$ & $7$ & ${\cal R}^{(3)} (0,1,2,5)$ \\
& ${\cal V} (2)$ & $6$ & $9$ & ${\cal V} (1/8)$ \\
& ${\cal V} (1)$ & $5$ & $7$ & ${\cal V} (5/8) \oplus {\cal V} (-1/24)$ \\
& ${\cal V} (7)$ & $6$ & $8$ & ${\cal V} (21/8)$ \\
& ${\cal V} (5)$ & $6$ & $7$ & ${\cal V} (33/8) \oplus {\cal V} (35/24)$ \\ \hline
${\cal V} (-1/24)$ & ${\cal V} (-1/24)$ & $5$ & $7$ & ${\cal R}^{(3)} (0,0,2,2) \oplus {\cal R}^{(2)} (1/3,1/3)$ \\
& ${\cal V} (21/8)$ & $5$ & $7$ & ${\cal R}^{(3)} (0,1,2,5)$ \\
& ${\cal V} (35/24)$ & $4$ & $6$ & $\hl{{\cal R}^{(3)} (0,1,2,7)} \oplus {\cal R}^{(2)} (1/3,10/3)$ \\
& ${\cal V} (5)$ & $5$ & $7$ & ${\cal R}^{(2)} (5/8,21/8)$ \\
& ${\cal V} (143/24)$ & $0$ & $6$ & $\hl{{\cal R}^{(3)} (1,5,7,15)} \oplus \hl{{\cal R}^{(2)} (10/3,28/3)}$ \\ \hline
${\cal V} (33/8)$ & ${\cal V} (33/8)$ & $6$ & $8$ & ${\cal R}^{(2)} (0,2)_{\hls{7}} \oplus \hl{{\cal R}^{(2)} (7,15)}$ \\
& ${\cal V} (10/3)$ & $5$ & $7$ & ${\cal V} (-1/24) \oplus {\cal V} (143/24)$ \\
& ${\cal V} (-1/24)$ & $5$ & $7$ & ${\cal R}^{(2)} (1/3,10/3)$ \\
& ${\cal V} (21/8)$ & $5$ & $7$ & ${\cal R}^{(2)} (0,1)_{\hls{5}} \oplus \hl{{\cal R}^{(2)} (5,12)}$ \\
& ${\cal V} (35/24)$ & $6$ & $7$ & ${\cal R}^{(2)} (1/3,1/3) \oplus {\cal R}^{(2)} (10/3,28/3)$ \\
& ${\cal V} (1)$ & $6$ & $9$ & ${\cal V} (21/8)$ \\
& ${\cal V} (7)$ & $5$ & $7$ & ${\cal V} (5/8) \oplus {\cal V} (85/8)$ \\
& ${\cal V} (5)$ & $6$ & $8$ & ${\cal V} (1/8) \oplus {\cal V} (65/8)$ \\ \hline
${\cal V} (10/3)$ & ${\cal V} (10/3)$ & $5$ & $7$ & ${\cal R}^{(2)} (0,2)_{\hls{5}} \oplus \hl{{\cal R}^{(2)} (5,15)} \oplus {\cal V} (1/3) \oplus {\cal V} (28/3)$ \\
& ${\cal V} (-1/24)$ & $5$ & $7$ & ${\cal R}^{(2)} (1/8,33/8) \oplus {\cal V} (35/24)$ \\
& ${\cal V} (21/8)$ & $5$ & $7$ & ${\cal R}^{(2)} (1/8,1/8) \oplus {\cal R}^{(2)} (33/8,65/8)$ \\
& ${\cal V} (35/24)$ & $4$ & $6$ & ${\cal R}^{(2)} (5/8,5/8) \oplus \hl{{\cal R}^{(2)} (21/8,85/8)} \oplus {\cal V} (-1/24) \oplus {\cal V} (143/24)$ \\
& ${\cal V} (5)$ & $3$ & $9$ & ${\cal R}^{(2)} (0,1)_{\hls{7}} \oplus \hl{{\cal R}^{(2)} (7,12)}$ \\ \hline
${\cal V} (21/8)$ & ${\cal V} (21/8)$ & $4$ & $6$ & ${\cal R}^{(2)} (1/3,1/3) \oplus \hl{{\cal R}^{(2)} (10/3,28/3)} \oplus {\cal R}^{(2)} (0,2)_{\hls{7}} \oplus \hl{{\cal R}^{(2)} (7,15)}$ \\
& ${\cal V} (35/24)$ & $4$ & $6$ & ${\cal R}^{(3)} (0,0,1,1) \oplus \hl{{\cal R}^{(3)} (2,5,7,12)}$ \\ \hline
${\cal V} (35/24)$ & ${\cal V} (35/24)$ & $3$ & $5$ & $ {\cal R}^{(3)} (0,0,2,2) \oplus \hl{{\cal R}^{(3)} (1,5,7,15)}  \oplus {\cal R}^{(2)} (1/3,1/3)$ \\
\multicolumn{2}{c|}{} &&& $\qquad \oplus \hl{{\cal R}^{(2)} (10/3,28/3)}$  \\ \hline
${\cal V} (2)$ & ${\cal V} (-1/24)$ & $6$ & $9$ & ${\cal V} (-1/24)$ \\
& ${\cal V} (33/8)$ & $6$ & $8$ & ${\cal V} (33/8)$ \\
& ${\cal V} (10/3)$ & $6$ & $7$ & ${\cal V} (10/3)$ \\
& ${\cal V} (21/8)$ & $6$ & $9$ & ${\cal V} (21/8)$ \\
& ${\cal V} (35/24)$ & $5$ & $7$ & ${\cal V} (35/24)$ \\
& ${\cal V} (2)$ & $6$ & $9$ & ${\cal R}^{(1)} (0)_2$ \\
& ${\cal V} (1)$ & $6$ & $9$ & ${\cal R}^{(1)} (0)_1$ \\
& ${\cal V} (7)$ & $6$ & $9$ & ${\cal V} (7)$ \\
& ${\cal V} (5)$ & $6$ & $9$ & ${\cal V} (5)$ \\
& ${\cal R}^{(1)} (0)_2$ & $6$ & $9$ & ${\cal V} (2)$ \\
& ${\cal R}^{(1)} (0)_1$ & $6$ & $9$ & ${\cal V} (1)$ \\ \hline
\end{tabular}

\renewcommand{\arraystretch}{1.2}
\begin{tabular}{c @{$\; \otimesf \; $} c | c | c || l}
\multicolumn{2}{c|}{} & $L$ & $\tilde{L}_{\mbox{\footnotesize max}}$ & Fusion product \\ \hline
${\cal V} (1)$ & ${\cal V} (-1/24)$ & $5$ & $7$ & ${\cal R}^{(2)} (1/8,1/8)$ \\
& ${\cal V} (10/3)$ & $5$ & $7$ & ${\cal R}^{(2)} (2,5)$ \\
& ${\cal V} (21/8)$ & $5$ & $7$ & ${\cal V} (33/8) \oplus{\cal V} (35/24)$ \\
& ${\cal V} (35/24)$ & $5$ & $7$ & ${\cal R}^{(2)} (5/8,21/8)$ \\
& ${\cal V} (1)$ & $6$ & $8$ & ${\cal R}^{(1)} (0)_2 \oplus {\cal V} (1/3)$ \\
& ${\cal V} (7)$ & $5$ & $7$ & ${\cal V} (5)$ \\
& ${\cal V} (5)$ & $5$ & $7$ & ${\cal V} (7) \oplus {\cal V} (10/3)$ \\ \hline
${\cal V} (7)$ & ${\cal V} (-1/24)$ & $5$ & $7$ & ${\cal V} (35/24)$ \\
& ${\cal V} (10/3)$ & $5$ & $7$ & ${\cal V} (1/3) \oplus {\cal V} (28/3)$ \\
& ${\cal V} (21/8)$ & $5$ & $7$ & ${\cal V} (1/8) \oplus {\cal V} (65/8)$ \\
& ${\cal V} (35/24)$ & $4$ & $6$ & ${\cal V} (-1/24) \oplus {\cal V} (143/24)$ \\
& ${\cal V} (7)$ & $5$ & $7$ & ${\cal R}^{(1)} (0)_2 \oplus {\cal V} (15)$ \\
& ${\cal V} (5)$ & $5$ & $7$ & ${\cal R}^{(1)} (0)_1 \oplus {\cal V} (12)$ \\ \hline
${\cal V} (5)$ & ${\cal V} (21/8)$ & $3$ & $6$ & ${\cal V} (5/8) \oplus {\cal V} (-1/24) \oplus {\cal V} (85/8) \oplus {\cal V} (143/24)$ \\
& ${\cal V} (35/24)$ & $4$ & $6$ & ${\cal R}^{(2)} (1/8,1/8) \oplus {\cal R}^{(2)} (33/8,65/8)$ \\
& ${\cal V} (5)$ & $5$ & $7$ & ${\cal R}^{(1)} (0)_2 \oplus {\cal V} (15) \oplus {\cal V} (1/3) \oplus {\cal V} (28/3)$ \\ \hline
${\cal R}^{(1)} (0)_2$ & ${\cal V} (5/8)$ & $6$ & $9$ & ${\cal V} (5/8)$ \\
& ${\cal V} (1/3)$ & $6$ & $9$ & ${\cal V} (1/3)$ \\
& ${\cal V} (1/8)$ & $6$ & $9$ & ${\cal V} (1/8)$ \\
& ${\cal V} (-1/24)$ & $6$ & $9$ & ${\cal V} (-1/24)$ \\
& ${\cal V} (33/8)$ & $6$ & $9$ & ${\cal V} (33/8)$ \\
& ${\cal V} (10/3)$ & $6$ & $9$ & ${\cal V} (10/3)$ \\
& ${\cal V} (21/8)$ & $6$ & $9$ & ${\cal V} (21/8)$ \\
& ${\cal V} (35/24)$ & $6$ & $9$ & ${\cal V} (35/24)$ \\
& ${\cal V} (2)$ & $6$ & $9$ & ${\cal V} (2)$ \\
& ${\cal V} (1)$ & $6$ & $9$ & ${\cal V} (1)$ \\
& ${\cal V} (7)$ & $6$ & $9$ & ${\cal V} (7)$ \\
& ${\cal V} (5)$ & $6$ & $9$ & ${\cal V} (5)$ \\
& ${\cal R}^{(1)} (0)_2$ & $6$ & $9$ & ${\cal R}^{(1)} (0)_2$ \\
& ${\cal R}^{(1)} (0)_1$ & $6$ & $9$ & ${\cal R}^{(1)} (0)_1$ \\ \hline
${\cal R}^{(1)} (0)_1$ & ${\cal V} (5/8)$ & $6$ & $9$ & ${\cal V} (1/8)$ \\
& ${\cal V} (1/3)$ & $6$ & $9$ & ${\cal R}^{(2)} (0,1)_7$ \\
& ${\cal V} (1/8)$ & $6$ & $9$ & ${\cal V} (5/8) \oplus{\cal V} (-1/24)$ \\
& ${\cal V} (-1/24)$ & $6$ & $9$ & ${\cal R}^{(2)} (1/8,1/8)$ \\
& ${\cal V} (33/8)$ & $6$ & $9$ & ${\cal V} (21/8)$ \\
& ${\cal V} (10/3)$ & $6$ & $9$ & ${\cal R}^{(2)} (2,5)$ \\
& ${\cal V} (21/8)$ & $6$ & $9$ & ${\cal V} (33/8) \oplus{\cal V} (35/24)$ \\
& ${\cal V} (35/24)$ & $6$ & $9$ & ${\cal R}^{(2)} (5/8,21/8)$ \\
& ${\cal V} (2)$ & $6$ & $9$ & ${\cal V} (1)$ \\
& ${\cal V} (1)$ & $6$ & $9$ & ${\cal V} (2) \oplus {\cal V} (1/3)$ \\
& ${\cal V} (7)$ & $6$ & $9$ & ${\cal V} (5)$ \\
& ${\cal V} (5)$ & $6$ & $9$ & ${\cal V} (7) \oplus {\cal V} (10/3)$ \\
& ${\cal R}^{(1)} (0)_2$ & $6$ & $9$ & ${\cal R}^{(1)} (0)_1$ \\
& ${\cal R}^{(1)} (0)_1$ & $6$ & $9$ & ${\cal R}^{(1)} (0)_2 \oplus {\cal V} (1/3)$ \\ \hline
\end{tabular}

\newpage
\normalsize

The next table lists our results for the fusion of
irreducible and rank $1$ representations with
rank $2$ representations. The notation stays as above.

\footnotesize
\vspace{0.8cm}
\renewcommand{\arraystretch}{1.2}
\begin{tabular}{c @{$\; \otimesf \; $} c | c | c || l}
\multicolumn{2}{c|}{} & $L$ & $\tilde{L}_{\mbox{\footnotesize max}}$ & Fusion product \\ \hline
${\cal V} (5/8)$ & ${\cal R}^{(2)} (5/8,5/8)$ & $5$ & $7$ & ${\cal R}^{(3)} (0,0,2,2)$ \\
& ${\cal R}^{(2)} (1/3,1/3)$ & $5$ & $7$ & $2\, {\cal V} (-1/24) \oplus{\cal V} (35/24)$ \\
& ${\cal R}^{(2)} (1/8,1/8)$ & $5$ & $7$ & ${\cal R}^{(3)} (0,0,1,1)$ \\
& ${\cal R}^{(2)} (5/8,21/8)$ & $5$ & $8$ & ${\cal R}^{(3)} (0,1,2,5)$ \\
& ${\cal R}^{(2)} (1/3,10/3)$ & $4$ & $6$ & ${\cal V} (-1/24) \oplus 2\, {\cal V} (35/24) \oplus {\cal V} (143/24) $ \\
& ${\cal R}^{(2)} (1/8,33/8)$ & $4$ & $6$ & $\hl{{\cal R} (0,1,2,7)}$ \\
& ${\cal R}^{(2)} (0,1)_5$ & $4$ & $6$ & $2\, {\cal V} (1/8) \oplus {\cal V} (21/8)$ \\
& ${\cal R}^{(2)} (0,1)_7$ & $4$ & $6$ & ${\cal R}^{(2)} (1/8,1/8)$ \\
& ${\cal R}^{(2)} (0,2)_5$ & $4$ & $6$ & ${\cal R}^{(2)} (5/8,5/8)$ \\
& ${\cal R}^{(2)} (0,2)_7$ & $4$ & $7$ & $2\, {\cal V} (5/8) \oplus {\cal V} (33/8)$ \\
& ${\cal R}^{(2)} (1,5)$ & $4$ & $6$ & ${\cal V} (1/8) \oplus 2\, {\cal V} (21/8) \oplus {\cal V} (65/8)$ \\
& ${\cal R}^{(2)} (2,5)$ & $4$ & $6$ & ${\cal R}^{(2)} (5/8,21/8) \oplus {\cal V} (65/8)$ \\
& ${\cal R}^{(2)} (1,7)$ & $4$ & $8$ & ${\cal R}^{(2)} (1/8,33/8)$ \\
& ${\cal R}^{(2)} (2,7)$ & $4$ & $7$ & ${\cal V} (5/8) \oplus 2\, {\cal V} (33/8) \oplus {\cal V} (85/8)$ \\ \hline
${\cal V} (1/3)$ & ${\cal R}^{(2)} (5/8,5/8)$ & $5$ & $7$ & ${\cal R}^{(2)} (5/8,21/8) \oplus 2\, {\cal V} (-1/24)$ \\
& ${\cal R}^{(2)} (1/3,1/3)$ & $5$ & $7$ & ${\cal R}^{(3)} (0,0,2,2) \oplus {\cal R}^{(2)} (1/3,1/3)$ \\
& ${\cal R}^{(2)} (1/8,1/8)$ & $4$ & $7$ & $2\, {\cal R}^{(2)} (1/8,1/8) \oplus {\cal V} (35/24)$ \\
& ${\cal R}^{(2)} (5/8,21/8)$ & $3$ & $10$ & $2\, {\cal R}^{(2)} (5/8,21/8) \oplus {\cal V} (-1/24) \oplus {\cal V} (143/24)$ \\
& ${\cal R}^{(2)} (1/3,10/3)$ & $3$ & $6$ & $\hl{{\cal R}^{(3)} (0,1,2,7)} \oplus {\cal R}^{(2)} (1/3,10/3)$ \\
& ${\cal R}^{(2)} (1/8,33/8)$ & $4$ & $7$ & ${\cal R}^{(2)} (1/8,1/8) \oplus {\cal R}^{(2)} (33/8,65/8) \oplus 2\, {\cal V} (35/24)$ \\
& ${\cal R}^{(2)} (0,1)_5$ & $4$ & $6$ & ${\cal R}^{(3)} (0,0,1,1)$ \\
& ${\cal R}^{(2)} (0,1)_7$ & $4$ & $6$ & $2\, {\cal R}^{(2)} (0,1)_7 \oplus {\cal V} (10/3)$ \\
& ${\cal R}^{(2)} (0,2)_5$ & $4$ & $7$ & ${\cal R}^{(2)} (2,5) \oplus 2\, {\cal V} (1/3)$ \\
& ${\cal R}^{(2)} (0,2)_7$ & $4$ & $6$ & ${\cal R}^{(2)} (1/3,1/3)$ \\
& ${\cal R}^{(2)} (1,5)$ & $5$ & $9$ & ${\cal R}^{(3)} (0,1,2,5)$ \\
& ${\cal R}^{(2)} (2,5)$ & $4$ & $6$ & $2\, {\cal R}^{(2)} (2,5) \oplus {\cal V} (1/3) \oplus {\cal V} (28/3)$ \\
& ${\cal R}^{(2)} (1,7)$ & $4$ & $8$ & ${\cal R}^{(2)} (0,1)_{\hls{7}} \oplus \hl{{\cal R}^{(2)} (7,12)} \oplus 2\, {\cal V} (10/3)$ \\
& ${\cal R}^{(2)} (2,7)$ & $4$ & $7$ & ${\cal R}^{(2)} (1/3,10/3)$ \\ \hline
${\cal V} (1/8)$ & ${\cal R}^{(2)} (5/8,5/8)$ & $5$ & $7$ & ${\cal R}^{(3)} (0,0,1,1) \oplus {\cal R}^{(2)} (1/3,10/3)$ \\
& ${\cal R}^{(2)} (1/3,1/3)$ & $5$ & $7$ & $2\, {\cal R}^{(2)} (1/8,1/8) \oplus {\cal R}^{(2)} (5/8,21/8)$ \\
& ${\cal R}^{(2)} (1/8,1/8)$ & $5$ & $7$ & ${\cal R}^{(3)} (0,0,2,2) \oplus 2\, {\cal R}^{(2)} (1/3,1/3)$ \\
& ${\cal R}^{(2)} (5/8,21/8)$ & $3$ & $6$ & $\hl{{\cal R}^{(3)} (0,1,2,7)} \oplus 2\, {\cal R}^{(2)} (1/3,10/3)$ \\
& ${\cal R}^{(2)} (1/3,10/3)$ & $3$ & $6$ & ${\cal R}^{(2)} (1/8,1/8) \oplus 2\, {\cal R}^{(2)} (5/8,21/8) \oplus \hl{{\cal R}^{(2)} (33/8,65/8)}$ \\
& ${\cal R}^{(2)} (1/8,33/8)$ & $3$ & $6$ & $\hl{{\cal R}^{(3)} (0,1,2,5)} \oplus {\cal R}^{(2)} (1/3,1/3) \oplus \hl{{\cal R}^{(2)} (10/3,28/3)}$ \\
& ${\cal R}^{(2)} (0,1)_5$ & $4$ & $6$ & $2\, {\cal V} (5/8) \oplus  {\cal V} (33/8) \oplus 2\, {\cal V} (-1/24) \oplus {\cal V} (35/24)$ \\
& ${\cal R}^{(2)} (0,1)_7$ & $4$ & $6$ & ${\cal R}^{(2)} (5/8,5/8) \oplus 2\, {\cal V} (-1/24)$ \\
& ${\cal R}^{(2)} (0,2)_5$ & $4$ & $6$ & ${\cal R}^{(2)} (1/8,1/8) \oplus {\cal V} (35/24)$ \\
& ${\cal R}^{(2)} (0,2)_7$ & $4$ & $7$ & $2\, {\cal V} (1/8) \oplus {\cal V} (21/8)$ \\
\end{tabular}

\footnotesize
\renewcommand{\arraystretch}{1.2}
\begin{tabular}{c @{$\; \otimesf \; $} c | c | c || l}
\multicolumn{2}{c|}{} & $L$ & $\tilde{L}_{\mbox{\footnotesize max}}$ & Fusion product \\ \hline
${\cal V} (1/8)$ & ${\cal R}^{(2)} (1,5)$ & $3$ & $6$ & ${\cal V} (5/8) \oplus 2\, {\cal V} (33/8) \oplus  {\cal V} (85/8) \oplus {\cal V} (-1/24) $ \\
\multicolumn{2}{c|}{} &&& $\qquad \oplus 2\, {\cal V} (-35/24) \oplus {\cal V} (143/24)$ \\
& ${\cal R}^{(2)} (2,5)$ & $4$ & $6$ & ${\cal R}^{(2)} (1/8,33/8) \oplus 2\, {\cal V} (35/24)$ \\
& ${\cal R}^{(2)} (1,7)$ & $3$ & $8$ & ${\cal R}^{(2)} (5/8,21/8) \oplus {\cal V} (-1/24) \oplus {\cal V} (143/24)$ \\
& ${\cal R}^{(2)} (2,7)$ & $4$ & $7$ & ${\cal V} (1/8) \oplus 2\, {\cal V} (21/8) \oplus {\cal V} (65/8)$ \\ \hline
${\cal V} (-1/24)$ & ${\cal R}^{(2)} (5/8,5/8)$ & $4$ & $6$ & $\hl{{\cal R}^{(3)} (0,1,2,5)} \oplus 2\, {\cal R}^{(2)} (1/3,1/3)$ \\
& ${\cal R}^{(2)} (1/3,1/3)$ & $4$ & $6$ & $2\, {\cal R}^{(2)} (5/8,5/8) \oplus {\cal R}^{(2)} (1/8,33/8) \oplus 2\, {\cal V} (-1/24) \oplus {\cal V} (35/24)$ \\
& ${\cal R}^{(2)} (1/8,1/8)$ & $4$ & $6$ & $2\, {\cal R}^{(3)} (0,0,1,1) \oplus {\cal R}^{(2)} (1/3,10/3)$ \\
& ${\cal R}^{(2)} (5/8,21/8)$ & $2$ & $5$ & $2\, \hl{{\cal R}^{(3)} (0,1,2,5)} \oplus {\cal R}^{(2)} (1/3,1/3) \oplus \hl{{\cal R}^{(2)} (10/3,28/3)}$ \\
& ${\cal R}^{(2)} (1/3,10/3)$ & $2$ & $5$ & ${\cal R}^{(2)} (5/8,5/8) \oplus 2\, \hl{{\cal R}^{(2)} (1/8,33/8)} \oplus  \hl{{\cal R}^{(2)} (21/8,85/8)}$ \\
\multicolumn{2}{c|}{} &&& $\qquad \oplus {\cal V} (-1/24) \oplus 2\, {\cal V} (35/24) \oplus {\cal V} (143/24) $ \\
& ${\cal R}^{(2)} (1/8,33/8)$ & $3$ & $6$ & ${\cal R} (0,0,1,1) \oplus \hl{{\cal R} (2,5,7,12)} \oplus 2\, {\cal R}^{(2)} (1/3,10/3)$ \\
& ${\cal R}^{(2)} (0,1)_5$ & $4$ & $6$ & $2\, {\cal R}^{(2)} (1/8,1/8) \oplus{\cal R}^{(2)} (5/8,21/8)$ \\
& ${\cal R}^{(2)} (0,1)_7$ & $4$ & $6$ & $2\, {\cal R}^{(2)} (1/8,1/8) \oplus {\cal V} (35/24)$ \\
& ${\cal R}^{(2)} (0,2)_5$ & $4$ & $7$ & ${\cal R}^{(2)} (5/8,21/8) \oplus 2\, {\cal V} (-1/24)$ \\
& ${\cal R}^{(2)} (0,2)_7$ & $4$ & $7$ & $2\, {\cal V} (-1/24) \oplus {\cal V} (35/24)$ \\
& ${\cal R}^{(2)} (1,5)$ & $2$ & $8$ & ${\cal R}^{(2)} (1/8,1/8) \oplus 2\, {\cal R}^{(2)} (5/8,21/8) \oplus \hl{{\cal R}^{(2)} (33/8,65/8)}$ \\
& ${\cal R}^{(2)} (2,5)$ & $3$ & $6$ & $2\, {\cal R}^{(2)} (5/8,21/8) \oplus {\cal V} (-1/24) \oplus {\cal V} (143/24)$ \\
& ${\cal R}^{(2)} (1,7)$ & $2$ & $8$ & ${\cal R}^{(2)} (1/8,1/8) \oplus \hl{{\cal R}^{(2)} (33/8,65/8)} \oplus 2\, {\cal V} (35/24)$ \\
& ${\cal R}^{(2)} (2,7)$ & $3$ & $6$ & ${\cal V} (-1/24) \oplus 2\, {\cal V} (35/24) \oplus {\cal V} (143/24)$ \\ \hline
${\cal V} (33/8)$ & ${\cal R}^{(2)} (5/8,5/8)$ & $3$ & $9$ & $\hl{{\cal R}^{(3)} (0,1,2,7)}$ \\
& ${\cal R}^{(2)} (1/3,1/3)$ & $3$ & $9$ & ${\cal V} (-1/24) \oplus 2\, {\cal V} (35/24) \oplus {\cal V} (143/24)$ \\
& ${\cal R}^{(2)} (1/8,1/8)$ & $3$ & $7$ & $\hl{{\cal R}^{(3)} (0,1,2,5)}$ \\
& ${\cal R}^{(2)} (5/8,21/8)$ & $2$ & $8$ & ${\cal R}^{(3)} (0,0,1,1) \oplus \hl{{\cal R}^{(3)} (2,5,7,12)}$ \\
& ${\cal R}^{(2)} (1/3,10/3)$ & $1$ & $9$ & $2 \, {\cal V} (-1/24) \oplus 2\, {\cal V} (35/24) \oplus 2\, {\cal V} (143/24) \oplus {\cal V} (323/24)$ \\
& ${\cal R}^{(2)} (1/8,33/8)$ & $2$ & $10$ & ${\cal R} (0,0,2,2) \oplus \hl{{\cal R}^{(3)} (1,5,7,15)}$ \\
& ${\cal R}^{(2)} (0,1)_5$ & $3$ & $9$ & ${\cal V} (1/8) \oplus 2\, {\cal V} (21/8) \oplus {\cal V} (65/8)$ \\
& ${\cal R}^{(2)} (0,1)_7$ & $3$ & $7$ & ${\cal R}^{(2)} (5/8,21/8)$ \\
& ${\cal R}^{(2)} (0,2)_5$ & $3$ & $9$ & $\hl{{\cal R}^{(2)} (1/8,33/8)}$ \\
& ${\cal R}^{(2)} (0,2)_7$ & $3$ & $9$ & ${\cal V} (5/8) \oplus 2\, {\cal V} (33/8) \oplus {\cal V} (85/8)$ \\
& ${\cal R}^{(2)} (1,5)$ & $2$ & $9$ & $2\, {\cal V} (1/8) \oplus 2\, {\cal V} (21/8) \oplus 2\, {\cal V} (65/8) \oplus {\cal V} (133/8)$ \\
& ${\cal R}^{(2)} (2,5)$ & $2$ & $9$ & ${\cal R}^{(2)} (1/8,1/8) \oplus \hl{{\cal R}^{(2)} (33/8,65/8)}$ \\
& ${\cal R}^{(2)} (1,7)$ & $2$ & $8$ & ${\cal R}^{(2)} (5/8,5/8) \oplus \hl{{\cal R}^{(2)} (21/8,85/8)}$ \\
& ${\cal R}^{(2)} (2,7)$ & $2$ & $9$ & $2\, {\cal V} (5/8) \oplus 2\, {\cal V} (33/8) \oplus 2\, {\cal V} (85/8) \oplus {\cal V} (161/8)$ \\ \hline
${\cal V} (10/3)$ & ${\cal R}^{(2)} (5/8,5/8)$ & $2$ & $8$ & ${\cal R}^{(2)} (1/8,1/8) \oplus \hl{{\cal R}^{(2)} (33/8,65/8)} \oplus 2\, {\cal V} (35/24)$ \\
& ${\cal R}^{(2)} (1/3,1/3)$ & $2$ & $8$ & $\hl{{\cal R}^{(3)} (0,1,2,7)} \oplus \hl{{\cal R}^{(2)} (1/3,10/3)}$ \\
& ${\cal R}^{(2)} (1/8,1/8)$ & $2$ & $8$ & $2\, {\cal R}^{(2)} (5/8,21/8) \oplus {\cal V} (-1/24) \oplus {\cal V} (143/24)$ \\
& ${\cal R}^{(2)} (0,1)_5$ & $2$ & $8$ & $\hl{{\cal R}^{(3)} (0,1,2,5)}$ \\
& ${\cal R}^{(2)} (0,1)_7$ & $2$ & $8$ & $2\, \hl{{\cal R}^{(2)} (2,5)} \oplus {\cal V} (1/3) \oplus {\cal V} (28/3)$ \\
& ${\cal R}^{(2)} (0,2)_5$ & $2$ & $8$ & ${\cal R}^{(2)} (0,1)_{\hls{7}} \oplus \hl{{\cal R}^{(2)} (7,12)} \oplus 2\, {\cal V} (10/3)$ \\
& ${\cal R}^{(2)} (0,2)_7$ & $2$ & $8$ & $\hl{{\cal R}^{(2)} (1/3,10/3)}$ \\ \hline
\end{tabular}

\footnotesize
\renewcommand{\arraystretch}{1.2}
\begin{tabular}{c @{$\; \otimesf \; $} c | c | c || l}
\multicolumn{2}{c|}{} & $L$ & $\tilde{L}_{\mbox{\footnotesize max}}$ & Fusion product \\ \hline
${\cal V} (21/8)$ & ${\cal R}^{(2)} (5/8,5/8)$ & $2$ & $8$ & $\hl{{\cal R}^{(3)} (0,1,2,5)} \oplus {\cal R}^{(2)} (1/3,1/3) \oplus \hl{{\cal R}^{(2)} (10/3,28/3)}$ \\
& ${\cal R}^{(2)} (1/3,1/3)$ & $2$ & $8$ & ${\cal R}^{(2)} (1/8,1/8) \oplus 2\, {\cal R}^{(2)} (5/8,21/8) \oplus \hl{{\cal R}^{(2)} (33/8,65/8)}$ \\
& ${\cal R}^{(2)} (1/8,1/8)$ & $2$ & $8$ & $\hl{{\cal R}^{(3)} (0,1,2,7)} \oplus 2\, \hl{{\cal R}^{(2)} (1/3,10/3)}$ \\
& ${\cal R}^{(2)} (0,1)_5$ & $2$ & $8$ & ${\cal V} (5/8) \oplus 2\, {\cal V} (33/8) \oplus {\cal V} (85/8) \oplus {\cal V} (-1/24) \oplus  2\, {\cal V} (35/24)$ \\
\multicolumn{2}{c|}{} &&& $\qquad \oplus {\cal V} (143/24)$ \\
& ${\cal R}^{(2)} (0,1)_7$ & $2$ & $8$ & $\hl{{\cal R}^{(2)} (1/8,33/8)} \oplus 2\, {\cal V} (35/24)$ \\
& ${\cal R}^{(2)} (0,2)_5$ & $2$ & $8$ & ${\cal R}^{(2)} (5/8,21/8) \oplus {\cal V} (-1/24) \oplus {\cal V} (143/24)$ \\
& ${\cal R}^{(2)} (0,2)_7$ & $2$ & $8$ & ${\cal V} (1/8) \oplus 2\, {\cal V} (21/8) \oplus {\cal V} (65/8)$ \\ \hline
${\cal V} (2)$ & ${\cal R}^{(2)} (5/8,5/8)$ & $5$ & $7$ & ${\cal R}^{(2)} (5/8,5/8)$ \\
& ${\cal R}^{(2)} (1/3,1/3)$ & $5$ & $7$ & ${\cal R}^{(2)} (1/3,1/3)$ \\
& ${\cal R}^{(2)} (1/8,1/8)$ & $5$ & $7$ & ${\cal R}^{(2)} (1/8,1/8)$ \\
& ${\cal R}^{(2)} (5/8,21/8)$ & $4$ & $7$ & ${\cal R}^{(2)} (5/8,21/8)$ \\
& ${\cal R}^{(2)} (1/3,10/3)$ & $4$ & $7$ & ${\cal R}^{(2)} (1/3,10/3)$ \\
& ${\cal R}^{(2)} (1/8,33/8)$ & $4$ & $7$ & ${\cal R}^{(2)} (1/8,33/8)$ \\
& ${\cal R}^{(2)} (0,1)_5$ & $4$ & $6$ & ${\cal R}^{(2)} (0,1)_5$ \\
& ${\cal R}^{(2)} (0,1)_7$ & $4$ & $6$ & ${\cal R}^{(2)} (0,1)_7$ \\
& ${\cal R}^{(2)} (0,2)_5$ & $4$ & $6$ & ${\cal R}^{(2)} (0,2)_5$ \\
& ${\cal R}^{(2)} (0,2)_7$ & $4$ & $6$ & ${\cal R}^{(2)} (0,2)_7$ \\
& ${\cal R}^{(2)} (1,5)$ & $4$ & $6$ & ${\cal R}^{(2)} (1,5)$ \\
& ${\cal R}^{(2)} (2,5)$ & $4$ & $6$ & ${\cal R}^{(2)} (2,5)$ \\
& ${\cal R}^{(2)} (1,7)$ & $6$ & $8$ & ${\cal R}^{(2)} (1,7)$ \\
& ${\cal R}^{(2)} (2,7)$ & $6$ & $8$ & ${\cal R}^{(2)} (2,7)$ \\ \hline
${\cal V} (1)$ & ${\cal R}^{(2)} (5/8,5/8)$ & $5$ & $7$ & ${\cal R}^{(2)} (1/8,1/8) \oplus {\cal V} (35/24)$ \\
& ${\cal R}^{(2)} (1/3,1/3)$ & $5$ & $7$ & ${\cal R}^{(3)} (0,0,1,1)$ \\
& ${\cal R}^{(2)} (1/8,1/8)$ & $5$ & $7$ & ${\cal R}^{(2)} (5/8,5/8) \oplus 2\, {\cal V} (-1/24)$ \\
& ${\cal R}^{(2)} (5/8,21/8)$ & $4$ & $7$ & ${\cal R}^{(2)} (1/8,33/8) \oplus 2\, {\cal V} (35/24)$ \\
& ${\cal R}^{(2)} (1/3,10/3)$ & $3$ & $6$ & $\hl{{\cal R}^{(3)} (0,1,2,5)}$ \\
& ${\cal R}^{(2)} (1/8,33/8)$ & $4$ & $5$ & ${\cal R}^{(2)} (5/8,21/8) \oplus {\cal V} (-1/24) \oplus {\cal V} (143/24)$ \\
& ${\cal R}^{(2)} (0,1)_5$ & $4$ & $6$ & ${\cal R}^{(2)} (0,2)_7 \oplus {\cal R}^{(2)} (1/3,1/3)$ \\
& ${\cal R}^{(2)} (0,1)_7$ & $4$ & $6$ & ${\cal R}^{(2)} (0,2)_5 \oplus 2\, {\cal V} (1/3)$ \\
& ${\cal R}^{(2)} (0,2)_5$ & $4$ & $7$ & ${\cal R}^{(2)} (0,1)_7 \oplus {\cal V} (10/3)$ \\
& ${\cal R}^{(2)} (0,2)_7$ & $4$ & $7$ & ${\cal R}^{(2)} (0,1)_5$ \\
& ${\cal R}^{(2)} (1,5)$ & $4$ & $6$ & $\hl{{\cal R}^{(2)} (2,7)} \oplus {\cal R}^{(2)} (1/3,10/3)$ \\
& ${\cal R}^{(2)} (2,5)$ & $4$ & $6$ & $\hl{{\cal R}^{(2)} (1,7)} \oplus 2\, {\cal V} (10/3)$ \\
& ${\cal R}^{(2)} (1,7)$ & $4$ & $8$ & ${\cal R}^{(2)} (2,5) \oplus {\cal V} (1/3) \oplus {\cal V} (28/3)$ \\
& ${\cal R}^{(2)} (2,7)$ & $4$ & $7$ & ${\cal R}^{(2)} (1,5)$ \\ \hline
${\cal V} (7)$ & ${\cal R}^{(2)} (5/8,5/8)$ & $4$ & $8$ & ${\cal R}^{(2)} (1/8,33/8)$ \\
& ${\cal R}^{(2)} (1/3,1/3)$ & $4$ & $8$ & ${\cal R}^{(2)} (1/3,10/3)$ \\
& ${\cal R}^{(2)} (1/8,1/8)$ & $4$ & $9$ & ${\cal R}^{(2)} (5/8,21/8)$ \\
& ${\cal R}^{(2)} (5/8,21/8)$ & $3$ & $7$ & ${\cal R}^{(2)} (1/8,1/8) \oplus \hl{{\cal R}^{(2)} (33/8,65/8)}$ \\
& ${\cal R}^{(2)} (1/3,10/3)$ & $3$ & $6$ & ${\cal R}^{(2)} (1/3,1/3) \oplus \hl{{\cal R}^{(2)} (10/3,28/3)}$ \\
& ${\cal R}^{(2)} (1/8,33/8)$ & $3$ & $7$ & ${\cal R}^{(2)} (5/8,5/8) \oplus \hl{{\cal R}^{(2)} (21/8,85/8)}$ \\
\end{tabular}

\footnotesize
\renewcommand{\arraystretch}{1.2}
\begin{tabular}{c @{$\; \otimesf \; $} c | c | c || l}
\multicolumn{2}{c|}{} & $L$ & $\tilde{L}_{\mbox{\footnotesize max}}$ & Fusion product \\ \hline
${\cal V} (7)$ & ${\cal R}^{(2)} (0,1)_5$ & $4$ & $6$ & ${\cal R}^{(2)} (1,5)$ \\
& ${\cal R}^{(2)} (0,1)_7$ & $4$ & $6$ & ${\cal R}^{(2)} (2,5)$ \\
& ${\cal R}^{(2)} (0,2)_5$ & $6$ & $7$ & ${\cal R}^{(2)} (1,7)$ \\
& ${\cal R}^{(2)} (0,2)_7$ & $5$ & $7$ & ${\cal R}^{(2)} (2,7)$ \\
& ${\cal R}^{(2)} (1,5)$ & $4$ & $6$ & ${\cal R}^{(2)} (0,1)_{\hls{5}} \oplus \hl{{\cal R}^{(2)} (5,12)}$ \\
& ${\cal R}^{(2)} (2,5)$ & $4$ & $6$ & ${\cal R}^{(2)} (0,1)_{\hls{7}} \oplus \hl{{\cal R}^{(2)} (7,12)}$ \\
& ${\cal R}^{(2)} (1,7)$ & $3$ & $6$ & ${\cal R}^{(2)} (0,2)_{\hls{5}} \oplus \hl{{\cal R}^{(2)} (5,15)}$ \\
& ${\cal R}^{(2)} (2,7)$ & $4$ & $7$ & ${\cal R}^{(2)} (0,2)_{\hls{7}} \oplus \hl{{\cal R}^{(2)} (7,15)}$ \\ \hline
${\cal V} (5)$ & ${\cal R}^{(2)} (5/8,5/8)$ & $3$ & $7$ & ${\cal R}^{(2)} (5/8,21/8) \oplus {\cal V} (-1/24) \oplus {\cal V} (143/24)$ \\
& ${\cal R}^{(2)} (1/3,1/3)$ & $3$ & $7$ & $\hl{{\cal R}^{(3)} (0,1,2,5)}$ \\
& ${\cal R}^{(2)} (1/8,1/8)$ & $3$ & $7$ & $\hl{{\cal R}^{(2)} (1/8,33/8)} \oplus 2\, {\cal V} (35/24)$ \\
& ${\cal R}^{(2)} (5/8,21/8)$ & $2$ & $8$ & ${\cal R}^{(2)} (5/8,5/8) \oplus \hl{{\cal R}^{(2)} (21/8,85/8)} \oplus 2\, {\cal V} (-1/24)$ \\
\multicolumn{2}{c|}{} &&& $\qquad \oplus 2\, {\cal V} (143/24)$ \\
& ${\cal R}^{(2)} (1/3,10/3)$ & $2$ & $8$ & ${\cal R}^{(3)} (0,0,1,1) \oplus \hl{{\cal R}^{(3)} (2,5,7,12)}$ \\
& ${\cal R}^{(2)} (1/8,33/8)$ & $2$ & $8$ & ${\cal R}^{(2)} (1/8,1/8) \oplus \hl{{\cal R}^{(2)} (33/8,65/8)} \oplus 2\, {\cal V} (35/24)$ \\
\multicolumn{2}{c|}{} &&& $\qquad  \oplus {\cal V} (323/24)$ \\
& ${\cal R}^{(2)} (0,1)_5$ & $3$ & $8$ & $\hl{{\cal R}^{(2)} (2,7)} \oplus {\cal R}^{(2)} (1/3,10/3)$ \\
& ${\cal R}^{(2)} (0,1)_7$ & $3$ & $9$ & $\hl{{\cal R}^{(2)} (1,7)} \oplus 2\, {\cal V} (10/3)$ \\
& ${\cal R}^{(2)} (0,2)_5$ & $3$ & $8$ & ${\cal R}^{(2)} (2,5) \oplus {\cal V} (1/3) \oplus {\cal V} (28/3)$ \\
& ${\cal R}^{(2)} (0,2)_7$ & $3$ & $8$ & $\hl{{\cal R}^{(2)} (1,5)}$ \\
& ${\cal R}^{(2)} (1,5)$ & $2$ & $8$ & ${\cal R}^{(2)} (0,2)_{\hls{7}} \oplus \hl{{\cal R}^{(2)} (7,15)} \oplus {\cal R}^{(2)} (1/3,1/3)$ \\
\multicolumn{2}{c|}{} &&& $\qquad \oplus \hl{{\cal R}^{(2)} (10/3,28/3)}$ \\
& ${\cal R}^{(2)} (2,5)$ & $2$ & $8$ & ${\cal R}^{(2)} (0,2)_{\hls{5}} \oplus \hl{{\cal R}^{(2)} (5,15)} \oplus 2\, {\cal V} (1/3) \oplus 2\, {\cal V} (28/3)$ \\
& ${\cal R}^{(2)} (1,7)$ & $2$ & $8$ & ${\cal R}^{(2)} (0,1)_{\hls{7}} \oplus \hl{{\cal R}^{(2)} (7,12)} \oplus 2\, {\cal V} (10/3) \oplus {\cal V} (55/3)$ \\
& ${\cal R}^{(2)} (2,7)$ & $2$ & $8$ & ${\cal R}^{(2)} (0,1)_{\hls{5}} \oplus \hl{{\cal R}^{(2)} (5,12)}$ \\ \hline
${\cal R}^{(1)} (0)_2$ & ${\cal R}^{(2)} (5/8,5/8)$ & $5$ & $7$ & ${\cal R}^{(2)} (5/8,5/8)$ \\
& ${\cal R}^{(2)} (1/3,1/3)$ & $5$ & $7$ & ${\cal R}^{(2)} (1/3,1/3)$ \\
& ${\cal R}^{(2)} (1/8,1/8)$ & $5$ & $7$ & ${\cal R}^{(2)} (1/8,1/8)$ \\
& ${\cal R}^{(2)} (5/8,21/8)$ & $4$ & $7$ & ${\cal R}^{(2)} (5/8,21/8)$ \\
& ${\cal R}^{(2)} (1/3,10/3)$ & $4$ & $7$ & ${\cal R}^{(2)} (1/3,10/3)$ \\
& ${\cal R}^{(2)} (1/8,33/8)$ & $4$ & $7$ & ${\cal R}^{(2)} (1/8,33/8)$ \\
& ${\cal R}^{(2)} (0,1)_5$ & $5$ & $7$ & ${\cal R}^{(2)} (0,1)_5$ \\
& ${\cal R}^{(2)} (0,1)_7$ & $5$ & $7$ & ${\cal R}^{(2)} (0,1)_7$ \\
& ${\cal R}^{(2)} (0,2)_5$ & $5$ & $7$ & ${\cal R}^{(2)} (0,2)_5$ \\
& ${\cal R}^{(2)} (0,2)_7$ & $5$ & $7$ & ${\cal R}^{(2)} (0,2)_7$ \\
& ${\cal R}^{(2)} (1,5)$ & $4$ & $7$ & ${\cal R}^{(2)} (1,5)$ \\
& ${\cal R}^{(2)} (2,5)$ & $4$ & $7$ & ${\cal R}^{(2)} (2,5)$ \\
& ${\cal R}^{(2)} (1,7)$ & $6$ & $7$ & ${\cal R}^{(2)} (1,7)$ \\
& ${\cal R}^{(2)} (2,7)$ & $6$ & $7$ & ${\cal R}^{(2)} (2,7)$ \\ \hline
\end{tabular}

\footnotesize
\renewcommand{\arraystretch}{1.2}
\begin{tabular}{c @{$\; \otimesf \; $} c | c | c || l}
\multicolumn{2}{c|}{} & $L$ & $\tilde{L}_{\mbox{\footnotesize max}}$ & Fusion product \\ \hline
${\cal R}^{(1)} (0)_1$ & ${\cal R}^{(2)} (5/8,5/8)$ & $5$ & $7$ & ${\cal R}^{(2)} (1/8,1/8) \oplus {\cal V} (35/24)$ \\
& ${\cal R}^{(2)} (1/3,1/3)$ & $5$ & $7$ & ${\cal R}^{(3)} (0,0,1,1)$ \\
& ${\cal R}^{(2)} (1/8,1/8)$ & $5$ & $7$ & ${\cal R}^{(2)} (5/8,5/8) \oplus 2\, {\cal V} (-1/24)$ \\
& ${\cal R}^{(2)} (5/8,21/8)$ & $4$ & $7$ & ${\cal R}^{(2)} (1/8,33/8) \oplus 2\, {\cal V} (35/24)$ \\
& ${\cal R}^{(2)} (1/3,10/3)$ & $4$ & $7$ & $\hl{{\cal R}^{(3)} (0,1,2,5)}$ \\
& ${\cal R}^{(2)} (1/8,33/8)$ & $4$ & $7$ & ${\cal R}^{(2)} (5/8,21/8) \oplus {\cal V} (-1/24) \oplus {\cal V} (143/24)$ \\
& ${\cal R}^{(2)} (0,1)_5$ & $5$ & $7$ & ${\cal R}^{(2)} (0,2)_7 \oplus {\cal R}^{(2)} (1/3,1/3)$ \\
& ${\cal R}^{(2)} (0,1)_7$ & $5$ & $7$ & ${\cal R}^{(2)} (0,2)_5 \oplus 2\, {\cal V} (1/3)$ \\
& ${\cal R}^{(2)} (0,2)_5$ & $5$ & $7$ & ${\cal R}^{(2)} (0,1)_7 \oplus {\cal V} (10/3)$ \\
& ${\cal R}^{(2)} (0,2)_7$ & $5$ & $7$ & ${\cal R}^{(2)} (0,1)_5$ \\
& ${\cal R}^{(2)} (1,5)$ & $4$ & $7$ & $\hl{{\cal R}^{(2)} (2,7)} \oplus {\cal R}^{(2)} (1/3,10/3)$ \\
& ${\cal R}^{(2)} (2,5)$ & $4$ & $7$ & $\hl{{\cal R}^{(2)} (1,7)} \oplus 2\, {\cal V} (10/3)$ \\
& ${\cal R}^{(2)} (1,7)$ & $4$ & $7$ & ${\cal R}^{(2)} (2,5) \oplus {\cal V} (1/3) \oplus {\cal V} (28/3)$ \\
& ${\cal R}^{(2)} (2,7)$ & $4$ & $7$ & ${\cal R}^{(2)} (1,5)$ \\ \hline
\end{tabular}

\vspace{1.0cm}
\normalsize
In order to extract the fusion of higher rank representations
we used the symmetry and associativity properties of the fusion product
along the lines described in section \ref{discussion_c0_consistency}.
For these calculations we applied our explicit results of the 
Nahm algorithm in the form stated in the tables above.
The results are listed below.

\footnotesize
\vspace{0.8cm}
\renewcommand{\arraystretch}{1.2}
\begin{tabular}{c @{$\; \otimesf \; $} c || l}
\multicolumn{2}{c||}{} & Fusion product \\ \hline
${\cal R}^{(2)} (5/8,5/8)$ & ${\cal R}^{(2)} (5/8,5/8)$ & $2\, {\cal R}^{(3)} (0,0,2,2) \oplus {\cal R}^{(3)} (0,1,2,5) \oplus {\cal R}^{(2)} (1/3,1/3)$ \\
\multicolumn{2}{c||}{} & $\qquad \oplus {\cal R}^{(2)} (10/3,28/3)$ \\
& ${\cal R}^{(2)} (1/3,1/3)$ & ${\cal R}^{(2)} (1/8,1/8) \oplus 2\, {\cal R}^{(2)} (5/8,21/8) \oplus {\cal R}^{(2)} (33/8,65/8)$ \\
\multicolumn{2}{c||}{} & $\qquad \oplus 4\, {\cal V} (-1/24) \oplus 2\, {\cal V} (35/24)$ \\
& ${\cal R}^{(2)} (1/8,1/8)$ & $2\, {\cal R}^{(3)} (0,0,1,1) \oplus {\cal R}^{(3)} (0,1,2,7) \oplus 2\, {\cal R}^{(2)} (1/3,10/3)$ \\
& ${\cal R}^{(2)} (5/8,21/8)$ & ${\cal R}^{(3)} (0,0,2,2) \oplus 2\, {\cal R}^{(3)} (0,1,2,5) \oplus {\cal R}^{(3)} (1,5,7,15)$ \\
\multicolumn{2}{c||}{} & $\qquad \oplus 2\, {\cal R}^{(2)} (1/3,1/3) \oplus 2\, {\cal R}^{(2)} (10/3,28/3)$ \\
& ${\cal R}^{(2)} (1/3,10/3)$ & $2\, {\cal R}^{(2)} (1/8,1/8) \oplus 2\, {\cal R}^{(2)} (5/8,21/8) \oplus 2\, {\cal R}^{(2)} (33/8,65/8)$ \\
\multicolumn{2}{c||}{} & $\qquad \oplus {\cal R}^{(2)} (85/8,133/8) \oplus 2\, {\cal V} (-1/24) \oplus 4\, {\cal V} (35/24)$ \\
\multicolumn{2}{c||}{} & $\qquad \oplus 2\, {\cal V} (143/24)$ \\
& ${\cal R}^{(2)} (1/8,33/8)$ & ${\cal R}^{(3)} (0,0,1,1) \oplus 2\, {\cal R}^{(3)} (0,1,2,7) \oplus {\cal R}^{(3)} (2,5,7,12)$ \\
\multicolumn{2}{c||}{} & $\qquad \oplus 2\, {\cal R}^{(2)} (1/3,10/3) \oplus  {\cal R}^{(2)} (28/3,55/3)$ \\
& ${\cal R}^{(2)} (0,1)_5$ & $2\, {\cal R}^{(2)} (1/8,1/8) \oplus {\cal R}^{(2)} (5/8,21/8) \oplus {\cal V} (-1/24) \oplus 2\, {\cal V} (35/24)$ \\
\multicolumn{2}{c||}{} & $\qquad \oplus {\cal V} (143/24)$ \\
& ${\cal R}^{(2)} (0,1)_7$ & $2\, {\cal R}^{(2)} (1/8,1/8) \oplus {\cal R}^{(2)} (1/8,33/8) \oplus 2\, {\cal V} (35/24)$ \\
& ${\cal R}^{(2)} (0,2)_5$ & $2\, {\cal R}^{(2)} (5/8,5/8) \oplus {\cal R}^{(2)} (5/8,21/8) \oplus {\cal V} (-1/24) \oplus {\cal V} (143/24)$ \\
& ${\cal R}^{(2)} (0,2)_7$ & $2\, {\cal R}^{(2)} (5/8,5/8) \oplus {\cal R}^{(2)} (1/8,33/8)$ \\ \hline
\end{tabular}

\footnotesize
\renewcommand{\arraystretch}{1.2}
\begin{tabular}{c @{$\; \otimesf \; $} c || l}
\multicolumn{2}{c||}{} & Fusion product \\ \hline
${\cal R}^{(2)} (1/3,1/3)$ & ${\cal R}^{(2)} (1/3,1/3)$ & $2\, {\cal R}^{(3)} (0,0,2,2) \oplus {\cal R}^{(3)} (0,1,2,7) \oplus 2\, {\cal R}^{(2)} (1/3,1/3)$ \\
\multicolumn{2}{c||}{} & $\qquad \oplus {\cal R}^{(2)} (1/3,10/3)$ \\
& ${\cal R}^{(2)} (1/8,1/8)$ & $4\, {\cal R}^{(2)} (1/8,1/8) \oplus 2\, {\cal R}^{(2)} (5/8,21/8) \oplus {\cal V} (-1/24)$ \\
\multicolumn{2}{c||}{} & $\qquad \oplus 2\, {\cal V} (35/24) \oplus {\cal V} (143/24)$ \\
& ${\cal R}^{(2)} (5/8,21/8)$ & $2\, {\cal R}^{(2)} (1/8,1/8) \oplus 4\, {\cal R}^{(2)} (5/8,21/8) \oplus 2\, {\cal R}^{(2)} (33/8,65/8)$ \\
\multicolumn{2}{c||}{} & $\qquad \oplus 2\, {\cal V} (-1/24) \oplus 2\, {\cal V} (35/24) \oplus 2\, {\cal V} (143/24) \oplus {\cal V} (323/24)$ \\
& ${\cal R}^{(2)} (1/3,10/3)$ & ${\cal R}^{(3)} (0,0,2,2) \oplus 2\, {\cal R}^{(3)} (0,1,2,7) \oplus {\cal R}^{(3)} (1,5,7,12)$ \\
\multicolumn{2}{c||}{} & $\qquad \oplus {\cal R}^{(2)} (1/3,1/3) \oplus 2\, {\cal R}^{(2)} (1/3,10/3) \oplus {\cal R}^{(2)} (10/3,28/3)$ \\
& ${\cal R}^{(2)} (1/8,33/8)$ & $2\, {\cal R}^{(2)} (1/8,1/8) \oplus 2\, {\cal R}^{(2)} (5/8,21/8) \oplus 2\, {\cal R}^{(2)} (33/8,65/8)$ \\
\multicolumn{2}{c||}{} & $\qquad \oplus {\cal R}^{(2)} (85/8,133/8) \oplus 2\, {\cal V} (-1/24) \oplus 4\, {\cal V} (35/24)$ \\
\multicolumn{2}{c||}{} & $\qquad \oplus 2\, {\cal V} (143/24)$ \\
& ${\cal R}^{(2)} (0,1)_5$ & $2\, {\cal R}^{(3)} (0,0,1,1) \oplus {\cal R}^{(3)} (0,1,2,5)$ \\
& ${\cal R}^{(2)} (0,1)_7$ & $2\, {\cal R}^{(3)} (0,0,1,1) \oplus {\cal R}^{(2)} (1/3,10/3)$ \\
& ${\cal R}^{(2)} (0,2)_5$ & ${\cal R}^{(3)} (0,1,2,5) \oplus 2\, {\cal R}^{(2)} (1/3,1/3)$ \\
& ${\cal R}^{(2)} (0,2)_7$ & $2\, {\cal R}^{(2)} (1/3,1/3) \oplus {\cal R}^{(2)} (1/3,10/3)$ \\ \hline
${\cal R}^{(2)} (1/8,1/8)$ & ${\cal R}^{(2)} (1/8,1/8)$ & $2\, {\cal R}^{(3)} (0,0,2,2) \oplus {\cal R}^{(3)} (0,1,2,5) \oplus 4\, {\cal R}^{(2)} (1/3,1/3)$ \\
& ${\cal R}^{(2)} (5/8,21/8)$ & ${\cal R}^{(3)} (0,0,1,1) \oplus 2\, {\cal R}^{(3)} (0,1,2,7) \oplus {\cal R}^{(3)} (2,5,7,12)$ \\
\multicolumn{2}{c||}{} & $\qquad \oplus 4\, {\cal R}^{(2)} (1/3,10/3)$ \\
& ${\cal R}^{(2)} (1/3,10/3)$ & $2\, {\cal R}^{(2)} (1/8,1/8) \oplus 4\, {\cal R}^{(2)} (5/8,21/8) \oplus 2\, {\cal R}^{(2)} (33/8,65/8)$ \\
\multicolumn{2}{c||}{} & $\qquad \oplus 2\, {\cal V} (-1/24) \oplus 2\, {\cal V} (35/24) \oplus 2\, {\cal V} (143/24) \oplus {\cal V} (323/24)$ \\
& ${\cal R}^{(2)} (1/8,33/8)$ & ${\cal R}^{(3)} (0,0,2,2) \oplus 2\, {\cal R}^{(3)} (0,1,2,5) \oplus {\cal R}^{(3)} (1,5,7,15)$ \\
\multicolumn{2}{c||}{} & $\qquad \oplus 2\, {\cal R}^{(2)} (1/3,1/3) \oplus 2\, {\cal R}^{(2)} (10/3,28/3)$ \\
& ${\cal R}^{(2)} (0,1)_5$ & $2\, {\cal R}^{(2)} (5/8,5/8) \oplus {\cal R}^{(2)} (1/8,33/8) \oplus 4\, {\cal V} (-1/24)$ \\
\multicolumn{2}{c||}{} & $\qquad \oplus 2\, {\cal V} (35/24)$ \\
& ${\cal R}^{(2)} (0,1)_7$ & $2\, {\cal R}^{(2)} (5/8,5/8) \oplus {\cal R}^{(2)} (5/8,21/8) \oplus 4\, {\cal V} (-1/24)$ \\
& ${\cal R}^{(2)} (0,2)_5$ & $2\, {\cal R}^{(2)} (1/8,1/8) \oplus {\cal R}^{(2)} (1/8,33/8) \oplus 2\, {\cal V} (35/24)$ \\
& ${\cal R}^{(2)} (0,2)_7$ & $2\, {\cal R}^{(2)} (1/8,1/8) \oplus {\cal R}^{(2)} (5/8,21/8)$ \\ \hline
${\cal R}^{(2)} (0,1)_5$  & ${\cal R}^{(2)} (0,1)_5$ & $2\, {\cal R}^{(2)} (0,2)_7 \oplus {\cal R}^{(2)} (2,7) \oplus 2\, {\cal R}^{(2)} (1/3,1/3)$ \\
\multicolumn{2}{c||}{} & $\qquad \oplus {\cal R}^{(2)} (1/3,10/3)$ \\
& ${\cal R}^{(2)} (0,1)_7$ & ${\cal R}^{(3)} (0,0,2,2) \oplus 2\, {\cal R}^{(2)} (1/3,1/3)$ \\
& ${\cal R}^{(2)} (0,2)_5$ & ${\cal R}^{(3)} (0,0,1,1) \oplus {\cal R}^{(2)} (1/3,10/3)$ \\
& ${\cal R}^{(2)} (0,2)_7$ & $2\, {\cal R}^{(2)} (0,1)_5 \oplus {\cal R}^{(2)} (1,5)$ \\ \hline
${\cal R}^{(2)} (0,1)_7$ & ${\cal R}^{(2)} (0,1)_7$ & $2\, {\cal R}^{(2)} (0,2)_5 \oplus {\cal R}^{(2)} (2,5) \oplus 4\, {\cal V} (1/3)$ \\
& ${\cal R}^{(2)} (0,2)_5$ & $2\, {\cal R}^{(2)} (0,1)_7 \oplus {\cal R}^{(2)} (1,7) \oplus 2\, {\cal V} (10/3)$ \\
& ${\cal R}^{(2)} (0,2)_7$ & ${\cal R}^{(3)} (0,0,1,1)$ \\ \hline
${\cal R}^{(2)} (0,2)_5$ & ${\cal R}^{(2)} (0,2)_5$ & $2\, {\cal R}^{(2)} (0,2)_5 \oplus {\cal R}^{(2)} (2,5) \oplus {\cal V} (1/3) \oplus {\cal V} (28/3)$ \\
& ${\cal R}^{(2)} (0,2)_7$ & ${\cal R}^{(3)} (0,0,2,2)$ \\ \hline
${\cal R}^{(2)} (0,2)_7$ & ${\cal R}^{(2)} (0,2)_7$ & $2\, {\cal R}^{(2)} (0,2)_7 \oplus {\cal R}^{(2)} (2,7)$ \\ \hline
\end{tabular}

\footnotesize
\renewcommand{\arraystretch}{1.2}
\begin{tabular}{c @{$\; \otimesf \; $} c || l}
\multicolumn{2}{c||}{} & Fusion product \\ \hline
${\cal R}^{(3)} (0,0,1,1)$ & ${\cal V} (5/8)$ & $2\, {\cal R}^{(2)} (1/8,1/8) \oplus {\cal R}^{(2)} (5/8,21/8)$ \\
& ${\cal V} (1/3)$ & $2\, {\cal R}^{(3)} (0,0,1,1) \oplus {\cal R}^{(2)} (1/3,10/3)$ \\
& ${\cal V} (1/8)$ & $2\, {\cal R}^{(2)} (5/8,5/8) \oplus {\cal R}^{(2)} (1/8,33/8) \oplus 4\, {\cal V} (-1/24) \oplus 2\, {\cal V} (35/24)$ \\
& ${\cal V} (-1/24)$ & $4\, {\cal R}^{(2)} (1/8,1/8) \oplus 2\, {\cal R}^{(2)} (5/8,21/8) \oplus {\cal V} (-1/24)$ \\
\multicolumn{2}{c||}{} & $\qquad \oplus 2\, {\cal V} (35/24) \oplus {\cal V} (143/24)$ \\
& ${\cal V} (2)$ & ${\cal R}^{(3)} (0,0,1,1)$ \\
& ${\cal V} (1)$ & ${\cal R}^{(3)} (0,0,2,2) \oplus 2\, {\cal R}^{(2)} (1/3,1/3)$ \\
& ${\cal V} (7)$ & ${\cal R}^{(3)} (0,1,2,5)$ \\
& ${\cal V} (5)$ & ${\cal R}^{(3)} (0,1,2,7) \oplus 2\, {\cal R}^{(2)} (1/3,10/3)$ \\
& ${\cal R}^{(1)} (0)_2$ & ${\cal R}^{(3)} (0,0,1,1)$ \\
& ${\cal R}^{(1)} (0)_1$ & ${\cal R}^{(3)} (0,0,2,2) \oplus 2\, {\cal R}^{(2)} (1/3,1/3)$ \\
& ${\cal R}^{(2)} (5/8,5/8)$ & ${\cal R}^{(2)} (5/8,5/8) \oplus 2\, {\cal R}^{(2)} (5/8,21/8) \oplus {\cal R}^{(2)} (21/8,85/8)$ \\
\multicolumn{2}{c||}{} & $\qquad \oplus 4\, {\cal R}^{(2)} (1/8,1/8) \oplus 2\, {\cal R}^{(2)} (1/8,33/8)$ \\
\multicolumn{2}{c||}{} & $\qquad \oplus 2\, {\cal V} (-1/24) \oplus 4\, {\cal V} (35/24) \oplus 2\, {\cal V} (143/24)$ \\
& ${\cal R}^{(2)} (1/3,1/3)$ & $4\, {\cal R}^{(3)} (0,0,1,1) \oplus 2\, {\cal R}^{(3)} (0,1,2,5) \oplus {\cal R}^{(2)} (1/3,1/3)$ \\
\multicolumn{2}{c||}{} & $\qquad \oplus 2\, {\cal R}^{(2)} (1/3,10/3) \oplus {\cal R}^{(2)} (10/3,28/3)$ \\
& ${\cal R}^{(2)} (1/8,1/8)$ & $4\, {\cal R}^{(2)} (5/8,5/8) \oplus 2\, {\cal R}^{(2)} (5/8,21/8) \oplus {\cal R}^{(2)} (1/8,1/8)$ \\
\multicolumn{2}{c||}{} & $\qquad \oplus 2\, {\cal R}^{(2)} (1/8,33/8) \oplus {\cal R}^{(2)} (33/8,65/8)$ \\
\multicolumn{2}{c||}{} & $\qquad \oplus 8\, {\cal V} (-1/24) \oplus 4\, {\cal V} (35/24)$ \\
& ${\cal R}^{(2)} (0,1)_5$ & $2\, {\cal R}^{(3)} (0,0,2,2) \oplus {\cal R}^{(3)} (0,1,2,7) \oplus 4\, {\cal R}^{(2)} (1/3,1/3)$ \\
\multicolumn{2}{c||}{} & $\qquad \oplus 2\, {\cal R}^{(2)} (1/3,10/3)$ \\
& ${\cal R}^{(2)} (0,1)_7$ & $2\, {\cal R}^{(3)} (0,0,2,2) \oplus {\cal R}^{(3)} (0,1,2,5) \oplus 4\, {\cal R}^{(2)} (1/3,1/3)$ \\
& ${\cal R}^{(2)} (0,2)_5$ & $2\, {\cal R}^{(3)} (0,0,1,1) \oplus {\cal R}^{(3)} (0,1,2,7) \oplus 2\, {\cal R}^{(2)} (1/3,10/3)$ \\
& ${\cal R}^{(2)} (0,2)_7$ & $2\, {\cal R}^{(3)} (0,0,1,1) \oplus {\cal R}^{(3)} (0,1,2,5)$ \\
& ${\cal R}^{(3)} (0,0,1,1)$ & $4\, {\cal R}^{(3)} (0,0,2,2) \oplus 2\, {\cal R}^{(3)} (0,1,2,5) \oplus {\cal R}^{(3)} (0,0,1,1)$ \\
\multicolumn{2}{c||}{} & $\qquad \oplus 2\, {\cal R}^{(3)} (0,1,2,7) \oplus {\cal R}^{(3)} (2,5,7,12)$ \\
\multicolumn{2}{c||}{} & $\qquad \oplus 8\, {\cal R}^{(2)} (1/3,1/3) \oplus 4\, {\cal R}^{(2)} (1/3,10/3)$ \\
& ${\cal R}^{(3)} (0,0,2,2)$ & ${\cal R}^{(3)} (0,0,2,2) \oplus 2\, {\cal R}^{(3)} (0,1,2,5) \oplus {\cal R}^{(3)} (1,5,7,15)$ \\
\multicolumn{2}{c||}{} & $\qquad \oplus 4\, {\cal R}^{(3)} (0,0,1,1) \oplus 2\, {\cal R}^{(3)} (0,1,2,7) \oplus 2\, {\cal R}^{(2)} (1/3,1/3)$ \\
\multicolumn{2}{c||}{} & $\qquad \oplus 4\, {\cal R}^{(2)} (1/3,10/3) \oplus 2\, {\cal R}^{(2)} (10/3,28/3)$ \\ \hline
${\cal R}^{(3)} (0,0,2,2)$ & ${\cal V} (5/8)$ & $2\, {\cal R}^{(2)} (5/8,5/8) \oplus {\cal R}^{(2)} (1/8,33/8)$ \\
& ${\cal V} (1/3)$ & ${\cal R}^{(3)} (0,1,2,5) \oplus 2\, {\cal R}^{(2)} (1/3,1/3)$ \\
& ${\cal V} (1/8)$ & $2\, {\cal R}^{(2)} (1/8,1/8) \oplus {\cal R}^{(2)} (5/8,21/8) \oplus {\cal V} (-1/24)$ \\
\multicolumn{2}{c||}{} & $\qquad \oplus 2\, {\cal V} (35/24) \oplus {\cal V} (143/24)$ \\
& ${\cal V} (-1/24)$ & ${\cal R}^{(2)} (1/8,1/8) \oplus 2\, {\cal R}^{(2)} (5/8,21/8) \oplus {\cal R}^{(2)} (33/8,65/8)$ \\
\multicolumn{2}{c||}{} & $\qquad \oplus 4\, {\cal V} (-1/24) \oplus 2\, {\cal V} (35/24)$ \\
& ${\cal V} (2)$ & ${\cal R}^{(3)} (0,0,2,2)$ \\
& ${\cal V} (1)$ & ${\cal R}^{(3)} (0,0,1,1) \oplus {\cal R}^{(2)} (1/3,10/3)$ \\
& ${\cal V} (7)$ & ${\cal R}^{(3)} (0,1,2,7)$ \\
& ${\cal V} (5)$ & ${\cal R}^{(3)} (0,1,2,5) \oplus {\cal R}^{(2)} (1/3,1/3) \oplus {\cal R}^{(2)} (10/3,28/3)$ \\
& ${\cal R}^{(1)} (0)_2$ & ${\cal R}^{(3)} (0,0,2,2)$ \\
& ${\cal R}^{(1)} (0)_1$ & ${\cal R}^{(3)} (0,0,1,1) \oplus {\cal R}^{(2)} (1/3,10/3)$ \\
\end{tabular}

\footnotesize
\renewcommand{\arraystretch}{1.2}
\begin{tabular}{c @{$\; \otimesf \; $} c || l}
\multicolumn{2}{c||}{} & Fusion product \\ \hline
${\cal R}^{(3)} (0,0,2,2)$ & ${\cal R}^{(2)} (5/8,5/8)$ & $4\, {\cal R}^{(2)} (5/8,5/8) \oplus 2\, {\cal R}^{(2)} (5/8,21/8) \oplus {\cal R}^{(2)} (1/8,1/8)$ \\
\multicolumn{2}{c||}{} & $\qquad \oplus 2\, {\cal R}^{(2)} (1/8,33/8) \oplus {\cal R}^{(2)} (33/8,65/8) \oplus 2\, {\cal V} (-1/24)$ \\
\multicolumn{2}{c||}{} & $\qquad \oplus 2\, {\cal V} (35/24) \oplus 2\, {\cal V} (143/24) \oplus {\cal V} (323/24)$ \\
& ${\cal R}^{(2)} (1/3,1/3)$ & ${\cal R}^{(3)} (0,0,1,1) \oplus 2\, {\cal R}^{(3)} (0,1,2,5) \oplus {\cal R}^{(3)} (2,5,7,12)$ \\
\multicolumn{2}{c||}{} & $\qquad \oplus 4\, {\cal R}^{(2)} (1/3,1/3) \oplus 2\, {\cal R}^{(2)} (1/3,10/3)$ \\
& ${\cal R}^{(2)} (1/8,1/8)$ & ${\cal R}^{(2)} (5/8,5/8) \oplus 2\, {\cal R}^{(2)} (5/8,21/8) \oplus {\cal R}^{(2)} (21/8,85/8)$ \\
\multicolumn{2}{c||}{} & $\qquad \oplus 4\, {\cal R}^{(2)} (1/8,1/8) \oplus 2\, {\cal R}^{(2)} (1/8,33/8)$ \\
\multicolumn{2}{c||}{} & $\qquad \oplus 2\, {\cal V} (-1/24) \oplus 4\, {\cal V} (35/24) \oplus 2\, {\cal V} (143/24)$ \\
& ${\cal R}^{(2)} (0,1)_5$ & $2\, {\cal R}^{(3)} (0,0,1,1) \oplus {\cal R}^{(3)} (0,1,2,5) \oplus {\cal R}^{(2)} (1/3,1/3)$ \\
\multicolumn{2}{c||}{} & $\qquad \oplus 2\, {\cal R}^{(2)} (1/3,10/3) \oplus {\cal R}^{(2)} (10/3,28/3)$ \\
& ${\cal R}^{(2)} (0,1)_7$ & $2\, {\cal R}^{(3)} (0,0,1,1) \oplus {\cal R}^{(3)} (0,1,2,7) \oplus 2\, {\cal R}^{(2)} (1/3,10/3)$ \\
& ${\cal R}^{(2)} (0,2)_5$ & $2\, {\cal R}^{(3)} (0,0,2,2) \oplus {\cal R}^{(3)} (0,1,2,5) \oplus {\cal R}^{(2)} (1/3,1/3) \oplus$ \\
\multicolumn{2}{c||}{} & $\qquad {\cal R}^{(2)} (10/3,28/3)$ \\
& ${\cal R}^{(2)} (0,2)_7$ & $2\, {\cal R}^{(3)} (0,0,2,2) \oplus {\cal R}^{(3)} (0,1,2,7)$ \\
& ${\cal R}^{(3)} (0,0,2,2)$ & $4\, {\cal R}^{(3)} (0,0,2,2) \oplus 2\, {\cal R}^{(3)} (0,1,2,5) \oplus {\cal R}^{(3)} (0,0,1,1)$ \\
\multicolumn{2}{c||}{} & $\qquad \oplus 2\, {\cal R}^{(3)} (0,1,2,7) \oplus {\cal R}^{(3)} (2,5,7,12) \oplus 2\, {\cal R}^{(2)} (1/3,1/3)$ \\
\multicolumn{2}{c||}{} & $\qquad \oplus 2\, {\cal R}^{(2)} (1/3,10/3) \oplus 2\, {\cal R}^{(2)} (10/3,28/3) \oplus {\cal R}^{(2)} (28/3,55/3)$ \\ \hline
${\cal R}^{(3)} (0,1,2,5)$ & ${\cal V} (5/8)$ & $2\, {\cal R}^{(2)} (5/8,21/8) \oplus {\cal R}^{(2)} (1/8,1/8) \oplus {\cal R}^{(2)} (33/8,65/8)$ \\
& ${\cal V} (1/3)$ & $2\, {\cal R}^{(3)} (0,1,2,5) \oplus {\cal R}^{(2)} (1/3,1/3) \oplus {\cal R}^{(2)} (10/3,28/3)$ \\
& ${\cal V} (1/8)$ & ${\cal R}^{(2)} (5/8,5/8) \oplus 2\, {\cal R}^{(2)} (1/8,33/8) \oplus {\cal R}^{(2)} (21/8,85/8)$ \\
\multicolumn{2}{c||}{} & $\qquad \oplus 2\, {\cal V} (-1/24) \oplus 4\, {\cal V} (35/24) \oplus 2\, {\cal V} (143/24)$ \\
& ${\cal V} (-1/24)$ & $2\, {\cal R}^{(2)} (1/8,1/8) \oplus 4\, {\cal R}^{(2)} (5/8,21/8) \oplus 2\, {\cal R}^{(2)} (33/8,65/8)$ \\
\multicolumn{2}{c||}{} & $\qquad \oplus 2\, {\cal V} (-1/24) \oplus 2\, {\cal V} (35/24) \oplus 2\, {\cal V} (143/24) \oplus {\cal V} (323/24)$ \\
& ${\cal V} (2)$ & ${\cal R}^{(3)} (0,1,2,5)$ \\
& ${\cal V} (1)$ & ${\cal R}^{(3)} (0,1,2,7) \oplus 2\, {\cal R}^{(2)} (1/3,10/3)$ \\
& ${\cal R}^{(1)} (0)_2$ & ${\cal R}^{(3)} (0,1,2,5)$ \\
& ${\cal R}^{(1)} (0)_1$ & ${\cal R}^{(3)} (0,1,2,7) \oplus 2\, {\cal R}^{(2)} (1/3,10/3)$ \\
& ${\cal R}^{(2)} (0,2)_7$ & ${\cal R}^{(3)} (0,0,1,1) \oplus 2\, {\cal R}^{(3)} (0,1,2,5) \oplus {\cal R}^{(3)} (2,5,7,12)$ \\ \hline
\end{tabular}
\normalsize

\vspace{1.0cm}
\newpage

%%%%%%%%%%%%%%%%%%%%%%%%%%%%%%%%%%%%%%%%%%%%%
%%%%%%%%%%%%%%%%%%%%%%%%%%%%%%%%%%%%%%%%%%%%%
%%%%%%%%%%%%%%%%%%%%%%%%%%%%%%%%%%%%%%%%%%%%%

\section{Explicit fusion rules for $\mathbf{c_{2,5}=-22/5}$\label{fusion_cminus22over5}}

In the following table we have collected the results 
of our explicit calculations of the
fusion product of irreducible representations in the
augmented $c_{2,5}=-22/5$ model. These results are certainly
not complete, but mainly serve to compute the lowest higher
rank representations as well as to check the proposed
fusion rules on a variety of examples. 
The notations are as before.

\footnotesize
\vspace{0.8cm}
\renewcommand{\arraystretch}{1.2}
\begin{tabular}{c @{$\; \otimesf \; $} c | c | c || l}
\multicolumn{2}{c|}{} & $L$ & $\tilde{L}_{\mbox{\footnotesize max}}$ & Fusion product \\ \hline
${\cal V} (11/8)$ & ${\cal V} (11/8)$ & $5$ & $7$ & ${\cal R}^{(2)} (0,4)_{13}$ \\
& ${\cal V} (27/40)$ & $5$ & $7$ & ${\cal R}^{(2)} (-1/5,14/5)_{11}$ \\
& ${\cal V} (2/5)$ & $5$ & $7$ & ${\cal V} (-9/40)$ \\
& ${\cal V} (7/40)$ & $5$ & $7$ & ${\cal R}^{(2)} (-1/5,9/5)_9$ \\
& ${\cal V} (-1/8)$ & $7$ & $8$ & ${\cal R}^{(2)} (0,1)_7$ \\
& ${\cal V} (-9/40)$ & $4$ & $6$ & ${\cal R}^{(2)} (2/5,2/5)$ \\ \hline
${\cal V} (27/40)$ & ${\cal V} (27/40)$ & $5$ & $7$ & ${\cal R}^{(2)} (0,4)_{13} \oplus {\cal R}^{(2)} (-1/5,9/5)_9$ \\
& ${\cal V} (2/5)$ & $5$ & $7$ & ${\cal R}^{(2)} (-1/8,-1/8)$ \\
& ${\cal V} (7/40)$ & $5$ & $7$ & ${\cal R}^{(2)} (0,1)_{7} \oplus {\cal R}^{(2)} (-1/5,14/5)_{11}$ \\
& ${\cal V} (-1/8)$ & $5$ & $7$ & $ {\cal R}^{(2)} (-1/5,9/5)_9 \oplus {\cal R}^{(2)} (2/5,2/5)$ \\
& ${\cal V} (-9/40)$ & $4$ & $6$ & ${\cal R}^{(3)} (0,0,1,1)$ \\ \hline
${\cal V} (2/5)$ & ${\cal V} (11/8)$ & $5$ & $7$ & ${\cal V} (-9/40)$ \\
& ${\cal V} (27/40)$ & $5$ & $7$ & ${\cal R}^{(2)} (-1/8,-1/8)$ \\
& ${\cal V} (2/5)$ & $5$ & $7$ & ${\cal R}^{(2)} (0,4)_{7} \oplus {\cal R}^{(2)} (-1/5,9/5)_{11} \oplus {\cal V} (2/5)$ \\ 
& ${\cal V} (7/40)$ & $4$ & $6$ & ${\cal R}^{(2)} (7/40,7/40) \oplus {\cal V} (-9/40)$ \\
& ${\cal V} (-1/8)$ & $4$ & $6$ & ${\cal R}^{(2)} (-1/8,-1/8) \oplus {\cal R}^{(2)} (27/40,27/40)$ \\
& ${\cal V} (-9/40)$ & $3$ & $5$ & ${\cal R}^{(2)} (7/40,7/40) \oplus {\cal R}^{(2)} (11/8,11/8) \oplus {\cal V} (-9/40)$ \\
& ${\cal R}^{(1)} (0)_1$ & $5$ & $7$ & ${\cal R}^{(2)} (0,1)_{13} \oplus {\cal R}^{(2)} (-1/5,14/5)_{9}$ \\
& ${\cal V} (1)$ & $3$ & $9$ & ${\cal R}^{(2)} (0,1)_{13} \oplus {\cal R}^{(2)} (-1/5,14/5)_{9}$ \\ \hline
${\cal V} (7/40)$ & ${\cal V} (7/40)$ & $4$ & $6$ & ${\cal R}^{(2)} (0,4)_{13} \oplus {\cal R}^{(2)} (-1/5,9/5)_{9} \oplus {\cal R}^{(2)} (2/5,2/5)$ \\
& ${\cal V} (-1/8)$ & $4$ & $6$ & $ {\cal R}^{(3)} (0,0,1,1) \oplus {\cal R}^{(2)} (-1/5,14/5)_{11}$ \\
& ${\cal V} (-9/40)$ & $4$ & $6$ & ${\cal R}^{(3)} (-1/5,-1/5,9/5,9/5) \oplus {\cal R}^{(2)} (2/5,2/5)$ \\ \hline
${\cal V} (-1/8)$ & ${\cal V} (-1/8)$ & $3$ & $5$ & $ {\cal R}^{(3)} (-1/5,-1/5,9/5,9/5) \oplus {\cal R}^{(2)} (0,4)_{13} $ \\
\multicolumn{2}{c|}{} &&& $\qquad \oplus {\cal R}^{(2)} (2/5,2/5)$ \\
& ${\cal V} (-9/40)$ & $4$ & $6$ & ${\cal R}^{(3)} (0,0,1,1) \oplus {\cal R}^{(3)} (-1/5,-1/5,14/5,14/5)$ \\ \hline
${\cal V} (-9/40)$ & ${\cal V} (-9/40)$ & $4$ & $5$ & ${\cal R}^{(3)} (0,0,4,4) \oplus {\cal R}^{(3)} (-1/5,-1/5,9/5,9/5)$ \\
\multicolumn{2}{c|}{} &&& $\qquad \oplus {\cal R}^{(2)} (2/5,2/5)$ \\ \hline
${\cal V} (14/5)$ & ${\cal V} (14/5)$ & $2$ & $8$ & ${\cal R}^{(1)} (0)_4 \oplus {\cal R}^{(1)} (-1/5)_2$ \\
& ${\cal R}^{(1)} (-1/5)_2$ & $2$ & $8$ & ${\cal V} (1) \oplus {\cal V} (14/5)$ \\
& ${\cal R}^{(1)} (-1/5)_3$ & $2$ & $8$ & ${\cal V} (4) \oplus {\cal V} (9/5)$ \\ \hline
${\cal V} (4)$ & ${\cal V} (4)$ & $2$ & $9$ & ${\cal R}^{(1)} (0)_4$ \\
& ${\cal R}^{(1)} (0)_1$ & $2$ & $9$ & ${\cal V} (1)$ \\
& ${\cal R}^{(1)} (0)_4$ & $2$ & $9$ & ${\cal V} (4)$ \\ \hline
\end{tabular}

\end{appendix}

%%%%%%%%%%%%%%%%%%%%%%%%%%%%%%%%%%%%%%%%%%%%%

\bibliographystyle{JHEP}
\bibliography{holgerJHEP}

%%%%%%%%%%%%%%%%%%%%%%%%%%%%%%%%%%%%%%%%%%%%%
\end{document}